\documentclass[%
 reprint,
 amsmath,amssymb,
 aps, 
prx,
]{revtex4-2}

\usepackage[version=4]{mhchem}
\usepackage{mathtools}
\usepackage{csquotes}
\usepackage{graphicx}
\usepackage{float}
\usepackage{dcolumn}
\usepackage{bm}
\usepackage{color}
\usepackage[table,xcdraw,dvipsnames]{xcolor}
\usepackage{multirow}
\usepackage{array}
\usepackage{booktabs}
\usepackage{siunitx}[=v2]
\usepackage[breaklinks]{hyperref}

\hypersetup{colorlinks=true, linkcolor=blue, citecolor=blue, filecolor=blue, urlcolor=blue}

\DeclareSIUnit{\angstrom}{\textup{\AA}}

\DeclareSIUnit\atom{atom}
\renewcommand{\Re}{\operatorname{Re}}

\DeclareMathOperator*{\argmax}{arg\,max}
\DeclareMathOperator*{\Tr}{Tr}

\DeclarePairedDelimiter{\pqty}{(}{)}
\DeclarePairedDelimiter{\bqty}{[}{]}

\DeclarePairedDelimiter{\Bqty}{\lbrace}{\rbrace}
\DeclarePairedDelimiterX{\braket}[2]{\langle}{\rangle}{#1\,\delimsize\vert\,\mathopen{}#2}
\DeclarePairedDelimiterX{\mel}[3]{\langle}{\rangle}{#1\,\delimsize\vert\,\mathopen{}#2\,\delimsize\vert\,\mathopen{}#3}
\DeclarePairedDelimiter{\norm}{\|}{\|}

\begin{document}

\preprint{APS/123-QED}

\title{Ab-initio heat transport in defect-laden quasi-1D systems from a symmetry-adapted perspective}

\author{Yu-Jie Cen}
\affiliation{Institute of Materials Chemistry, TU Wien, A-1060 Vienna, Austria}

\author{Sandro Wieser}
\affiliation{Institute of Materials Chemistry, TU Wien, A-1060 Vienna, Austria}

\author{Georg K. H. Madsen}
\affiliation{Institute of Materials Chemistry, TU Wien, A-1060 Vienna, Austria}

\author{Jesús Carrete}
\email{jcarrete@unizar.es}
\affiliation{Instituto de Nanociencia y Materiales de Aragón, CSIC-Universidad de Zaragoza, Zaragoza, Spain}

\begin{abstract}
Due to their aspect ratio and wide range of thermal conductivities, nanotubes hold significant promise as heat-management nanocomponents. Their practical use is, however, often limited by thermal resistance introduced by structural defects or material interfaces. An intriguing question is the role that structural symmetry plays in thermal transport through those defect-laden sections. To address this, we develop a framework that combines representation theory with the mode-resolved Green's function method, enabling a detailed, symmetry-resolved analysis of phonon transmission through defected segments of quasi-1D systems. To avoid artifacts inherent to formalisms developed for bulk 3D systems, we base our analysis on line groups, the appropriate description of the symmetries of quasi-1D structures. This categorization introduces additional quantum numbers that partition the phonon branches into smaller, symmetry-distinct subsets, enabling clearer mode classification. We employ an Allegro-based machine learning potential to obtain the force constants and phonons with near-ab-initio accuracy. We  calculate detailed phonon transmission profiles  for single- and multi-layer \ce{MoS2}-\ce{WS2} nanotubes and connect the transmission probability of each mode to structural symmetry. Surprisingly, we find that pronounced symmetry breaking can suppress scattering by relaxing selection rules and opening additional transmission channels. Molecular dynamics shows that the behavior persists even when anharmonicity is considered. The fact that higher disorder introduced through defects can enhance thermal transport, and not just suppress it, demonstrates the critical role of symmetry in deciphering the nuances of nanoscale thermal transport.
\end{abstract}

\maketitle


\section{Introduction}

Transition metal dichalcogenide (TMD) nanotubes have been extensively studied due to their unique physical properties. Their widely tunable band gap makes them highly suitable for optoelectronic devices, field-effect transistors \cite{Ghorbani-Asl2013, 10.1063/1.4894440}, and water-splitting photocatalysts \cite{Dyachkov2019, Piskunov2019}. In recent years, with the development of experimental techniques, heterostructures of TMD nanotubes have been synthesized successfully \cite{Yomogida2023}, showing multi-layer TMD nanotubes can exist as stable systems.

Quasi-1D materials are natural candidates for applications in heat dissipation nanodevices \cite{PhysRevLett.87.215502, PhysRevLett.95.065502}, for thermoelectric energy conversion \cite{Blackburn2018}, or in various other electronic devices \cite{Yadgarov2024}. Many relevant applications involve heat transfer, which makes it highly relevant to study the origins of the underlying transport processes.

Many nanotube systems can reach very high thermal conductivity values, sometimes even exceeding the values observed in the 2D monolayer counterparts, as has been shown for WS$_2$ nanotubes \cite{Wan2021} or, more extensively, for carbon nanotubes \cite{Kumanek2019}. Conversely, structural modifications such as specific chiralities, the length and thickness of the wire, and the introduction of defects, interfaces or other imperfections can drastically limit thermal transport channels. Experiment alone often faces confounding factors like different nanotube orientations or packing conditions, making it challenging to pinpoint the origin of differences in the observed properties \cite{Kumanek2019}. Computational studies can help by enabling the study of those important heat transport roadblocks in much more detail. Most simulations have tackled the influence of defects in the context of molecular dynamics (MD) \cite{Wan2021, Ohnishi2017, Ren2010}, which is usually costly to conduct at scale and often not easy to connect to the physics of thermal carriers. More comprehensive insights can be gained at a cheaper computational cost by specifically analyzing lattice-dynamical properties \cite{PhysRevB.74.125402, PhysRevB.86.235304, PhysRevB.83.064303}, which will be our primary focus in the following. 

The calculation of thermal resistances across defect-laden regions is an area with a long development history starting with simple continuum approximations for interfaces (the acoustic and diffuse mismatch models) and progressing towards predictive strategies based on an atomistic picture in the context of molecular or lattice dynamics \cite{RevModPhys.94.025002}. Among the latter, the atomistic Green's function (AGF) method is a powerful tool to study elastic phonon scattering by such breakdowns of periodicity. The first practical implementations of the methods targeted only the total energy transmission coefficient across defects \cite{PhysRevB.68.245406}. More recently, using the concept of the Bloch matrix, Ong et al. developed the mode-resolved phonon transmission that can discriminate between the contributions of individual phonon modes \cite{PhysRevB.91.174302, 10.1063/1.5048234}. An extension of the method has also been proposed that incorporates anharmonicity into the AGF by adding a many-body self-energy term in the calculations \cite{PhysRevB.101.041301}.

Quasi-1D systems pose specific difficulties for the applications of the AGF method. Common lattice-dynamics workflows, designed and developed for 3D systems, only take advantage of translational symmetry to classify the vibrational modes; thus, when applied to a nanotube they yield a very large number of uncategorized phonon bands \cite{C9CP00052F}. Likewise, they only partially account for the continuous symmetries of free space and therefore introduce significant artifacts into those bands, rendering the results ill-suited for subsequent calculations \cite{C9CP00052F, rotational_2016}. Finally, those workflows typically involve symmetry detectors developed for fully periodic systems, which cannot identify rotational orders other than $2$, $3$, $4$ and $6$. All of these issues can be avoided by basing the study of vibrations in quasi-1D systems on line groups \cite{damnjanovic2010line}, i.e., the products of a single generalized translation and a point group, rather than the more usual space groups. In particular, group projection techniques \cite{wigner2012group, M_Damnjanovic_1994, DAMNJANOVIC20151} can provide new quantum numbers that split the set of phonon branches into much smaller subsets, each of which  transforms according to a well defined  irreducible representation (or irrep) of the line group, avoiding cross-contamination artifacts. This classification, based on fundamental physical features that would otherwise stay hidden, can be carried over to other quantities derived from those vibrational properties.

For nanotubes, and especially multi-layer nanotubes, the large number of atoms in the structure is another limitation that restricts the use of workflows based on first principles. Machine-learning interatomic potentials (MLIPs) provide a powerful tool for overcoming those limitations \cite{Carrete_Nanoscale19}. Some of the cutting-edge MLIPs in terms of economy of training data and transferability are MACE \cite{Batatia2022mace}, NequIP \cite{Batzner2022} and Allegro \cite{Musaelian2023}, all using equivariant neural networks. The lowered cost of evaluating energies and forces using those potentials makes it possible to obtain relatively long MD trajectories using the same dynamical model as in the AGF method, and thus assess the effects of anharmonicity and temperature on the thermal transport coefficients.

In this paper, we train a machine-learning potential for multi-layer \ce{MoS2}-\ce{WS2} nanotubes and use it to obtain their interatomic force constants (IFCs) in pristine and defect-laden configurations. We then use the mode-resolved AGF method to calculate the phonon transmission across those defects and the associated thermal conductance. We enforce all continuum and line-group symmetries on the IFCs to remove any artifacts stemming from their violation and employ symmetry-adapted bases in our calculations to keep track of which irreps are involved in scattering processes and relate those observations to symmetry breakdowns. We show that less symmetric defects, even though intuitively introducing more disorder, can in fact strengthen thermal transport by opening up previously forbidden channels for phonons. We then carry out finite-temperature MD calculations to check that the systems retain their symmetry and that anharmonicity does not change our qualitative conclusions about the correlation between symmetry and thermal transport.

\section{Methods}

\subsection{Construction of \ce{MoS2}-\ce{WS2} nanotube structures}
\label{subsec:construction}

The starting configuration of \ce{MoS2}-type nanotubes can be obtained by rolling up a strip of the corresponding 2D hexagonal lattice. That strip is defined \cite{C9CP00052F} by two chiral indices $\pqty*{n,m}$ that also determine the line group of the structure, and we carefully constrain the resulting coordinates so that it indeed belongs to the desired line group \cite{nano13192699, damnjanovic2010line}. To further improve the quality of the result, we require that the distances between each chalcogenide atom and its three nearest-neighbor Mo atoms be equal to the bond length in the 2D monolayer structure. Since we apply this constraint directly to the symcell (the smallest unit that can generate the unit cell, and therefore the whole structure, through symmetry operations), it does not change the line group of the final structure. The starting configurations of multi-layer \ce{MoS2}-\ce{WS2} nanotubes are built by combining single-wall nanotubes of \ce{MoS2} and \ce{WS2} with different diameters. 
Specifically, since the W-S and Mo-S bond lengths in the 2D monolayer structures are very similar (approximately \cite{Curtarolo2012,Taylor2014} \SI{2.43}{\angstrom}), we use the \ce{MoS2} lattice vectors to build the initial structure of double-walled nanotubes, after which we relax the atomic positions using the software and parameters discussed in the next subsection. The code for constructing nanotubes is available as part of Pulgon-tools \cite{PulgonTools}. The effect of the relaxation is shown in the supplementary information.

\subsection{Density functional theory (DFT) calculations}

We use the projector-augmented-wave formalism \cite{PhysRevB.50.17953} as implemented in the 6.4.0 release of VASP \cite{PhysRevB.54.11169, KRESSE199615, PhysRevB.59.1758}, together with the Perdew–Burke–Ernzerhof approximation to the exchange and correlation energy \cite{PhysRevLett.77.3865} and the DFT-D3 method with a Becke-Johnson damping function \cite{10.1002/jcc.21759} to approximately account for van der Waals interactions. We choose $4p^{6}5s^{1}4d^{5}$, $6s^{2}5d^{4}$ and $3s^{2}3p^{4}$ as valence configurations for Mo, W and S respectively, a plane-wave cutoff energy of \SI{500}{\eV}, and $1\times 1\times 8$, $1\times 1\times 2$ and $2\times 2\times 1$ Monkhorst-Pack k-point grids for single-walled nanotubes, double-walled nanotubes and 2D structures respectively. We moreover set an electronic convergence criterion of \SI{e-8}{\eV} for all VASP runs. For structural relaxations, we use a threshold of \SI{e-2}{\eV\per\angstrom} for the forces.

\subsection{Machine-learning interatomic potential}

Our MLIP is based on Allegro \cite{Musaelian2023}. We use a force field with two hidden layers and take a cutoff radius $r_\mathrm{cut}=\SI{7}{\angstrom}$. For the training we choose an Adam \cite{ADAM} optimizer with a learning rate of \num{5e-4}. which we run for $2500$ epochs. The dimensions of the two-body latent multi-layer perceptron (MLP) are set to $\left[128, 256, 512\right]$. The latent and output MLPs contain a single layer each, with widths of $512$ and $32$ respectively. All three perceptrons use the Swish-1 activation function \cite{swish}.

\begin{figure}[htb!]
    \centering
    \includegraphics[width=1\linewidth]{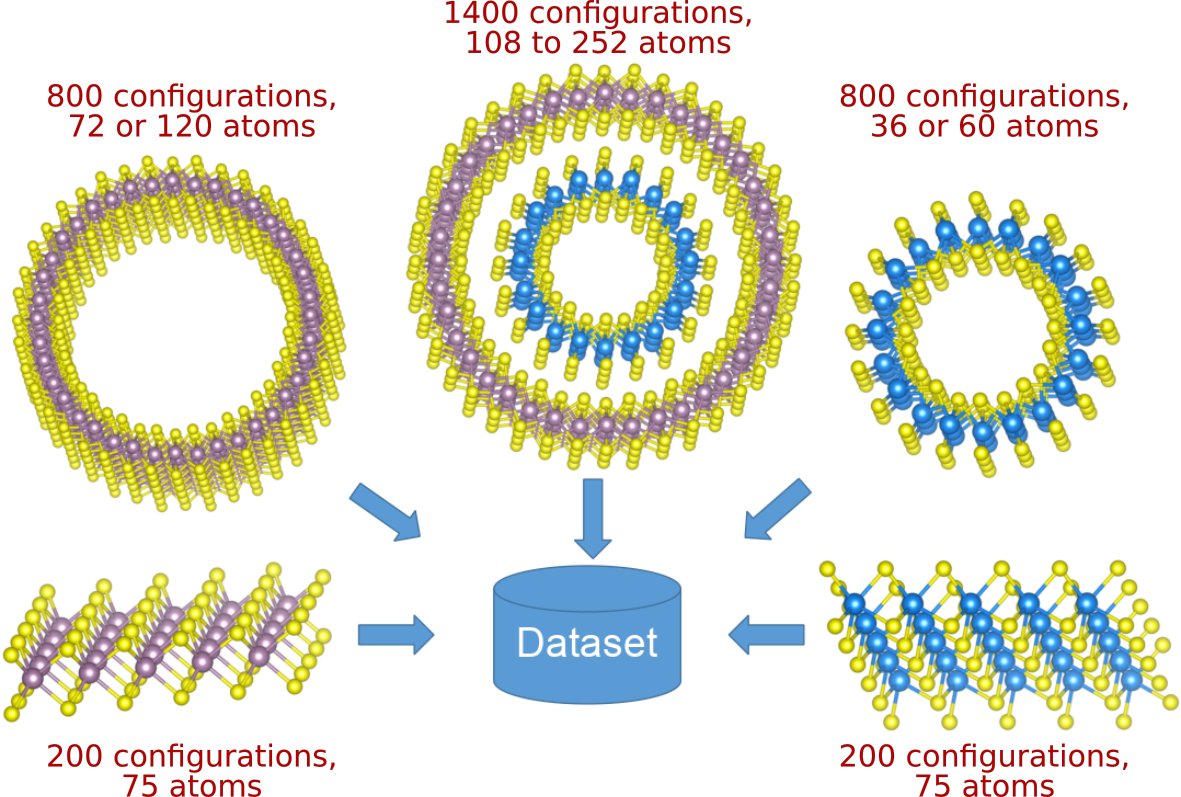}
    \caption{Structures, numbers of atoms, and numbers of configurations included in the ab-initio dataset.}
    \label{fig:datasets}
\end{figure}

The first-principles dataset used for the development of our MLIP contains configurations of the 2D monolayers as well as single- and double-wall nanotubes relaxed via DFT. Each structure is generated by starting from the ground-state configuration and applying random atomic displacements following Gaussian distributions with standard deviations of \SIrange{3e-2}{5e-2}{\angstrom}. Those displaced structures are then used in single-point DFT calculations to obtain their energies and the forces on each atom, which are added to the dataset. From a total of $3400$ configurations, we allocate $3000$ to the training set and $400$ to the validation set. The detailed breakdown is given in Fig.~\ref{fig:datasets}.

\subsection{Interatomic force constants}\label{sec:IFCs}

To obtain the second-order IFCs, i.e., the mass-reduced second-order derivatives of the potential energy that determine the vibrational modes of our structures, we first generate the original single- or double-walled nanotubes using the methods described in \ref{subsec:construction} and relax them using the trained MLIP. Secondly, we create a minimal set of displaced supercell structures using Phonopy \cite{phonopy-phono3py-JPCM,phonopy-phono3py-JPSJ}. Finally, we calculate the forces in those configurations with the trained MLIP and reconstruct the IFC tensor with Phonopy.

The second-order approximation to the potential energy landscape of a system defined by its harmonic IFCs cannot violate the conservation of linear and angular momentum. This imposes several sets of linear constraints on the IFCs, known as the acoustic sum rules (ASRs). For fully periodic 3D systems, generally only the translational sum rules need to be considered: their violation leads to unphysical nonzero frequencies for the acoustic phonon branches at the $\Gamma$ point. However, for subperiodic systems like multilayers and nanotubes the Born-Huang rotational sum rules \cite{10.1119/1.1934059} are also critical \cite{rotational_2016, Lin2022}. Specifically, in quasi-1D systems they ensure the presence of four acoustic branches (instead of three) and the quadratic character of two of those near $\Gamma$. There are several published strategies to enforce the ASRs. Some take the form of postprocessing steps, including a modified harmonic expansion in terms of explicitly scalar coordinates \cite{rotational_2016}, a projection of the \enquote{raw} harmonic IFCs onto an ASR-compliant reduced space (as implemented in Quantum ESPRESSO \cite{Lin2022}), and the addition of a correction term to the raw IFCs, minimized subject to the symmetry constraints (a strategy used by hiPhive \cite{Eriksson2019}). However, it is also possible to directly include the constraints as part of the set of linear equations that is solved to obtain the IFCs of all orders, as done by ALAMODE \cite{Tadano2014}. We implement our own version of the postprocessing approach, focusing on performance for large sets of second-order IFCs with many constraints, and on compatibility with the additional constraint of short range for the IFCs that the AGF formalism requires.

Specifically, we apply a post-processing step to the raw force constants to enforce the full set of constraints while deviating as little as possible from the original result. We do so by solving a linearly constrained quadratic optimization problem:

\begin{gather}
    \bm{H}_{\mathrm{corrected}}=\argmax_{\bm{H}_{\mathrm{corrected}}}\Bqty*{\norm*{\bm{H}_\mathrm{corrected} - \bm{H}_\mathrm{raw}}^2} \label{eqn:quadratic} \\\nonumber
    \bm{C} \cdot \bm{H}_{\mathrm{corrected}} = 0
\end{gather}

\noindent Here, $\bm{H}$ is the matrix of harmonic IFCs, with elements $H_{ij}^{\alpha\beta}$ (Latin subscripts denoting atom numbers and Greek superscripts representing Cartesian directions), and $\bm{C}$ is the matrix containing the coefficients of all the linear constraints. The list of constraints we impose comprises the basic matrix transposition symmetry ($\Phi_{ij}^{\alpha\beta} = \Phi_{ji}^{\beta \alpha}$), the usual acoustic translational sum rules, and the Born-Huang rotational rules \cite{10.1119/1.1934059}. On top of these physically motivated constraints, we add a second set of linear equations to ensure that the IFCs are short-sighted and thus conform to the assumptions of the AGF workflow that we implement \cite{10.1063/1.5048234}. In particular, this requires that the system be divided in translationally equivalent blocks along its periodic direction and that the blocks are long enough that only first-neighbor-block interactions need to be taken into account. We thus require that $\Phi_{ij}^{\alpha\beta}=0$ if atoms $i$ and $j$ are in different, non-adjacent blocks and that the IFC submatrix connecting blocks $b$ and $b+1$ is equal to that connecting blocks $b-1$ and $b$.

We solve the problem expressed by Eq.~\eqref{eqn:quadratic} using the highly efficient convex-optimization package CVXPY \cite{cvxpy}, which calls the OSQP \cite{osqp} operator-splitting solver. We use the resulting high-quality IFCs in all our calculations.

\subsection{Mode-resolved AGF calculations}

\begin{figure}[htb!]
    \centering
    \includegraphics[width=1\linewidth]{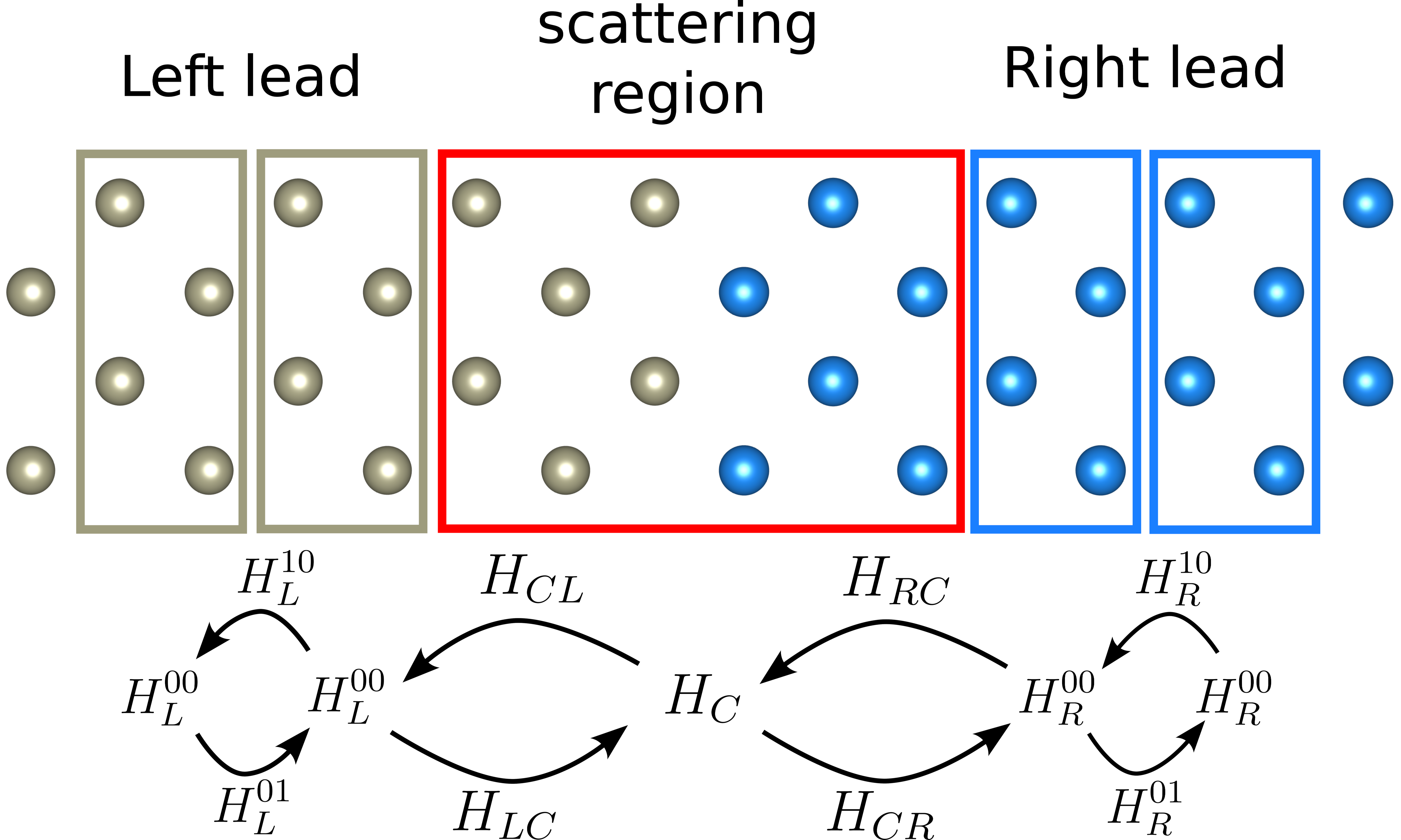}
    \caption{Schematic diagram of AGF's structural partitioning (Left lead, scattering region, and right lead), which determines the shape of the IFC matrices.}
    \label{fig:AGF_region}
\end{figure}

In the traditional AGF method \cite{PhysRevB.68.245406, decimation2}, the quasi-1D system is decomposed into three different parts: the left/right leads -- two semi-infinite regions with discrete translation symmetry -- and a scattering region where the interface is. This is schematically shown in Fig.~\ref{fig:AGF_region}. Accordingly, the IFC matrix $H$ can be divided into a finite set of blocks, $\bm{H}_{L}^{00}$, $\bm{H}_{C}$, $\bm{H}_{R}^{00}$, $\bm{H}_{\mathrm{L}}^{01}$, $\bm{H}_{\mathrm{L}}^{10}$, $\bm{H}_{\mathrm{LC}}$, $\bm{H}_{\mathrm{CL}}$, $\bm{H}_{\mathrm{CR}}$, $\bm{H}_{\mathrm{RC}}$, $\bm{H}_{\mathrm{R}}^{01}$, and $\bm{H}_{\mathrm{R}}^{10}$, which describe the interactions within a region or between different regions. The retarded surface Green's functions of the uncoupled leads are calculated using decimation, an efficient real-space iterative process \cite{decimation1, decimation2}. Those are:

\begin{subequations}
\begin{align}
    \bm{g}_{\alpha,-}^{\mathrm{ret}} &= \bqty*{\pqty*{\omega^2 + i\eta}\bm{1} - \bm{H}_{\alpha}^{00} - \bm{H}_{\alpha}^{10}\bm{g}_{\alpha,-}^{\mathrm{ret}}\bm{H}_{\alpha}^{01}}^{-1}\\
    \bm{g}_{\alpha,+}^{\mathrm{ret}} &= \bqty*{\pqty*{\omega^2 + i\eta}\bm{1} - \bm{H}_{\alpha}^{00} - \bm{H}_{\alpha}^{01}\bm{g}_{\alpha,+}^{\mathrm{ret}}\bm{H}_{\alpha}^{10}}^{-1}
\end{align}
\end{subequations}

\noindent and their Hermitian conjugates are $\bm{g}_{\alpha,-}^{\mathrm{adv}}$ and $\bm{g}_{\alpha,+}^{\mathrm{adv}}$, respectively. Here, $\omega$ is the vibration frequency of phonons, $\alpha$ is either $L$ or $R$ and refers to the left or right lead, and $\mathrm{ret}$/$\mathrm{adv}$ indicates whether this is the retarded or advanced Green's function, and $+/-$ describes whether the semi-infinite lead extends towards the right or the left, respectively. $\eta$ is a small positive number enforcing the choice between $\mathrm{ret}$ and $\mathrm{adv}$, and $\bm{1}$ is an identity matrix of the appropriate size.

The Green's function in the central region is defined as:
\begin{align}
    \bm{G}_{\mathrm{C}}^{\mathrm{ret}}=(\omega^{2}\bm{1} - \bm{H}_{\mathrm{C}} - \Sigma_{\mathrm{L}} - \Sigma_{\mathrm{R}})^{-1},
\end{align}
which involves the two self-energies $\Sigma_{L}=\bm{H}_{\mathrm{CL}}\bm{g}_{L,-}^{\mathrm{ret}}\bm{H}_{\mathrm{LC}}$ and $\Sigma_{R}=\bm{H}_{\mathrm{CR}}\bm{g}_{R,+}^{\mathrm{ret}}\bm{H}_{\mathrm{RC}}$.

The total transmission coefficient of the system can be calculated through the Caroli formula: 

\begin{equation}
    \Xi(\omega) = \Tr[\bm{\Gamma}_{\mathrm{R}}\bm{G}_{\mathrm{C}}^{\mathrm{ret}}\bm{\Gamma}_{\mathrm{L}}(\bm{G}_{\mathrm{C}}^{\mathrm{ret}})^{\dagger}]
     \label{eqn:caroli}
\end{equation}
where $\bm{\Gamma}_{\mathrm{L}}=i\bm{H}_{\mathrm{L}}^{10}(\bm{g}_{\mathrm{L},-}^{\mathrm{ret}}-\bm{g}_{\mathrm{L},-}^{\mathrm{adv}})\bm{H}_{\mathrm{L}}^{01}$ and $\bm{\Gamma}_{\mathrm{R}}=i\bm{H}_{\mathrm{R}}^{01}(\bm{g}_{\mathrm{R},+}^{\mathrm{ret}}-\bm{g}_{\mathrm{R},+}^{\mathrm{adv}})\bm{H}_{\mathrm{R}}^{10}$. 

The mode-resolved AGF \cite{PhysRevB.91.174302, 10.1063/1.5048234, PhysRevB.97.205306} method improves upon that traditional AGF approach in that it uses the Bloch matrices to model the transmission of individual phonon modes. This enables a detailed analysis of the contributions of specific phonon modes to interfacial thermal transport.

The relevant Bloch matrices are constructed as follows:
\begin{equation}
    \bm{F}_{\alpha}^{\mathrm{adv}/\mathrm{ret}}(-)^{-1} = \bm{g}_{\alpha, -}^{\mathrm{adv}/\mathrm{ret}}\bm{H}_{\alpha}^{01}
\end{equation}

The results are extracted from the solutions of the eigenvalue equations for the Bloch matrices,
\begin{equation}
    \bm{F}_{\alpha}^{\mathrm{adv}/\mathrm{ret}}(-)^{-1}\bm{U}_{\alpha}^{\mathrm{adv}/\mathrm{ret}}(-) = \bm{\Lambda}_{\alpha}^{\mathrm{adv}/\mathrm{ret}}(-)^{-1}\bm{U}_{\alpha}^{\mathrm{adv}/\mathrm{ret}}(-),
\end{equation}
where $\bm{U}_{\alpha}^{\mathrm{adv}/\mathrm{ret}}=(\bm{e}_{1},\bm{e}_{2},...,\bm{e}_{n})$ are the matrices of normalized eigenvectors of $\bm{F}_{\alpha}^{\mathrm{adv}/\mathrm{ret}}$ and $\bm{\Lambda}_{\alpha}^{\mathrm{adv}/\mathrm{ret}}$ represent the corresponding diagonal matrices with their eigenvalues.

The number of rightward-going phonon channels in a perfect lead at frequency $\omega$ is given by:
\begin{equation}
    N_{\alpha} = \mathrm{Tr}[\bm{V}^{\mathrm{adv}}_{\alpha}(-) \bm{\widetilde{V}}^{\mathrm{adv}}_{\alpha}(-)]
\end{equation}

\noindent where $\bm{V}^{\mathrm{adv}/\mathrm{ret}}_{\alpha}$ is the diagonal matrix that contains the group velocities of the eigenvectors in $\bm{U}^{\mathrm{adv}/\mathrm{ret}}_{\alpha}$ as its elements. $\bm{\widetilde{V}}^{\mathrm{adv}/\mathrm{ret}}_{\alpha}$ is also a diagonal matrix, whose elements are either the inverses of the corresponding diagonal elements of $\bm{V}^{\mathrm{adv}/\mathrm{ret}}_{\alpha}$ or zero (when those diagonal elements are also zero).

The mode-resolved transmission matrix from left to right is:
\begin{align}
    \bm{t}_{\mathrm{RL}}=\frac{2i\omega}{\sqrt{a_{\mathrm{R}}a_{\mathrm{L}}}}[\bm{V}_{\mathrm{R}}^{\mathrm{ret}}(+)]^{1/2}[\bm{U}_{\mathrm{R}}^{\mathrm{ret}}(+)]^{-1}\bm{G}_{\mathrm{RL}}^{\mathrm{ret}}\times \\ \notag
    \times[\bm{U}_{\mathrm{L}}^{\mathrm{adv}}(-)^{\dagger}]^{-1}[\bm{V}_{\mathrm{L}}^{\mathrm{adv}}(-)]^{1/2},
\end{align}
involving the coupled retarded Green's function $\bm{G}_{\mathrm{RL}}^{\mathrm{ret}}=[(\omega^{2}+i\eta)\bm{1}-\bm{H}_{\mathrm{L}}]^{-1}\bm{H}_{\mathrm{RC}}\bm{G}_{\mathrm{C}}^{\mathrm{ret}}\bm{H}_{\mathrm{CL}}[(\omega^{2}+i\eta)\bm{1}-\bm{H}_{\mathrm{R}}]^{-1}$.

The transmission coefficient for each phonon channel $n$ is written as:

\begin{align}
    \Xi_{\mathrm{L,n}}(\omega) = [\bm{t}_{\mathrm{RL}}^{\dagger}\bm{t}_{\mathrm{RL}}]_{nn}
    \label{eqn:transcoeff}
\end{align}

\subsection{Symmetry-group-projector method}
For each of the leads, the matrix of harmonic IFCs is block-diagonal on a basis of linear combinations of the atomic displacements adapted to the irreps of the line group. Since we only work with commensurate structures, it is possible to choose a set of irreps with well defined wave numbers ($q$) along the nanotube axis \footnote{Note, that an alternative choice of irreps, using a chiral wave number $\tilde{q}$, is also possible, and in fact applicable even to incommensurate systems. However, working in terms of $q$ enables a more natural adaptation of the methods designed for $2D$ systems.}. Thus, it is enough to focus on block-diagonalizing the dynamical matrix $\bm{D}\pqty*{q}$ at each value of $q$. To obtain the required basis, we first employ our line group symmetry analysis software, Pulgon-tools \cite{PulgonTools}, to identify the line group $\mathrm{L=Z\cdot P}$ and determine the character $\chi^{\beta}$ for each group element $g$. Here, $\mathrm Z$ denotes the generalized translation group, which may take the form of a helical group $\pqty*{\mathrm C_{Q}|f}$ or a glide plane group $\pqty*{\mathrm \sigma_{v}|a/2}$. The axial point group $\mathrm P$ belongs to one of the following seven types: $\mathrm C_{n}$, $\mathrm S_{2n}$, $\mathrm C_{nh}$, $\mathrm D_{n}$, $\mathrm C_{nv}$, $\mathrm D_{nd}$ and $\mathrm D_{nh}$. 

Next, we build the incomplete projection operator: 

\begin{equation}
    \bm{P}^{\beta} = \frac{d_{\beta}}{|G|}\sum_{g}\chi^{\beta}(g)^{*}\bm{S}(g)
\end{equation} 
where $d_{\beta}$ is the degeneracy of the irrep $\beta$, $|G|$ is the order of the symmetry group $G$, and $\chi^{(\beta)}(g)$ is the character of the group element $g$ in irrep $\beta$; $\bm{S}\pqty*{g}$ is the matrix for operation $g$, defined by the transformation matrix of the atomic displacements in the simulation box ($\bm{M}$) and two phase factors:

\begin{equation}
    \bm{S}_{ij}^{\alpha\beta}(g) = \bm{M}_{ij}^{\alpha\beta}(g) \cdot e^{iqT_{z}} \cdot e^{iq\cdot (r_{i}^{z} - r_{j}^{z})}.
\end{equation}
In turn, $\bm{M}$ is constructed by placing copies of the three-dimensional rotation matrix associated to $g$ at the rows and columns defined by the atom permutation induced by said symmetry operation. The first factor, $e^{iqT_{z}}$,  is associated with the translational component of the group operation $g$, and $T_{z}$ is the $z$ component of the translational vector (since we define the periodic direction to be $z$, the $x$ and $y$ components of the translation are always zero for line group operations). For integer translational vectors this term is equal to $1$ and can be ignored; however, in the case of helical symmetry operations in line groups, fractional translation vectors are possible, and the factor may become relevant. The second factor, $e^{iq\cdot \pqty*{r^z_{i} - r^z_{j}}}$, is related to the atomic positions of atom pairs $i$ and $j$ connected by the permutation part of $g$. Note that it is possible to include this phase in the expression of the dynamical matrix instead, in which case it does not need to be added here.

We then extract an orthonormal basis $\pqty*{\bm{b}_1, \bm{b}_2\ldots\bm{b}_{d_\beta}}$ for the column space of the projection operator $\bm{P}^{\beta}$. The symmetry-adapted basis that block-diagonalizes $\bm{D}\pqty*{q}$ is built by concatenating all those bases.

\begin{figure*}[htb!]
    \centering
    \includegraphics[width=1\linewidth]{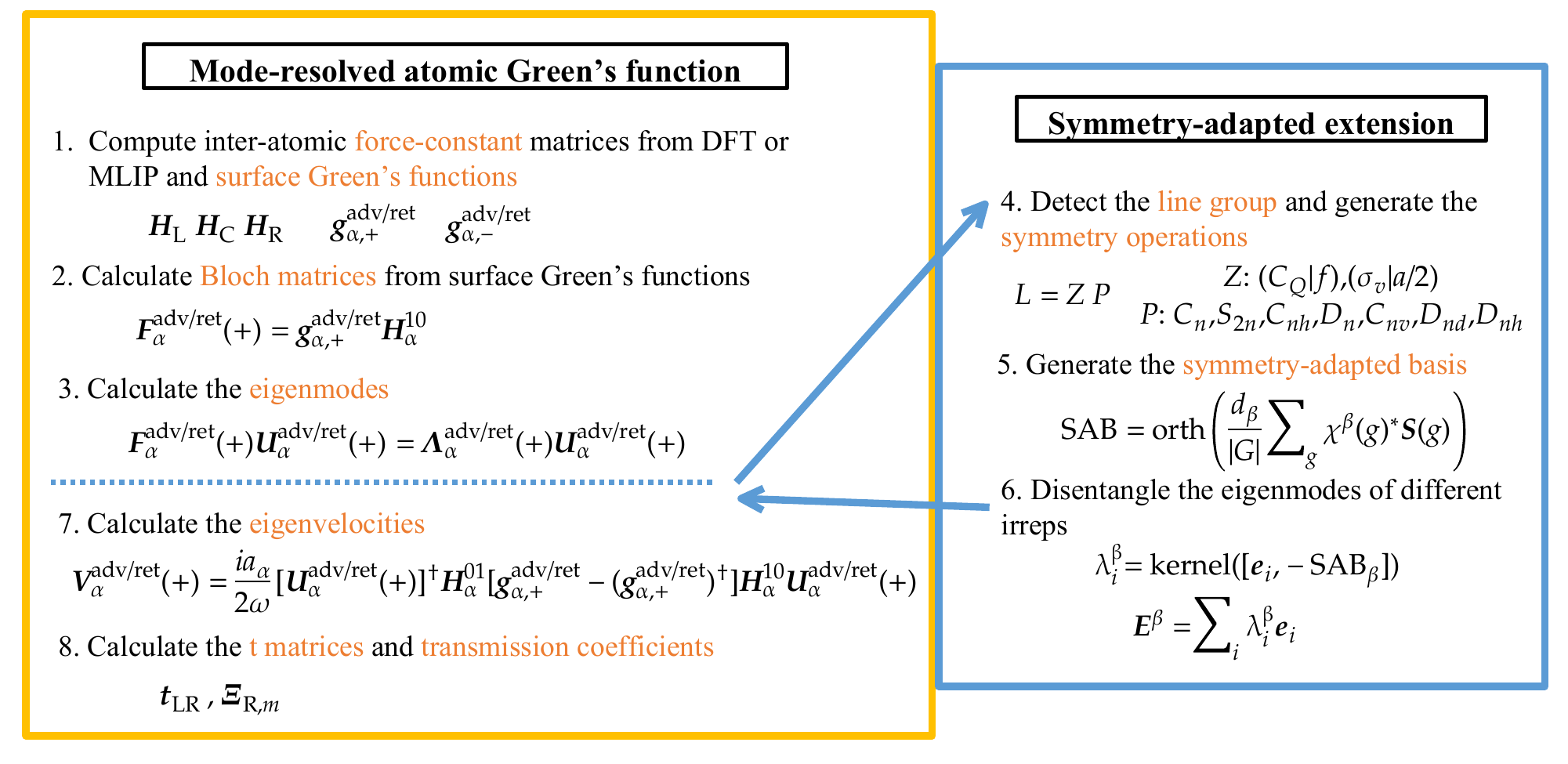}
    \caption{Flowchart illustrating the additional steps in the mode-resolved AGF method.}
    \label{fig:flowchart}
\end{figure*}

\subsection{Symmetry-adapted transmission}
\label{subsec:adapted-trans}

To develop a symmetry-adapted version of AGF we use the symmetry-adapted basis to separate the eigenmodes belonging to each irrep before proceeding to calculating the group-velocity matrices and the mode-resolved transmission $\bm{V}_{\alpha}^{\mathrm{adv}/\mathrm{ret}}$.

From the eigenvalues $\Lambda$ we can obtain the wave number $q_{n}$ for each mode:
\begin{equation}
    q_{n} = \frac{1}{a_{\alpha}}\cos^{-1}\Re[\bm{\Lambda}_{\alpha}^{\mathrm{adv}/\mathrm{ret}}]_{nn},
\end{equation}

\noindent where $a_{\alpha}$ is the cell length of the $\alpha$ ($\mathrm{L}$ or $\mathrm{R}$) lead. Degenerate modes are detected by grouping together numerically equal values of $q_{n}$ within a tolerance. While eigenvectors for non-degenerate modes belong to an irrep and do not require postprocessing, degenerate sets of modes can come out as linear combinations mixing different irreps. Therefore, it is necessary to disentangle them by calculating $\mathrm{kernel}([\bm{e}_{1},...,\bm{e}_{n}, -\bm{b}_{1}, ...,-\bm{b}_{d_{\beta}}])$. The first $n$ components of each element of an orthogonal basis of that kernel serve as the set of coefficients $\Bqty*{\lambda_i^\beta}$ for a linear combination of the $\Bqty*{\bm{e}_i}$. The new eigenvector $\bm{E^{\beta}}=\sum_{i}\lambda_i^\beta \bm{e_{i}}$ lies in the space spanned by the $\Bqty*{\bm{b}_j}$ which correspond to the irrep $\beta$.

A detailed scheme of the symmetry-adapted extension of the mode-resolved AGF technique is shown in Fig.~\ref{fig:flowchart}.

\subsection{Thermal conductance}

The thermal conductance for the material is calculated in the Landauer formalism as \cite{PhysRevLett.81.232, PhysRevB.74.125402}:

\begin{equation}
    \sigma(T) = \frac{1}{2\pi}\int_{0}^{\omega_{max}}\hbar\omega \frac{\partial n(\omega, T)}{\partial T} \Xi(\omega) d\omega.
    \label{eqn:thmocond}
\end{equation}
Here, $n(\omega, T)$ is the Bose-Einstein distribution and $\Xi(\omega)$ is the transmission coefficient.

\subsection{Equilibrium MD simulations}

Equilibrium MD simulations were employed to analyze the finite-temperature configurations and calculate the thermal conductivity of the systems. Specifically, to obtain the thermal conductivity, Green-Kubo simulations were employed at a temperature of \SI{300}{\kelvin} with a time step of \SI{1}{\femto\second}. For each system, 15 independent simulations were performed with different velocity initialization seeds and an equilibration period of \SI{20}{\pico\second} in an NVT ensemble guided by a Langevin thermostat. The heat flux was subsequently computed at every time step during a \SI{200}{\pico\second} period in NVE ensemble. The time integration was carried out by the atomic simulation environment (ASE) \cite{Hjorth_Larsen_2017} and the full heat flux was computed using

\begin{equation}
    \mathbf{J} = \frac{1}{V}\sum_{jk} \mathbf{r}_{kj} \left( \frac{\partial U}{\partial \mathbf{r}_{jk}}  \cdot \mathbf{v}_k \right) + \frac{1}{V}\sum_j E_j\mathbf{v}_j,
    \label{eqn:heat_flux_allegro}
\end{equation}

\noindent where $\mathbf{r}_{kj}$ is the distance vector between atom $j$ and $q$, $\mathbf{v}_k$ is the atomic velocity, $U$ the potential energy of the system, $V$ is the system volume, and $E_j$ is the energy of atom $j$. The derivative in the first term was obtained via automatic differentiation over the Allegro model, which has been suggested as a computationally efficient method to obtain the heat flux for local MLIPs \cite{langer_heat_2023}.

The autocorrelation function of the computed heat flux in the periodic direction is then used to evaluate the thermal conductivity $\kappa$ with its integral using the Green-Kubo formalism,

\begin{equation}
    \mathbf{\kappa} = \frac{V}{k_B T^2} \int_0^\infty dt \left< \mathbf{J}(t)\mathbf{J}(0)\right>.
    \label{eqn:green_kubo}
\end{equation}

However, Green-Kubo integrals can be very noisy and difficult to converge. Thus, we employ cepstral analysis for the evaluation as described in \cite{ercole_accurate_2017,ercole_sportran_2022,wieser_accelerating_2025}. This approach is based on denoising the power spectrum of the heat flux to obtain a more accurate thermal conductivity value at lower simulation times. The full details of our approach and the tools required to implement it can be found in Ref.~\cite{wieser_accelerating_2025}. For this application, a cutoff of the power spectrum has to be chosen and was set to \SI{0.5}{\tera\hertz} after careful consideration. The nanotube volume was computed based on the total volume occupied by the Van-der-Waals spheres around the individual atoms. For more details and convergence tests, see the supplementary information.

To investigate whether the defect systems maintain their symmetry at elevated temperature, we employed MD simulations using analogous settings and computed the average atomic positions during the NVE part of the simulation at \SI{300}{K} and \SI{500}{K}.

\section{Results and discussion}

In this section, we firstly demonstrate the reliability of the MLIP by evaluating the error ranges in its energy and force predictions. Next, we illustrate the significance of irreps and quantum numbers in phonon vibrational modes using a pristine single-walled nanotube as an example, and then we calculate the symmetry-adapted transmission of a pristine double-walled nanotube. Finally, we investigate defect-laden double-walled nanotubes and show that the breaking of symmetry can actually open up more transmission channels and enhance the thermal conductance. We analyze this situation in detail through our symmetry-adapted AGF approach and also check that the qualitative conclusions hold when the effects of temperature (including anharmonicity) are taken into account.

\subsection{Energy and force errors from our MLIP}

\begin{figure}[htb!]
    \centering
    \includegraphics[width=1\linewidth]{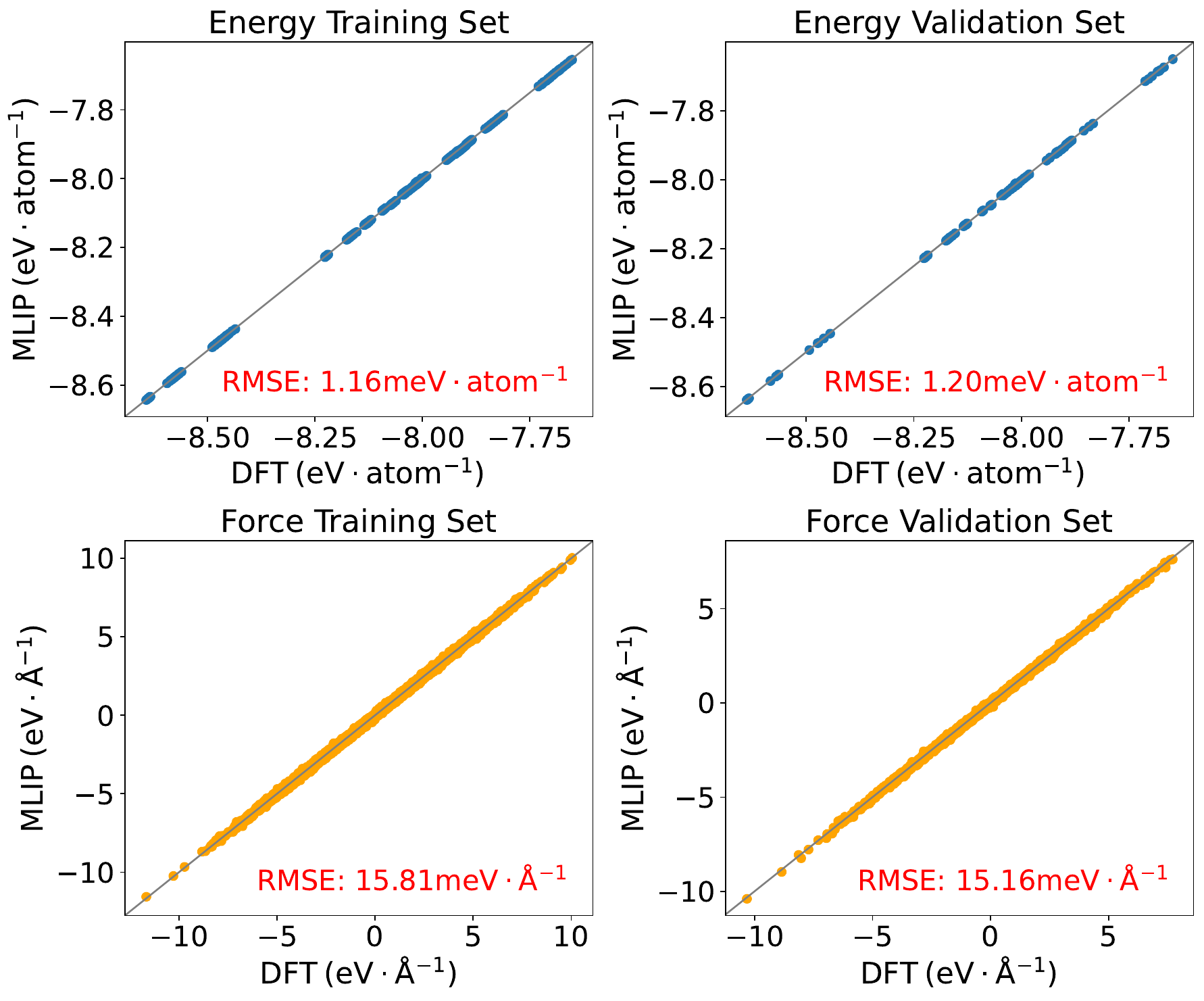}
    \caption{Parity plots for the total energy and the forces over the training and validation sets: MLIP vs DFT ground truth.}
    \label{fig:ref-pred}
\end{figure}

As shown in Fig.~\ref{fig:ref-pred}, the accuracy of the MLIP is reflected in root-mean-square errors (RMSE) of \SI{1.16}{\milli\electronvolt\per\atom} and \SI{15.81}{\milli\electronvolt\per\angstrom} for the training energies and training forces, respectively. Over the validation set, the RMSEs for energies and forces are \SI{1.20}{\milli\electronvolt\per\atom} and \SI{15.16}{\milli\electronvolt\per\angstrom}, which does not suggest any overfitting. Figure \ref{fig:ref-pred} shows a more detailed view, in the form of parity plots for the energies and forces over both sets. Representative errors in the forces are well below, for instance, typical discrepancies between different DFT implementations \cite{Larsen_PRB09}. Thus, those forces are suitable for calculating the IFCs, taking into account the symmetry corrections outlined in Section \ref{sec:IFCs}.

Since we use the same MLIP for our MD simulations, it is also important to assess its precision and accuracy in that setting. To do so, we sampled $50$ states from room-temperature trajectories of representative pristine and defect-laden systems, and calculated the energies and forces for those configurations using both the MLIP and direct DFT calculations. The agreement is still sufficient to evaluate the quality of the MLIP as DFT-like, with only a slight degradation in the prediction of the energies for the defect-laden case, in the form of a mostly configuration-independent offset that may originate in the DFT calculations themselves. The detailed results are presented in the supplementary information.

\subsection{Vibrational modes of a pristine single-walled \ce{WS2} nanotube}
The irrep that each mode belongs to reflects how it changes under each of the symmetry operations, and therefore identifies the nature of the atomic displacements in that mode. Besides the linear quasimomentum $q$, expressing how the phase of the eigenvector changes under linear translations along the nanotube axis, each irrep is also characterized by an integer value of $m$ describing its behavior under rotation. Irreps are further subdivided according to parities $\Pi_{\mathrm{V}}$.

\begin{figure}[htb!]
    \centering
    \includegraphics[width=1\linewidth]{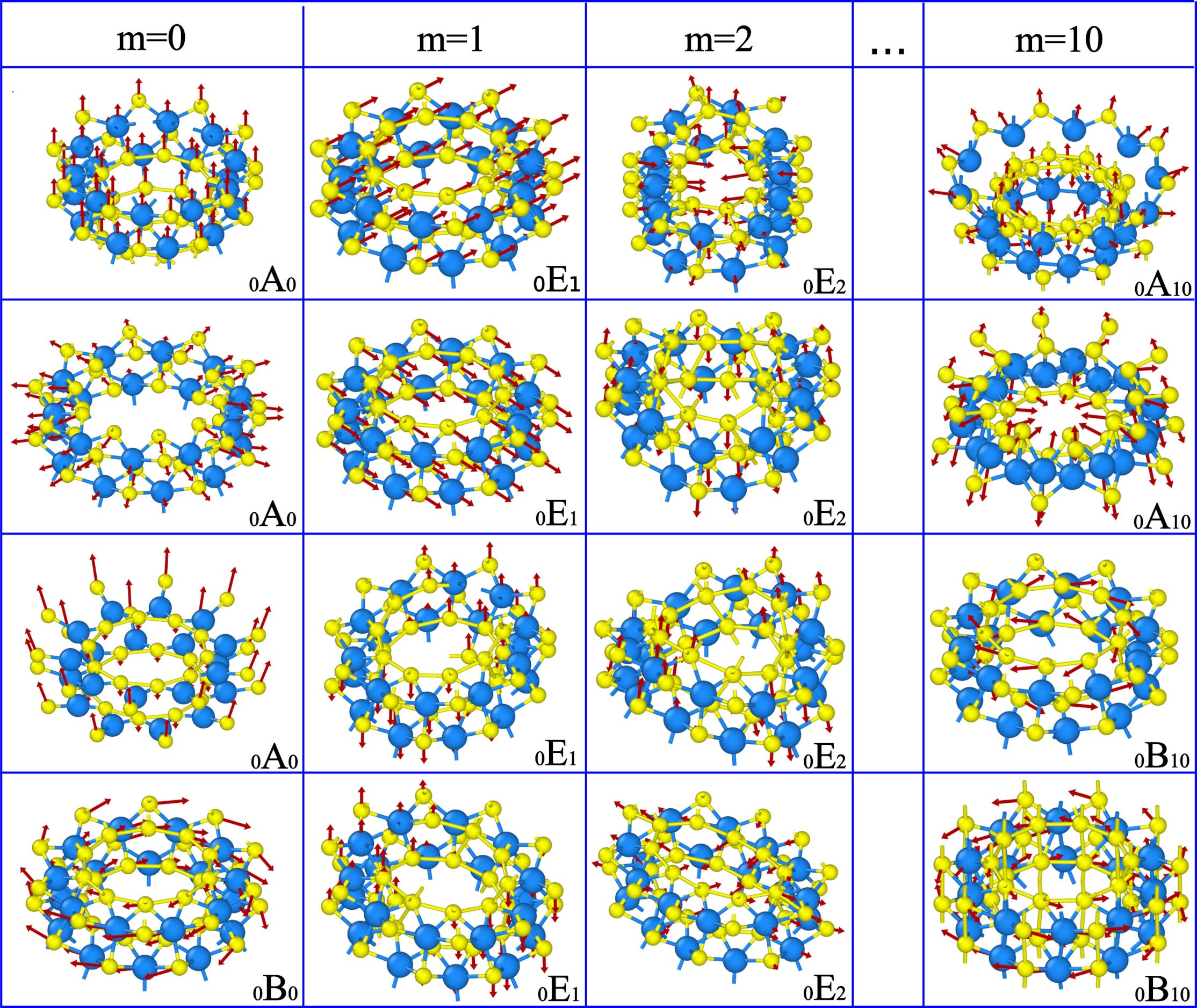}
    \caption{Phonon vibration modes associated to four different values of $m$ at the $\Gamma$ point $\pqty*{q=0}$ of a pristine single-walled \ce{WS2} nanotube.}
    \label{fig:vib_modes}
\end{figure}

To illustrate this more concretely, we consider the phonon modes of a \ce{WS2} nanotube with chirality (10,0), which belongs to the $L2n_{10}mc$ group, from the eighth of the thirteen line group families \cite{damnjanovic2010line}. Focusing on the $\Gamma$ point ($q=0$), we plot a selection of vibrational modes with different $m$ in Fig.~\ref{fig:vib_modes} There are three types of irreps at that point: $\prescript{}{0}{\mathrm{A}}_{0/10}$, $\prescript{}{0}{\mathrm{B}}_{0/10}$, and $\prescript{}{0}{\mathrm{E}}_{m}$ (with $1 \leq m \leq 9$).  Both $\prescript{}{0}{\mathrm{A}}_{0/10}$ and $\prescript{}{0}{\mathrm{B}}_{0/10}$ are non-degenerate one-dimensional irreps; A-type modes have even parity for reflections with respect to a vertical plane (i.e., one containing the $OZ$ axis), whereas B-type modes have odd parity. The first three vibration modes for $m=0$ and the first two for $m=10$ correspond to A-type irreps, combining axial expansion and contraction (breathing) with longitudinal vibrations; the last mode for $m=0$ and the last two for $m=10$ belong to B-type irreps and correspond to shear and torsion modes.  On the other hand, $\prescript{}{0}{\mathrm{E}}_{m}$ is a doubly degenerate two-dimensional irrep that arises in systems with $C_{n}$ rotational symmetry. The degeneracy reflects vibrations along two possible orthogonal directions. All vibrational modes for $1 \leq m \leq 9$ are bidirectional stretching modes belonging to E-type irreps.

As the analysis above exemplifies, the additional quantum numbers assigned to vibrational modes when viewed through the lens of the line group of the system are not merely a mathematical tool to reduce the size of the matrix blocks involved in the workflow: they provide physical insight into the modes they label. Just as $q$ indexes the linear quasimomentum (conjugate to the Cartesian coordinate along the longitudinal axis) by characterizing the transformation under a translation, $m$ indexes the angular quasimomentum of the mode (conjugate to the angular coordinate around the same axis) by characterizing the transformation under a pure rotation. To further highlight this physical meaning, it is worth mentioning that for groups with a helical generator there exists an alternative set of quantum numbers that eschew the usual linear quasimomentum in favor of bands indexed by a helical quasimomentum $\tilde{q}$ and a pure angular quasimomentum $\tilde{m}$ associated only to the operations of the point subgroup \cite{damnjanovic2010line}. For commensurate systems like those treated here, there is a correspondence between the sets of irreps defined by $\pqty*{q, m}$ and those defined by $\pqty*{\tilde{q}, \tilde{m}}$, but incommensurate helical systems only admit a description based on the latter. In those cases, instead of phonons propagating along the nanotube axis with a wave number taken from the usual Brillouin zone, the system's vibrations are conceptualized as \enquote{helical phonons} traveling along helical arcs, and $\tilde{q}$ takes values within a helical Brillouin zone. Note that in this work we always opt for the $\pqty*{q, m}$ description because it integrates better with supercell-based calculations.

\subsection{Spectrum and transmission of a pristine double-walled \ce{WS2}-\ce{MoS2} nanotube}

The translational block of our model double-walled \ce{WS2}-\ce{MoS2} nanotube contains $40$ \ce{Mo}, $20$ \ce{W} and $120$ \ce{S} atoms and has chirality  $(10,0)-(20,0)$. It belongs to the $Lnmm$ line group, part of the sixth family. There is no interface scattering, so the transmission coefficient at a given frequency equals the number of available phonon modes. By employing our symmetry-adapted AGF approach, the phonons can be categorized into different irreps, allowing a detailed analysis of each irrep's contribution to the transmission coefficient.

\begin{figure}
    \centering
    \includegraphics[width=1\linewidth]{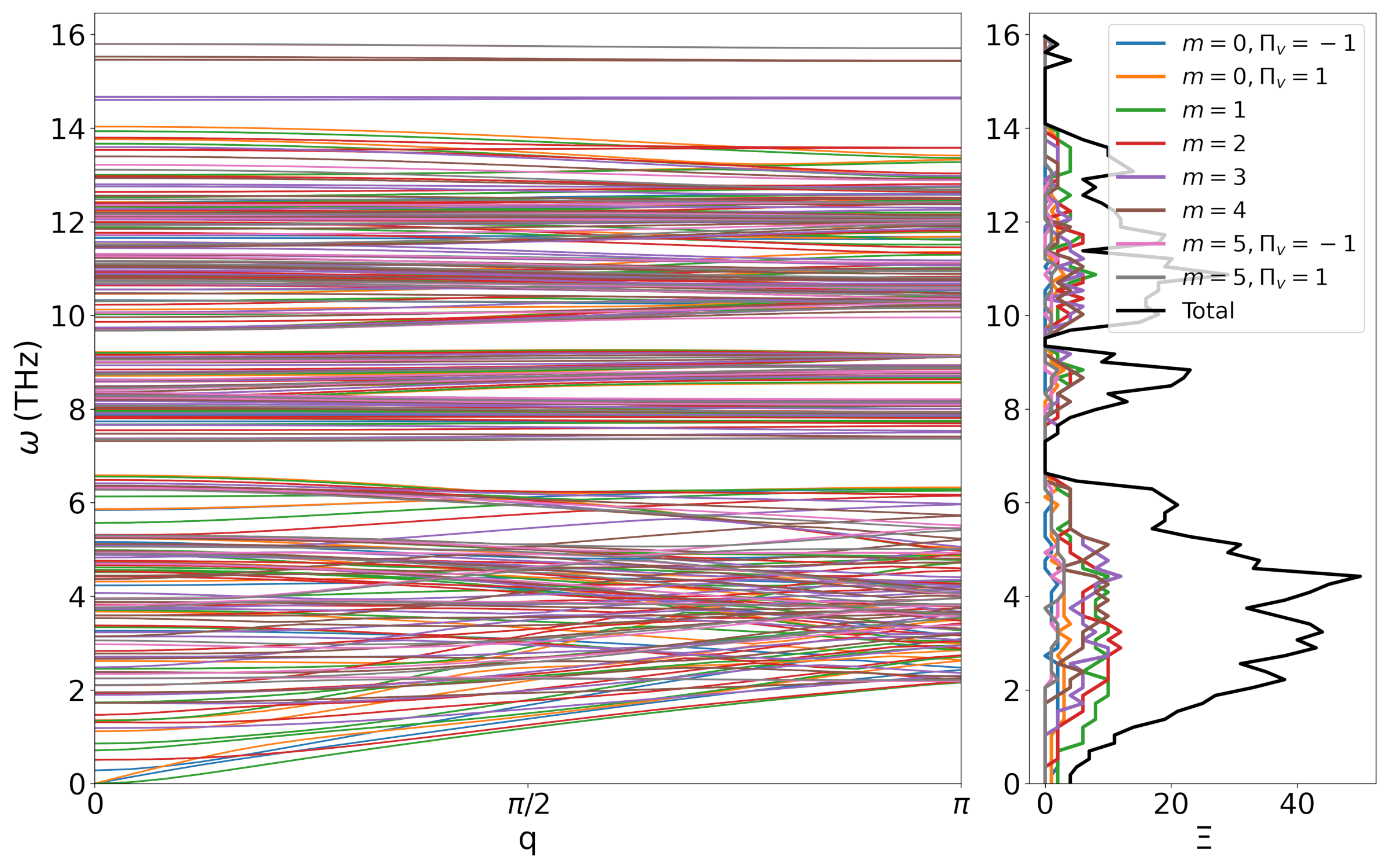}
    \caption{Symmetry-adapted phonon dispersions (left) and transmission (right) for a pristine double-walled  \ce{WS2}-\ce{MoS2} nanotube. Different colors correspond to different irreps, classified by angular quasimomentum $m$ and parity $\Pi_{\mathrm{V}}$. }
    \label{fig:phonon_trans}
\end{figure}

The results are presented in Fig.~\ref{fig:phonon_trans}. There are two degenerate quadratic ZA branches, which belong to $m=1$ and irrep $\prescript{}{q}{\mathrm{E}}_{1}$. The two remaining acoustic branches are linear and correspond to $m=0$. The irrep corresponding to the LA branch (with a larger sound velocity) is $\prescript{}{q}{\mathrm{A}}_{0}$ and the irrep corresponding to the TA branch (with a smaller sound velocity) is $\prescript{}{q}{\mathrm{B}}_{0}$. The presence of four acoustic branches, two of which are quadratic, indicates that the symmetrization of the force constants was performed correctly, since otherwise only three acoustic branches would have emerged from the bulk-like workflow \cite{C9CP00052F}. 

Analogously to how, by breaking translational symmetry in a bulk system, a defect can cause scattering between phonon modes with different wave vectors, the breakdown of the symmetries making up the line group can cause scattering between irreps, potentially involving not only $q$ but also $m$ and the parities if applicable. We explore this issue in the next subsection.

\subsection{Defect-laden double-walled \ce{WS2}-\ce{MoS2} nanotube}

\begin{figure}
    \centering
    \includegraphics[width=1\linewidth]{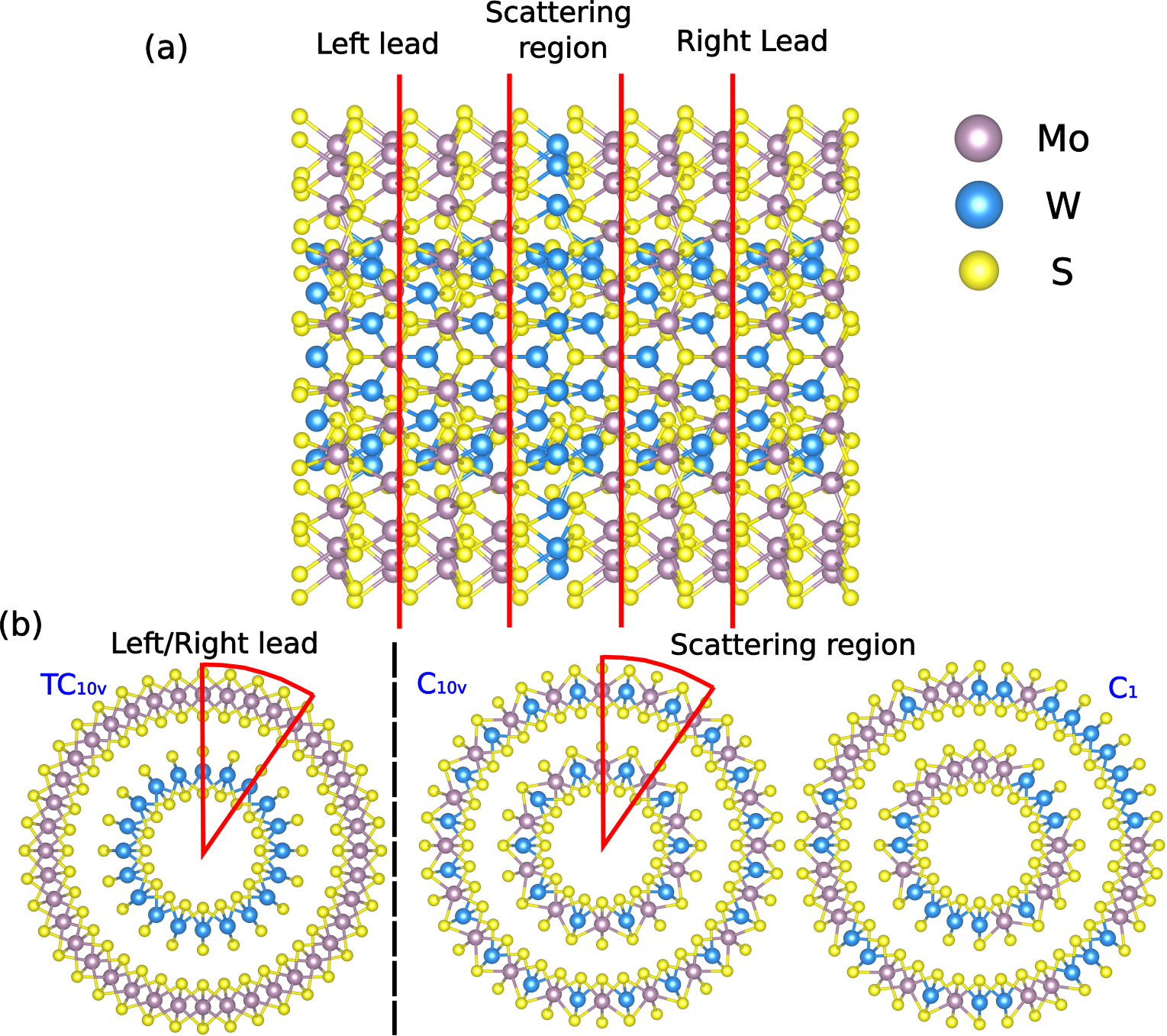}
    \caption{a) Longitudinal view of a defect-laden double-walled nanotube and the segments used in the AGF scattering calculations. b) Cross-sectional views of the pristine leads and of the two defect-laden configurations with different symmetries.}
    \label{fig:strcut}
\end{figure}

For a double-layer \ce{WS2}-\ce{MoS2} nanotube with combination of chiralities (10,0)-(20,0), the line group is $TC_{10v}$. In this application of the symmetry-adapted AGF method, both the left and right leads are composed of the same pristine structure (40 \ce{Mo}, 20 \ce{W} and 120 \ce{S} atoms each slide), while the scattering region contains the defects. To investigate how symmetry influences heat transfer, we compare two defect-laden configurations with the same number of substitutions but different symmetries in the scattering region. As shown in Fig.~\ref{fig:strcut}, we replace 20 Mo atoms with 20 W atoms in the outer-layer nanotube and 10 W atoms with 10 Mo atoms in the inner-layer nanotube. We specifically identify sets of equivalent atoms so as to create a configuration that preserves the $C_{10v}$ symmetry (bottom center). Meanwhile, we generate a completely asymmetric $C_{1}$ configuration (bottom right) by random replacement. To ensure the comparison is affected only by symmetry, we choose exactly the same number of substitutional atoms in each nanotube in both cases. Access links to the corresponding structures are provided in the \enquote{Data availability} section.

The second-order IFC matrices $\bm{H}$ for these two defect configurations are computed employing the MLIP and used as inputs to the AGF. The total transmission coefficient $\Xi(\omega)$ and the thermal conductances $\sigma(T)$ can be calculated by Eq.~\eqref{eqn:caroli} and Eq.~\eqref{eqn:thmocond} respectively.

\begin{figure}[htb!]
    \centering
    \includegraphics[width=1\linewidth]{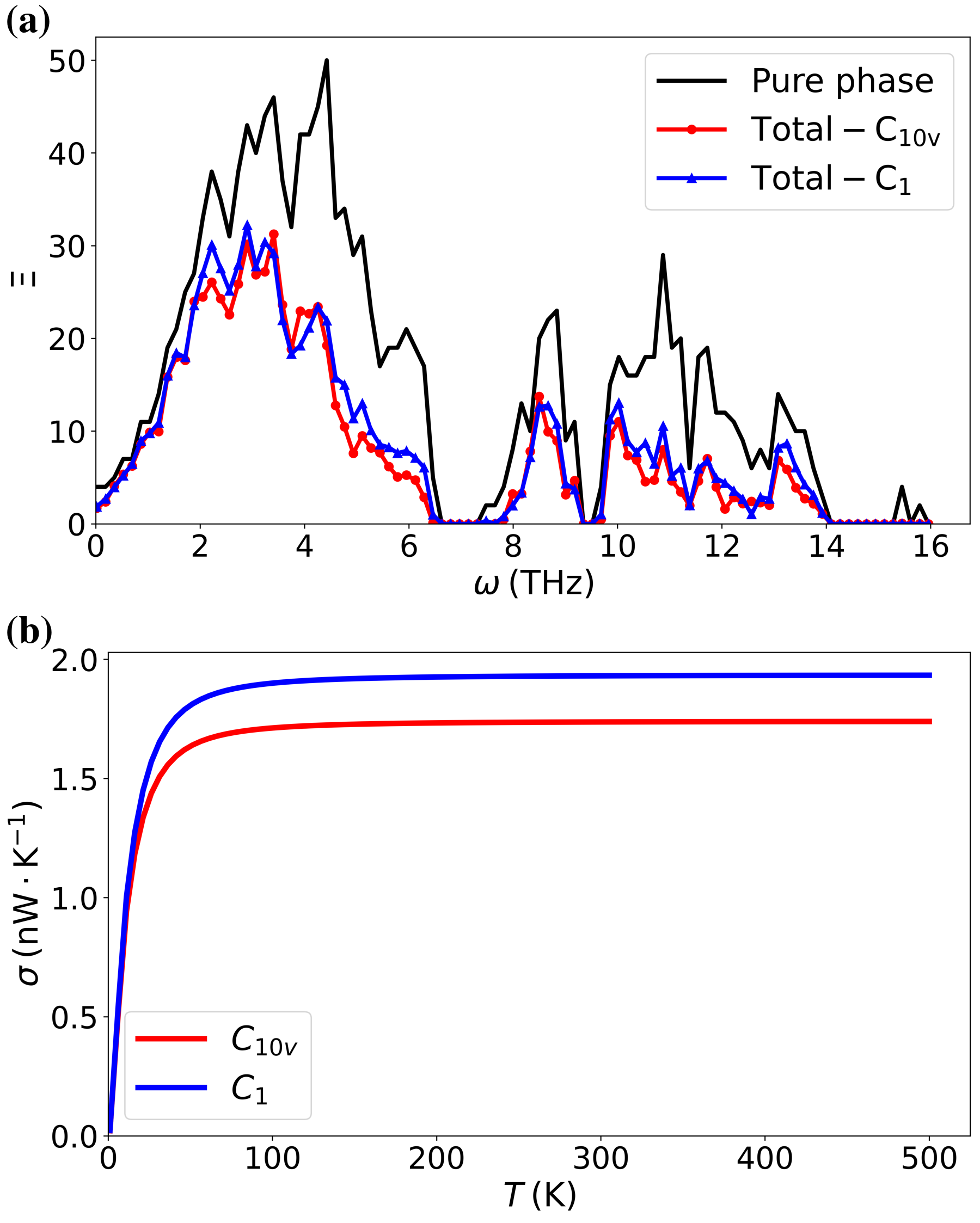}
    \caption{a) Transmission coefficient spectra for (10,0)-(20,0) \ce{WS2}-\ce{MoS2} double-walled nanotubes excluding or including a defect-laden segment with either $C_{10v}$ or $C_{1}$ symmetries. b) Temperature dependence of the thermal conductances for the two defect-laden structures.}
    \label{fig:total_trans}
\end{figure}

As shown in Fig.~\ref{fig:total_trans}(a), we find that the transmission coefficient of the $C_{10v}$ structure is lower than that of the $C_{1}$ structure in most of the frequency range. Using Eq.~\eqref{eqn:thmocond}, the thermal conductance for these two configurations is calculated and presented in Fig.~\ref{fig:total_trans}(b). The results align with the transmission coefficients, with the thermal conductance of the $C_{1}$ also higher than that of the $C_{10v}$ structure. This result is somewhat counterintuitive, as it is generally expected that increased disorder at the interface will lead to a stronger scattering effect and, consequently, lower thermal conductance. The mode-resolved information is key to analyzing this surprising result.

\begin{figure*}[htb!]
    \centering
    \includegraphics[width=1\linewidth]{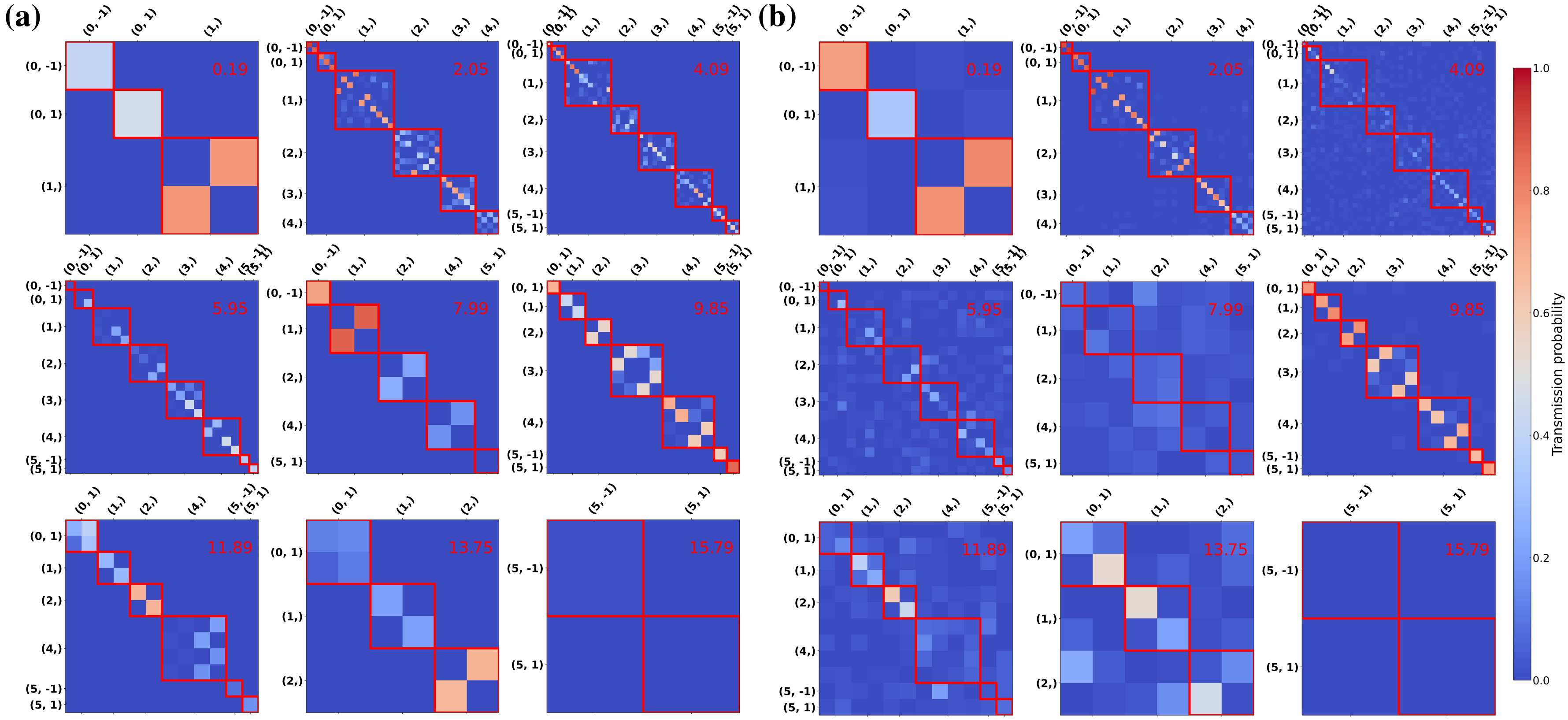}
    \caption{ (a) Mode-to-mode transmission matrix of the structure containing a $C_{10v}$ defect-laden segment. (b) Transmission matrix of the structure containing a $C_{1}$ defect-laden segment. The red number in the upper-right corner of each matrix corresponds to the angular frequency of the incident and scattered phonons, in $\si{\radian\per\pico\second}$. The element $t_{ij}$ in the transmission matrix represent the transmission probability from phonon mode $i$ in the left lead to the mode $j$ in the right lead. The pair $(m, \Pi_{\mathrm{V}})$ on the axes indicates the irrep, and the red square outlines groups together matrix elements corresponding to incident and transmitted modes within the same irrep.}
    \label{fig:trans-matrix}
\end{figure*}

The symmetry-adapted transmission matrix from the left to the right lead can be calculated using Eq.~\eqref{eqn:transcoeff} if we use take the step of adapting the eigenmodes $\bm{U}_{\mathrm{L/R}}^{\mathrm{ret/adv}}$ to the symmetry. The results are shown in Fig.~\ref{fig:trans-matrix}, where each element $i,j$ in $|\bm{t}_{\mathrm{LR}}|^{2}_{ij}$ represents the transmission probability from phonon mode $i$ in the right lead transferring to phonon mode $j$ in the left lead. The diagonal blocks (outlined with red squares) correspond to incident and transmitted modes belonging to the same irrep. If the interface (scattering region) does not break the symmetry, the harmonic matrix $\bm{H}_C$ can be block-diagonalized based on the irrep. This implies that there is no interaction between different irreps and the heat can not be transferred between channels with differing irreps. However, when the symmetry is broken, this restriction does not exist anymore. This behavior is evident in Fig.~\ref{fig:trans-matrix}: the non-zero transmission values are confined within the red squares for the $C_{10v}$ interface [Fig.~\ref{fig:trans-matrix}(a)]. In contrast, for the $C_{1}$ interface [Fig.~\ref{fig:trans-matrix}(b)], some transmission appears outside the red square.

The most salient of the symmetry-induced selection rules evidenced by Fig.~\ref{fig:trans-matrix} through their fulfillment and breakdown is the conservation of $m$, or of the complete angular quasimomentum. While $C_{10v}$-symmetric defect arrangement respects the rotational part of the line group and therefore does not mix irreps with different $m$, the $C_1$ configuration completely destroys the rotational symmetry and is thus incompatible with normal modes with a well defined $m$. This is apparent, for instance, in the transmission matrix for phonons of frequency $\SI{5.95}{\radian\per\pico\second}$ in the less symmetric defect-laden configuration, with numerous visibly nonzero elements connecting blocks with different values of $m$.

\begin{figure}[htb!]
    \centering
    \includegraphics[width=1\linewidth]{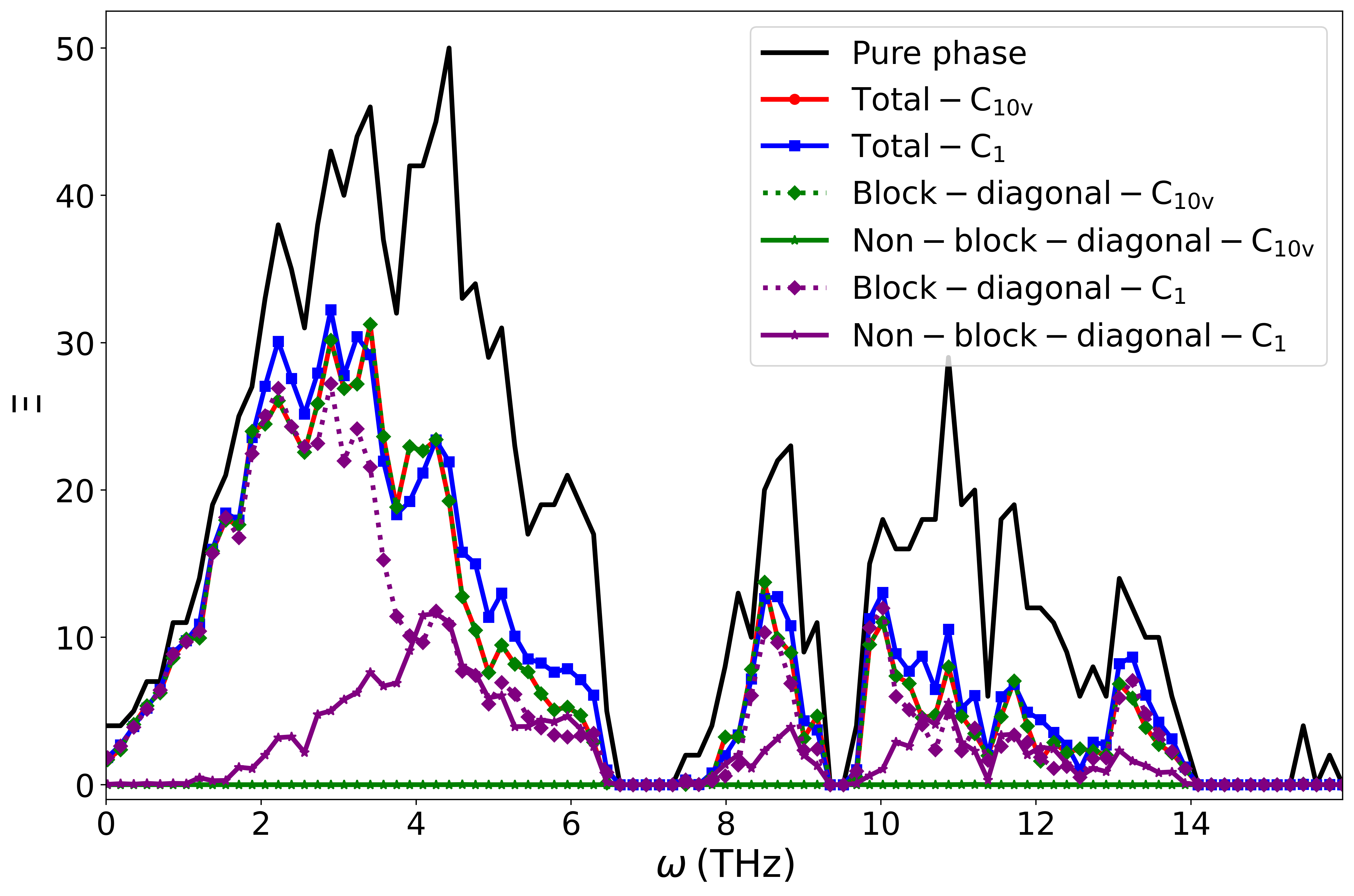}
    \caption{Contributions to the transmission from diagonal and off-diagonal blocks in the transmission matrix for different pristine or defect-laden configurations.}
    \label{fig:block-trans}
\end{figure}

For structures with broken symmetry, energy can access more transmission channels instead of being confined to the same irrep. In this way, although the irrep-preserving transmission probability decreases, the overall transmission  increases due to compensation from the off-diagonal terms. As shown in Fig.~\ref{fig:block-trans}, the transmission for the structure maintaining $C_{10v}$ symmetry shows no contribution from the off-diagonal terms. In contrast, for the broken-symmetry structure $C_{1}$, the off-diagonal terms contribute significantly and are even comparable to the diagonal terms in some frequency ranges. We can see that in these areas with off-diagonal compensation, the total transmission coefficient of the $C_{1}$ structure is higher than that of the $C_{10v}$ structure.

\begin{figure}[htb!]
    \centering
    \includegraphics[width=1\linewidth]{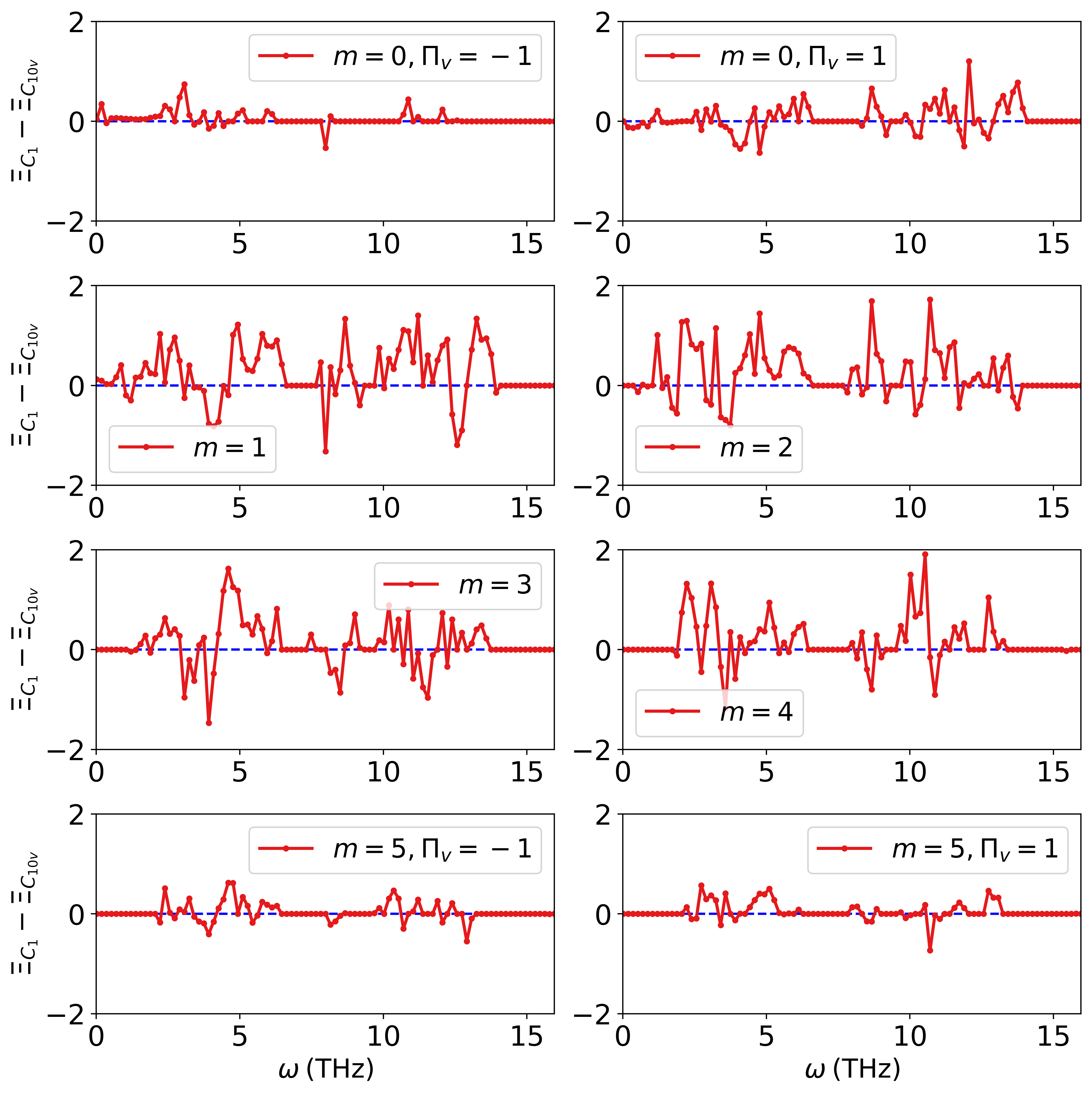}
    \caption{ The transmittance differences $\Xi_{C_{1}}-\Xi_{C_{10v}}$ between the systems with symmetry-breaking and symmetry-conserving defects for different irreps. }
    \label{fig:Irreps_difference}
\end{figure}

In Fig.~\ref{fig:Irreps_difference} we compare the transmissions of the $C_{1}$ and $C_{10v}$ structures by plotting their differences irrep by irrep. In the low-frequency range, transmission is predominantly influenced by modes with low $m$ numbers, whereas larger values of $m$ correspond to modes in higher frequency ranges. The difference between $\Xi_{C_{1}}$ and $\Xi_{C_{10v}}$ confirms that, across most frequency ranges, the more disordered structure exhibits a higher transmission probability.

Eight more comparisons between lower- and higher-symmetry configurations of this  system are provided as part of the supplementary information, showing that these results are not fortuitous, and supporting our analysis in terms of the off-diagonal blocks of the transmission matrix.

\subsection{Finite-temperature results}

The calculations presented so far represent a purely harmonic level of description, and therefore fail to account for either the progressively more important effect of anharmonicity at higher temperatures or possible temperature-induced structural distortions causing deviations from symmetry. Here, we discuss the impact of those finite temperature effects on the basis of our MD simulations, which can capture anharmonicity to all orders, to assess the robustness of our observations.

A first class of problems could emerge at finite temperatures due to the breaking of structural symmetries. However, our MD simulations revealed that the time-averaged atomic positions at temperatures of \SI{300}{K} and \SI{500}{K} maintain the original symmetry to a tolerance of \SI{0.01}{\angstrom}. At higher temperatures, the double-walled \ce{WS2}-\ce{MoS2} system will occasionally exhibit a relative rotation between the weakly interacting inner and outer tubes, but this only causes a transition between two equivalent configurational minima with the same symmetry.

Next, we investigate the full thermal conductivity using Green-Kubo simulations. They reveal thermal conductivity values of \SI{9.8 \pm 0.8}{\watt\per\meter\kelvin} and \SI{6.1 \pm 0.6}{\watt\per\meter\kelvin} for the least and more symmetric defect-laden versions of the double-walled nanotube. This result confirms our findings from the transmission calculations and shows an even more significant difference in the total thermal conductivity when including higher-order scattering processes. While these MD simulations do not afford a similarly detailed look into the origin of those larger differences, their results do show that asymmetric defects can lead to an enhanced conductivity at finite temperatures compared to symmetric defects.

\section{Conclusions}

In summary, we have developed a symmetry-resolved AGF method, which enables detailed  computational analysis of how symmetry variations affect phonon transmission. By applying the symmetry projector technique, phonon branches are decomposed into distinct irreps, splitting the problem into many simpler ones and avoiding numerical artifacts. This decomposition yields real phonon modes categorized by irreps that can be directly used to compute the symmetry-adapted transmission.

We have applied this method to a multi-layer \ce{WS2}-\ce{MoS2} system where we have modulated the structural symmetry by altering the substitutional defect atom types, \ce{W} and \ce{Mo}, in the scattering region. Our findings show that high-symmetry defects constrain phonon scattering to occur within the same irreps, consistent with symmetry-imposed selection rules. In contrast, as the defect-induced symmetry is significantly reduced, additional phonon transmission channels, previously forbidden by selection rules, become possible, leading to an increase in both total transmission and thermal conductance.  

\section{Data availability}
The DFT dataset, the Allegro MLIP model and the double-walled nanotube structures and are available on Zenodo (\href{https://doi.org/10.5281/zenodo.17279460}{https://doi.org/10.5281/zenodo.17279460}).

\section{Code availability}
The code for our symmetry-adapted AGF is available at \href{https://github.com/pulgon-project/transmission}{https://github.com/pulgon-project/transmission}. Related line group symmetry detection tools for quasi-1D systems are available at \href{https://github.com/pulgon-project/pulgon_tools}{https://github.com/pulgon-project/pulgon\_tools}. The code to compute the heat flux for the Allegro model is available at \href{https://github.com/pulgon-project/nequip-flux-calculator}{https://github.com/pulgon-project/nequip-flux-calculator}.

\section{Author contributions}
Y.-J.C. was the main code developer, generated and analyzed most of the data and contributed to the writing process. S.W. collaborated in coding and writing, and led the finite-temperature simulations. G.K.H.M. participated in analysis, writing and project management. J.C. designed the workflow, planned and supervised the work, secured funding, contributed to the code and analysis and composed the first version of the main manuscript text. All authors reviewed the manuscript.

\section{Competing interests}
The authors declare that they have no competing interests.

\begin{acknowledgments}
 This research was funded in whole or in part by the Austrian Science Fund (FWF) [10.55776/P36129]. For open access purposes, the author has applied a CC BY public copyright license to any author-accepted manuscript version arising from this submission. It was also supported by MCIN with funding from the European Union NextGenerationEU (PRTR-C17.I1) promoted by the Government of Aragon. J.C. acknowledges Grant CEX2023-001286-S funded by MICIU/AEI /10.13039/501100011033.  
\end{acknowledgments}

\bibliography{manuscript}  

\end{document}


\title{Supplementary information for \textit{Ab-initio heat transport in defect-laden quasi-1D systems from a symmetry-adapted perspective}}

\author{Yu-Jie Cen}
\affiliation{Institute of Materials Chemistry, TU Wien, A-1060 Vienna, Austria}

\author{Sandro Wieser}
\affiliation{Institute of Materials Chemistry, TU Wien, A-1060 Vienna, Austria}

\author{Georg K. H. Madsen}
\affiliation{Institute of Materials Chemistry, TU Wien, A-1060 Vienna, Austria}

\author{Jesús Carrete}
\email{jcarrete@unizar.es}
\affiliation{Instituto de Nanociencia y Materiales de Aragón, CSIC-Universidad de Zaragoza, Zaragoza, Spain}


\maketitle

\section{Additional comparisons between defect-laden systems with different symmetries}
\FloatBarrier

\begin{figure}[h]
    \centering
    \begin{subfigure}[b]{0.48\linewidth}
        \centering
        \includegraphics[width=\linewidth]{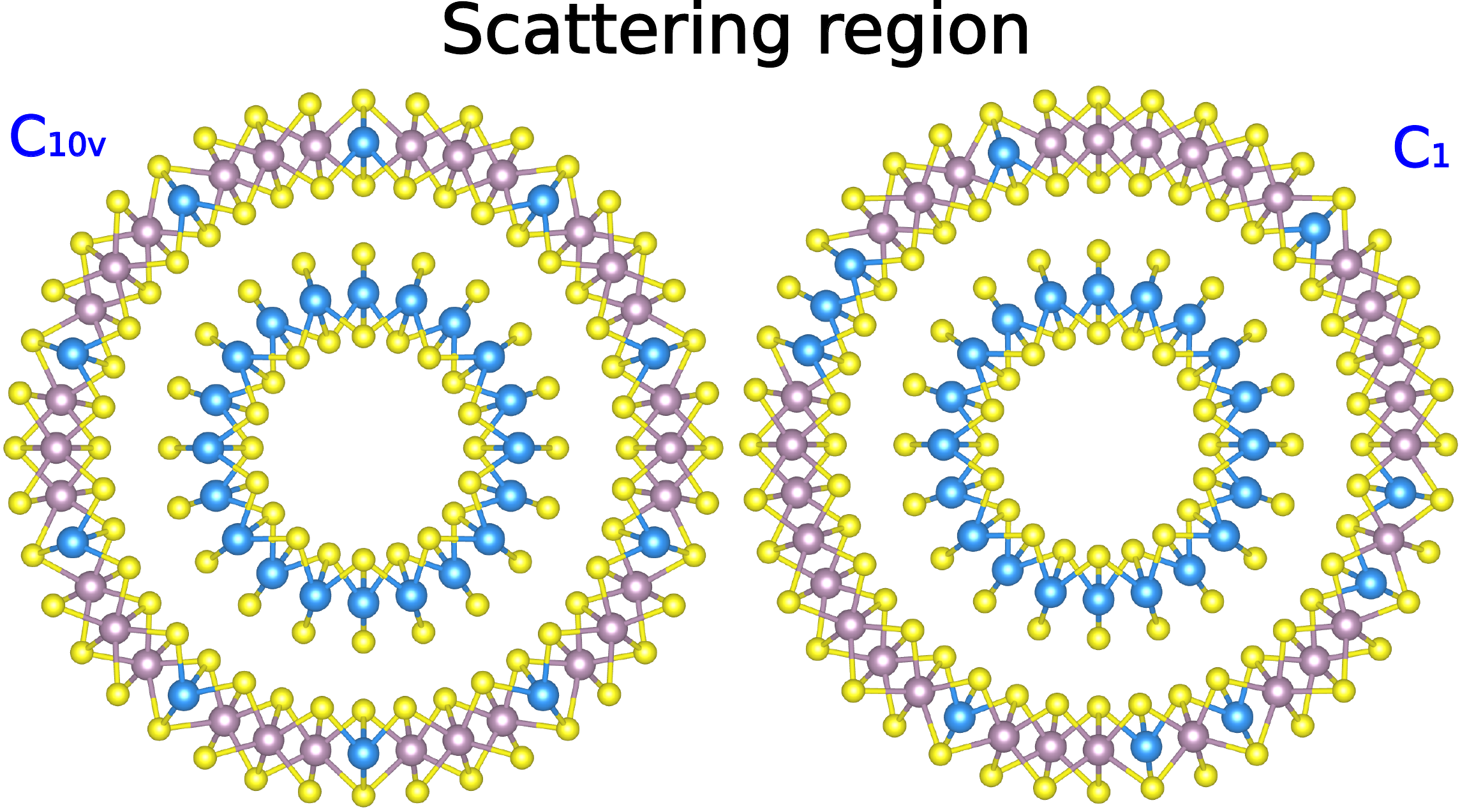}
        \caption{Cross-sectional views of the two defect-laden configurations with different symmetries.}
    \end{subfigure}
    \hfill
    \begin{subfigure}[b]{0.48\linewidth}
        \centering
        \includegraphics[width=\linewidth]{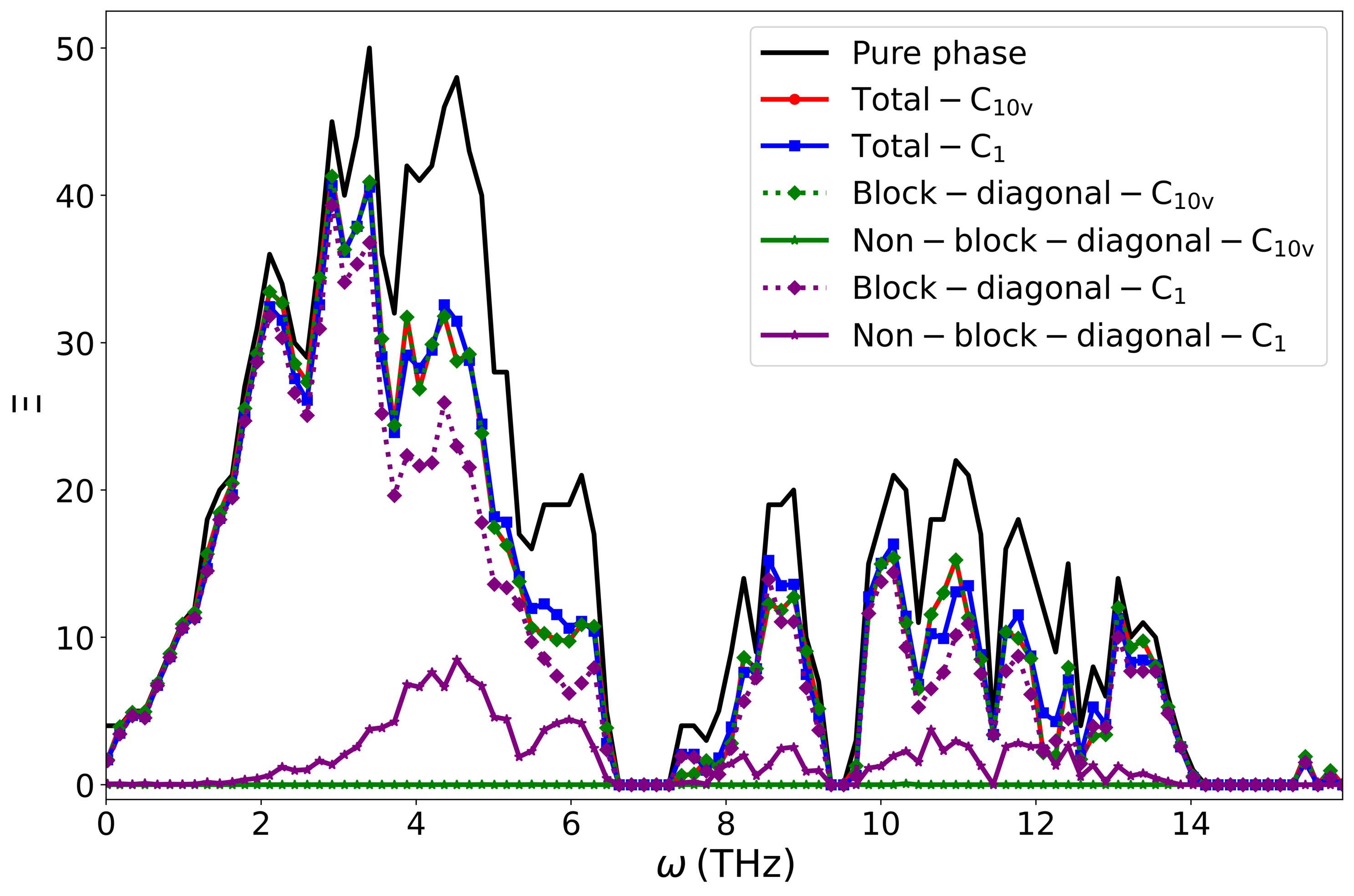}
        \caption{Contributions to the transmission from diagonal and off-diagonal blocks in the transmission matrix for different pristine or defect-laden configurations.}
    \end{subfigure}

    \vskip\baselineskip
        \begin{subfigure}{0.6\textwidth}
    \centering
    \includegraphics[width=1\linewidth]{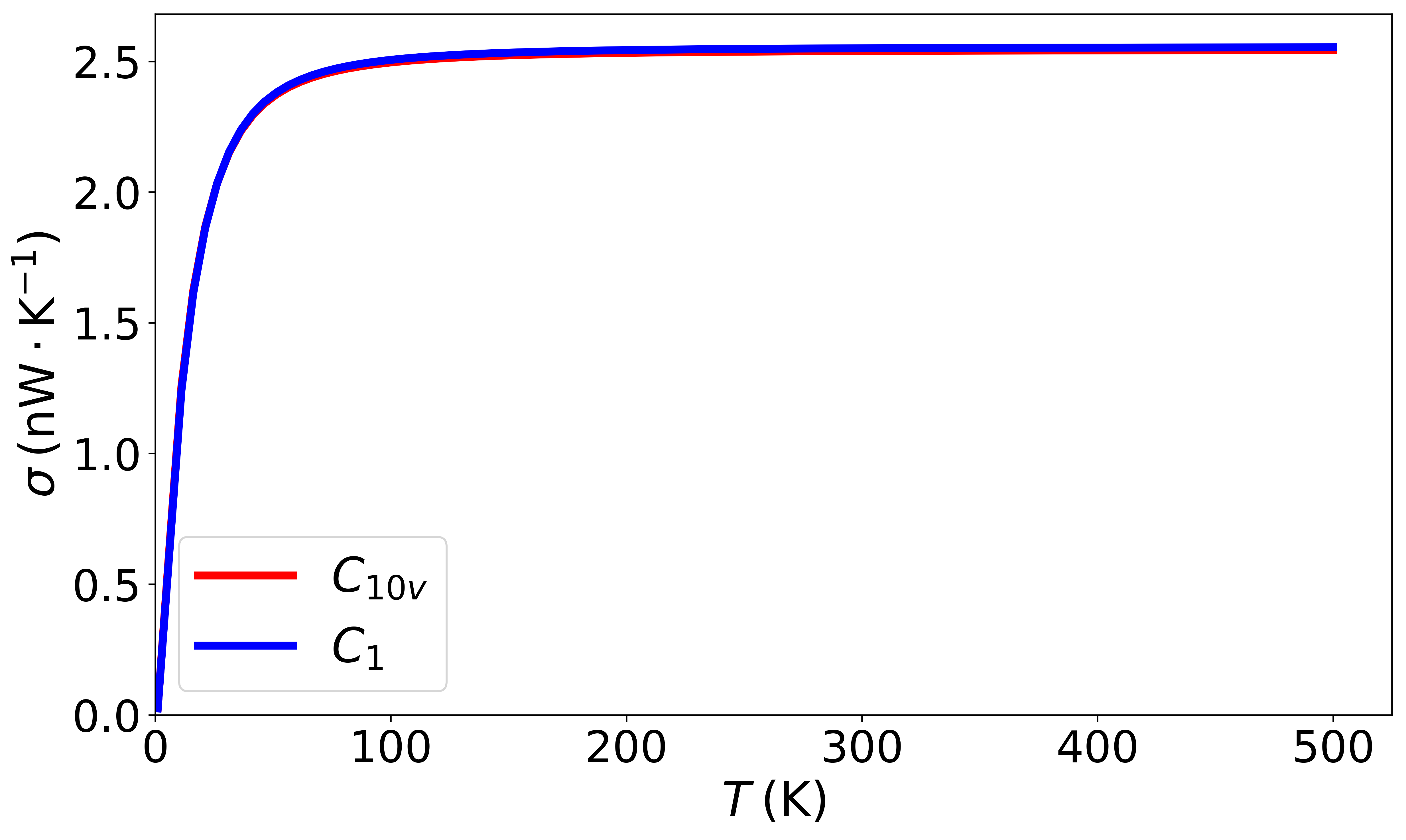}
    \caption{Temperature dependence of the thermal conductances for the two defect-laden structures.}
    \end{subfigure}
    \caption{Additional case 1: 10 Mo atoms are substituted by W atoms in the outer-layer nanotube to form the $C_{10v}$ and $C_{1}$ configurations.}
    \label{fig:case1}
\end{figure}

\begin{figure}[htbp]
    \centering
    \begin{subfigure}[b]{0.48\linewidth}
        \centering
        \includegraphics[width=\linewidth]{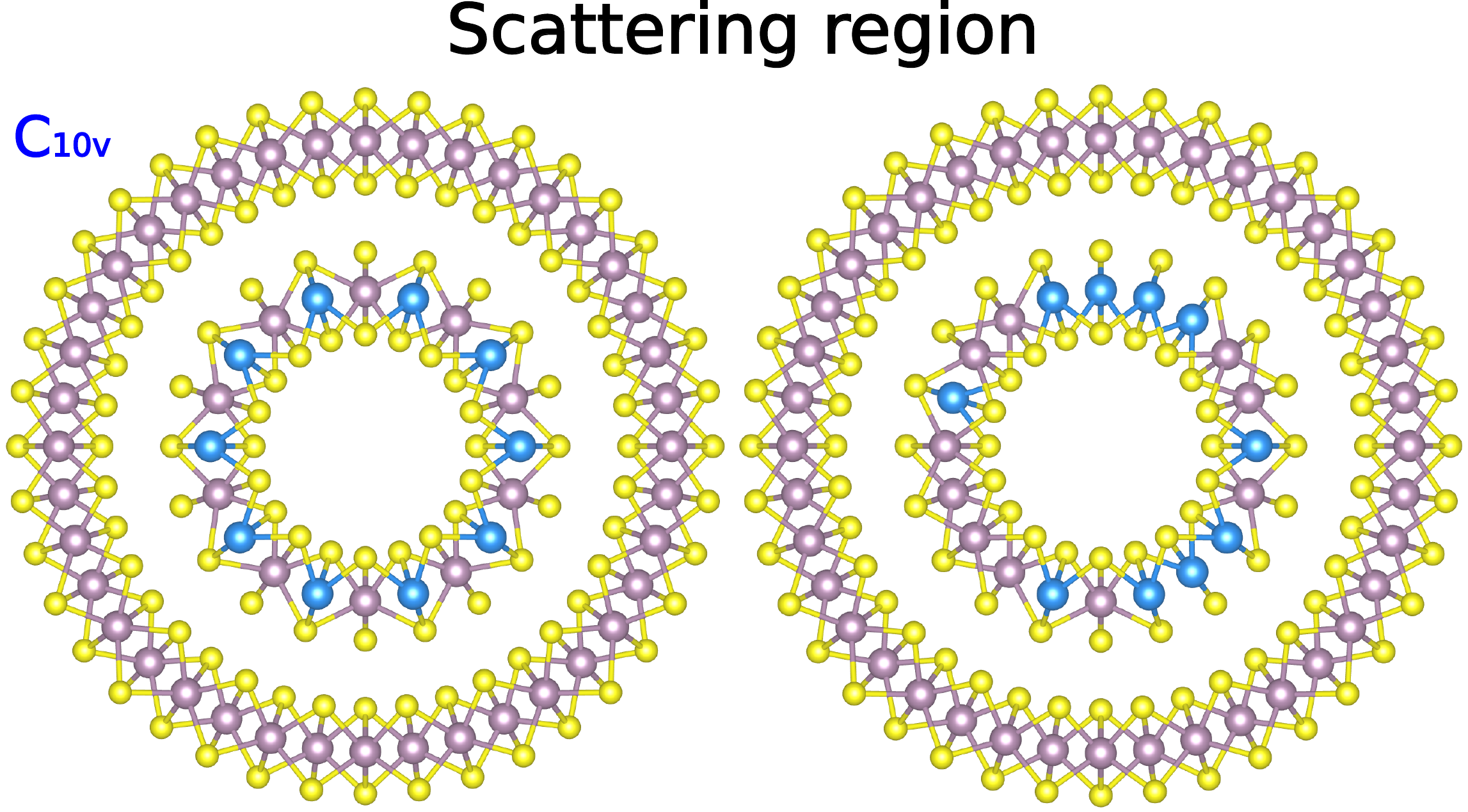}
        \caption{Cross-sectional views of the two defect-laden configurations with different symmetries.}
    \end{subfigure}
    \hfill
    \begin{subfigure}[b]{0.48\linewidth}
        \centering
        \includegraphics[width=\linewidth]{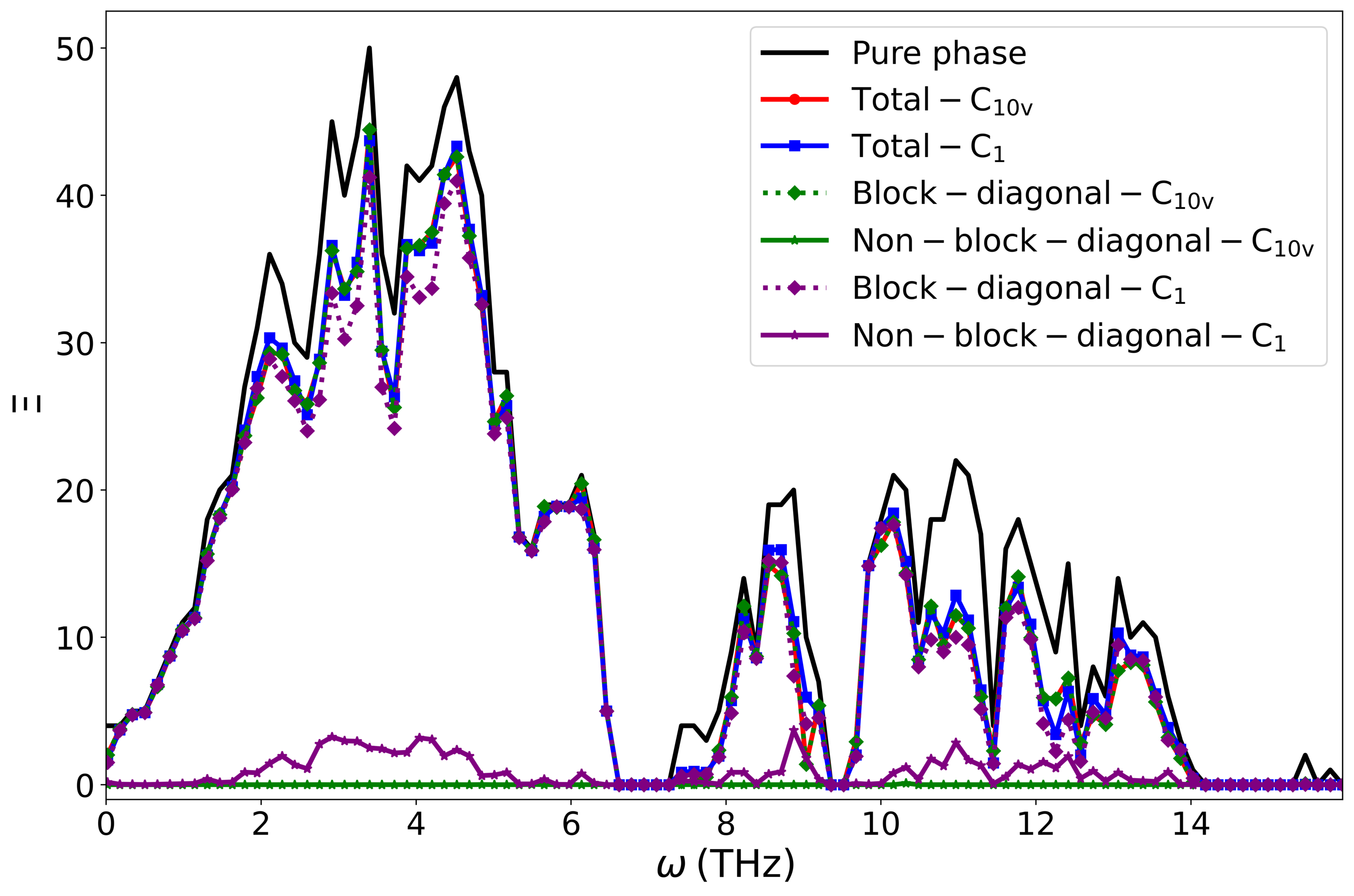}
        \caption{Contributions to the transmission from diagonal and off-diagonal blocks in the transmission matrix for different pristine or defect-laden configurations.}
    \end{subfigure}
    \vskip\baselineskip
        \begin{subfigure}{0.6\textwidth}
    \centering
    \includegraphics[width=1\linewidth]{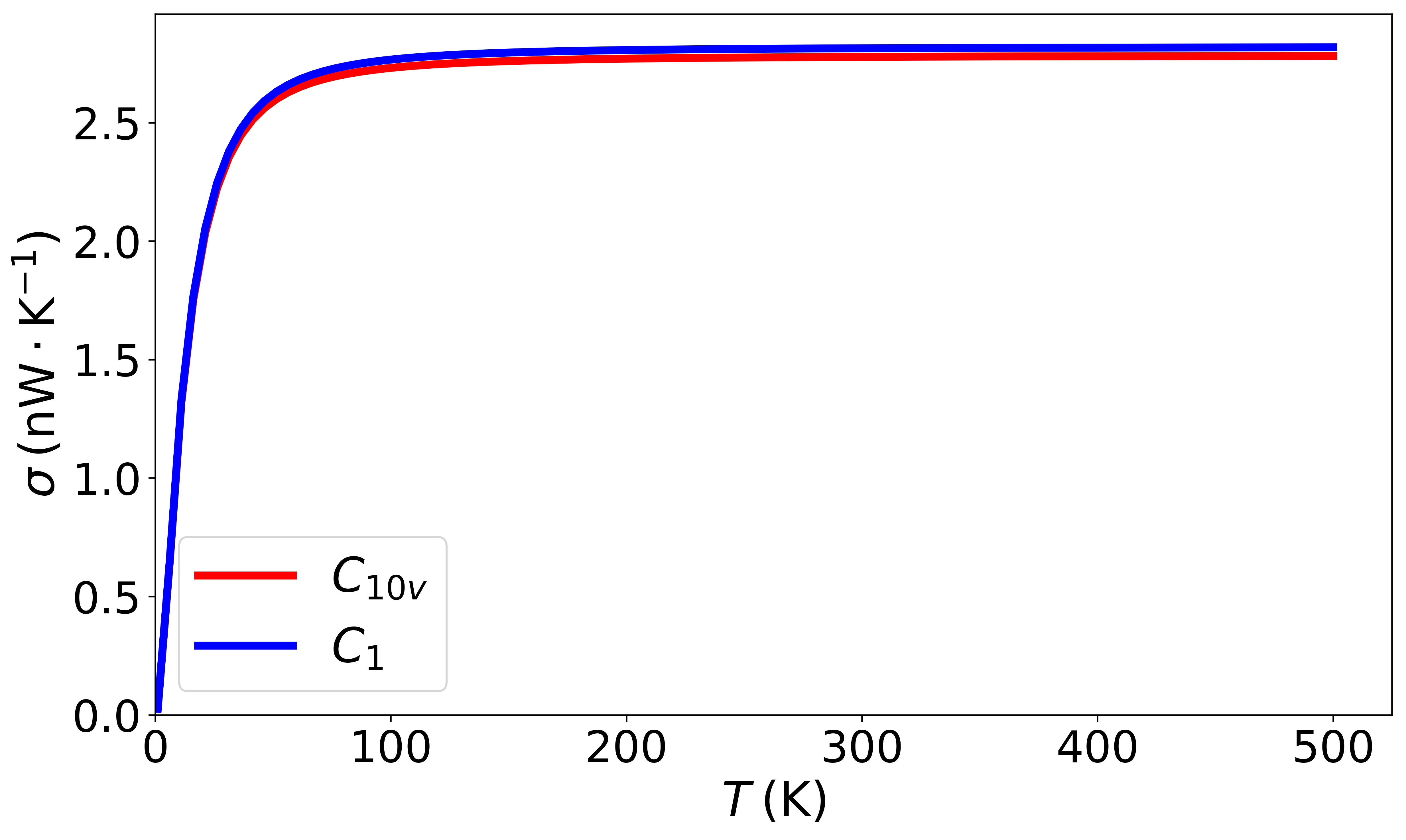}
    \caption{Temperature dependence of the thermal conductances for the two defect-laden structures.}
    \end{subfigure}
    \caption{Additional case 2: 10 W atoms are substituted by Mo atoms in the inner-layer nanotube to form the $C_{10v}$ and $C_{1}$ configurations.}
    \label{fig:case2}
\end{figure}

\begin{figure}[htbp]
    \centering
    \begin{subfigure}[b]{0.48\linewidth}
        \centering
        \includegraphics[width=\linewidth]{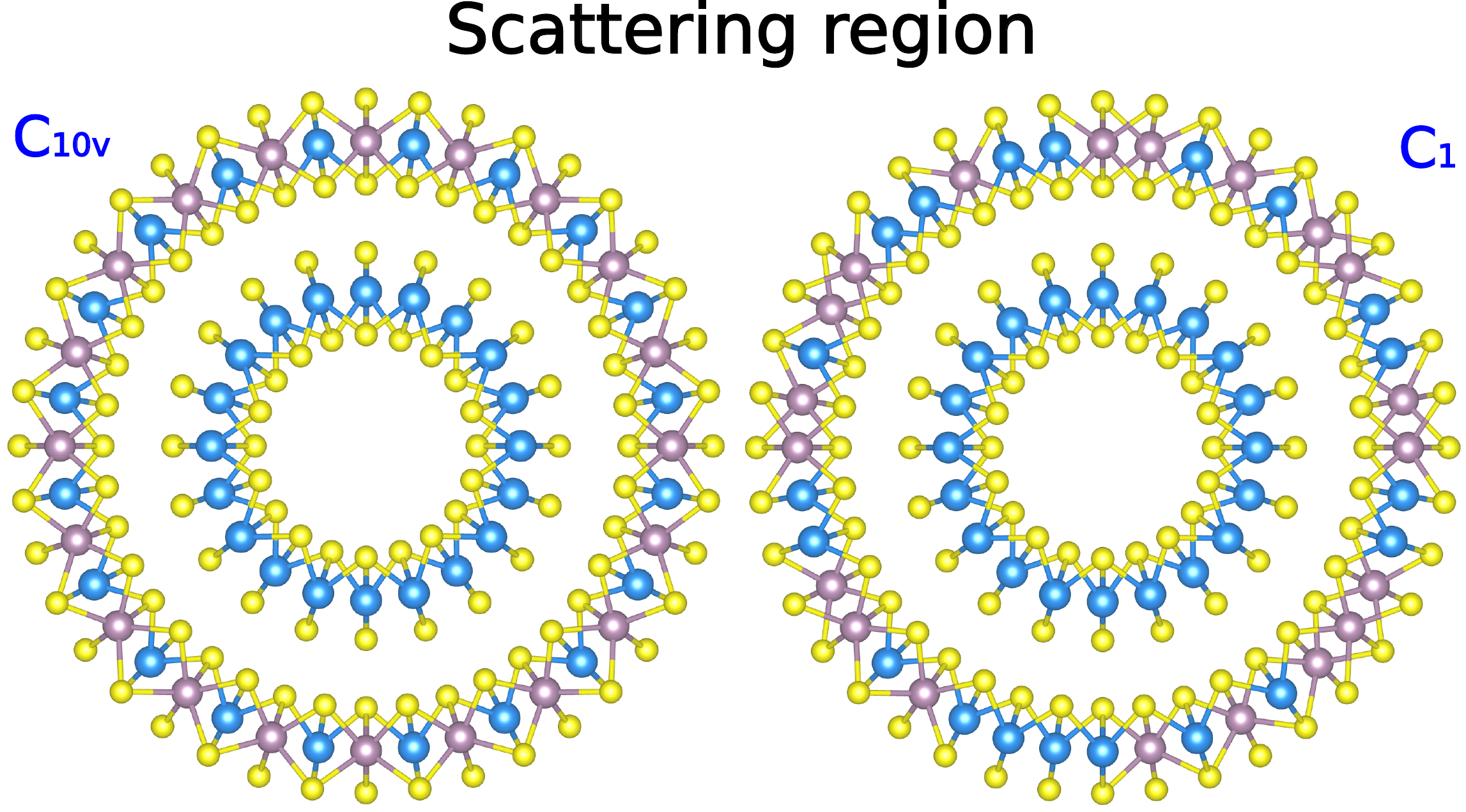}
        \caption{Cross-sectional views of the two defect-laden configurations with different symmetries.}
    \end{subfigure}
    \hfill
    \begin{subfigure}[b]{0.48\linewidth}
        \centering
        \includegraphics[width=\linewidth]{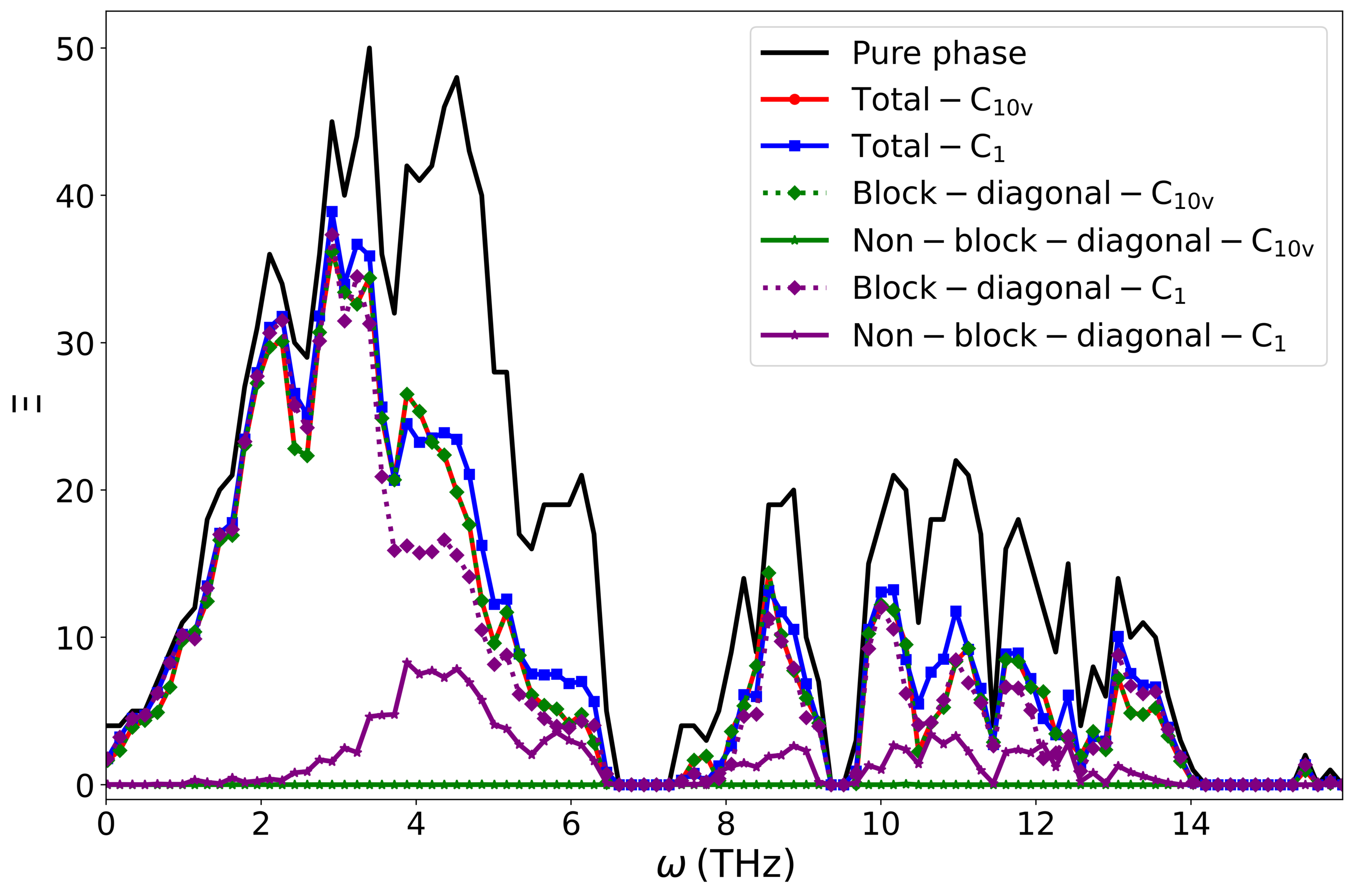}
        \caption{Contributions to the transmission from diagonal and off-diagonal blocks in the transmission matrix for different pristine or defect-laden configurations.}
    \end{subfigure}
    \vskip\baselineskip
        \begin{subfigure}{0.6\textwidth}
    \centering
    \includegraphics[width=1\linewidth]{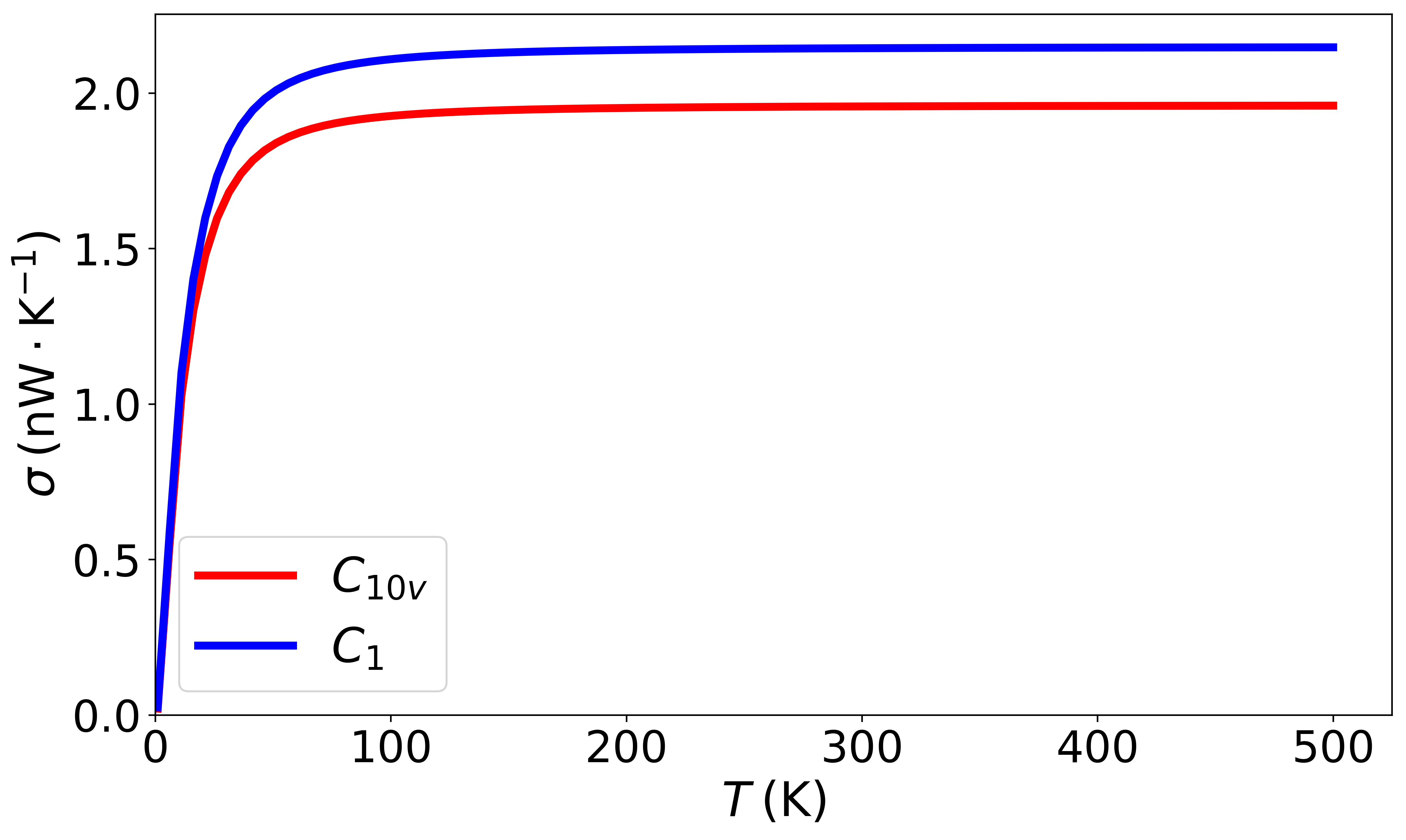}
    \caption{Temperature dependence of the thermal conductances for the two defect-laden structures.}
    \end{subfigure}
    \caption{Additional case 3: 20 Mo atoms are substituted by W atoms in the outer-layer nanotube to form the $C_{10v}$ and $C_{1}$ configurations.}
    \label{fig:case3}
\end{figure}

\begin{figure}[htbp]
    \centering
    \begin{subfigure}[b]{0.48\linewidth}
        \centering
        \includegraphics[width=\linewidth]{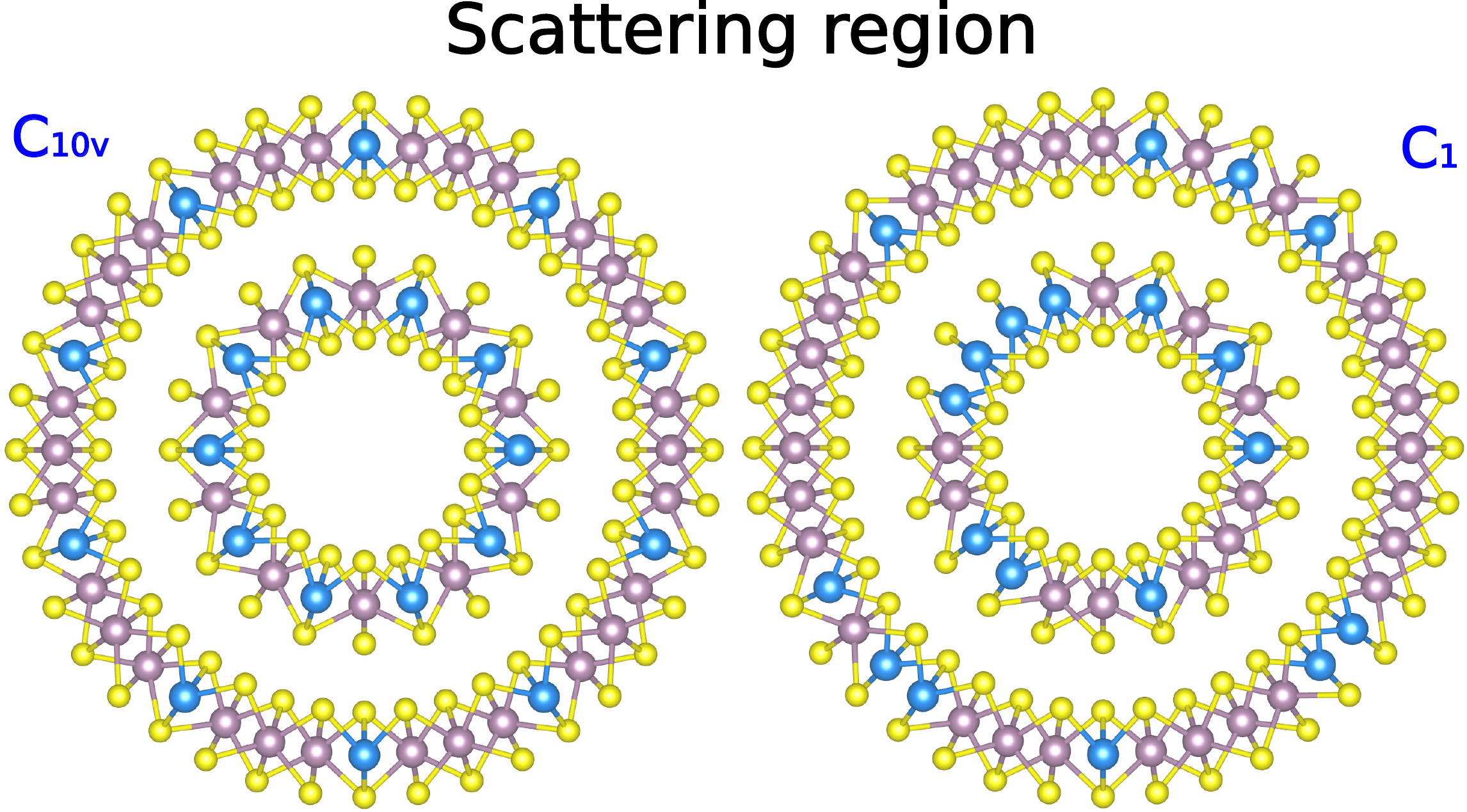}
        \caption{Cross-sectional views of the two defect-laden configurations with different symmetries.}
    \end{subfigure}
    \hfill
    \begin{subfigure}[b]{0.48\linewidth}
        \centering
        \includegraphics[width=\linewidth]{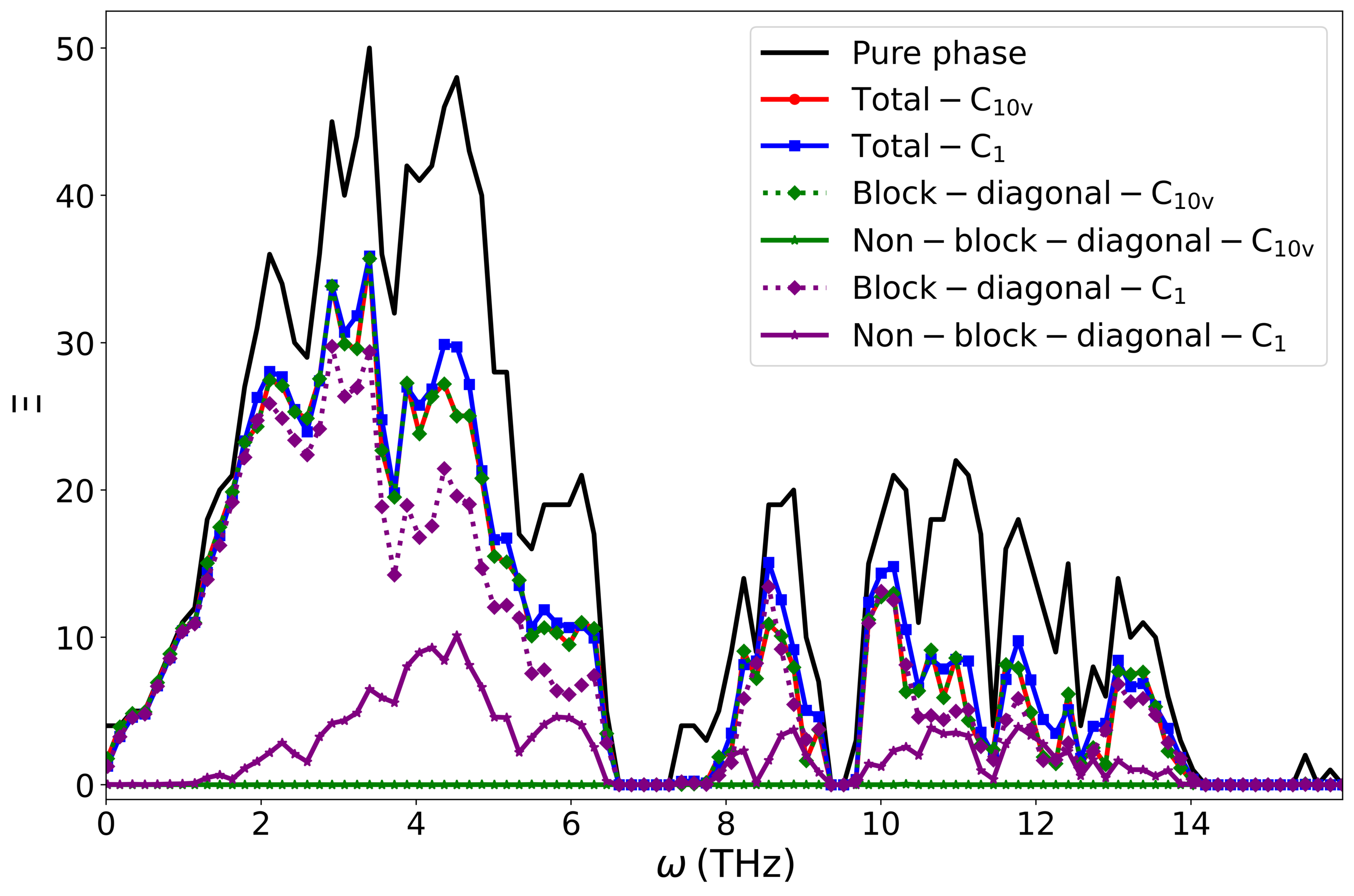}
        \caption{Contributions to the transmission from diagonal and off-diagonal blocks in the transmission matrix for different pristine or defect-laden configurations.}
    \end{subfigure}
    \vskip\baselineskip
        \begin{subfigure}{0.6\textwidth}
    \centering
    \includegraphics[width=1\linewidth]{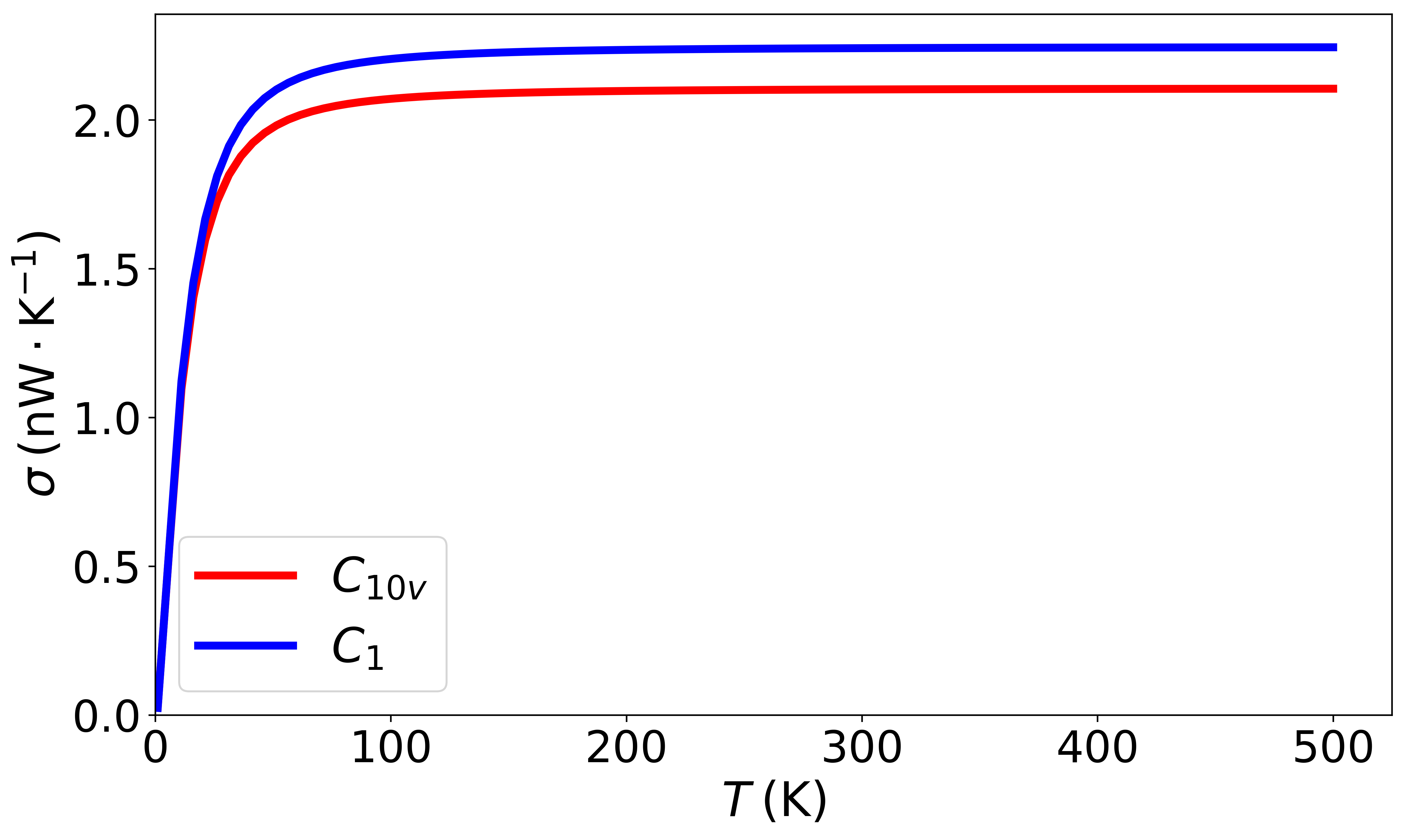}
    \caption{Temperature dependence of the thermal conductances for the two defect-laden structures.}
    \end{subfigure}
    \caption{Additional case 4: 10 Mo atoms are substituted by W atoms in the outer-layer nanotube and 10 W atoms are substituted by Mo atoms in the inner-layer nanotube to form the $C_{10v}$ and $C_{1}$ configurations.}
    \label{fig:case4}
\end{figure}

\begin{figure}[htbp]
    \centering
    \begin{subfigure}[b]{0.48\linewidth}
        \centering
        \includegraphics[width=\linewidth]{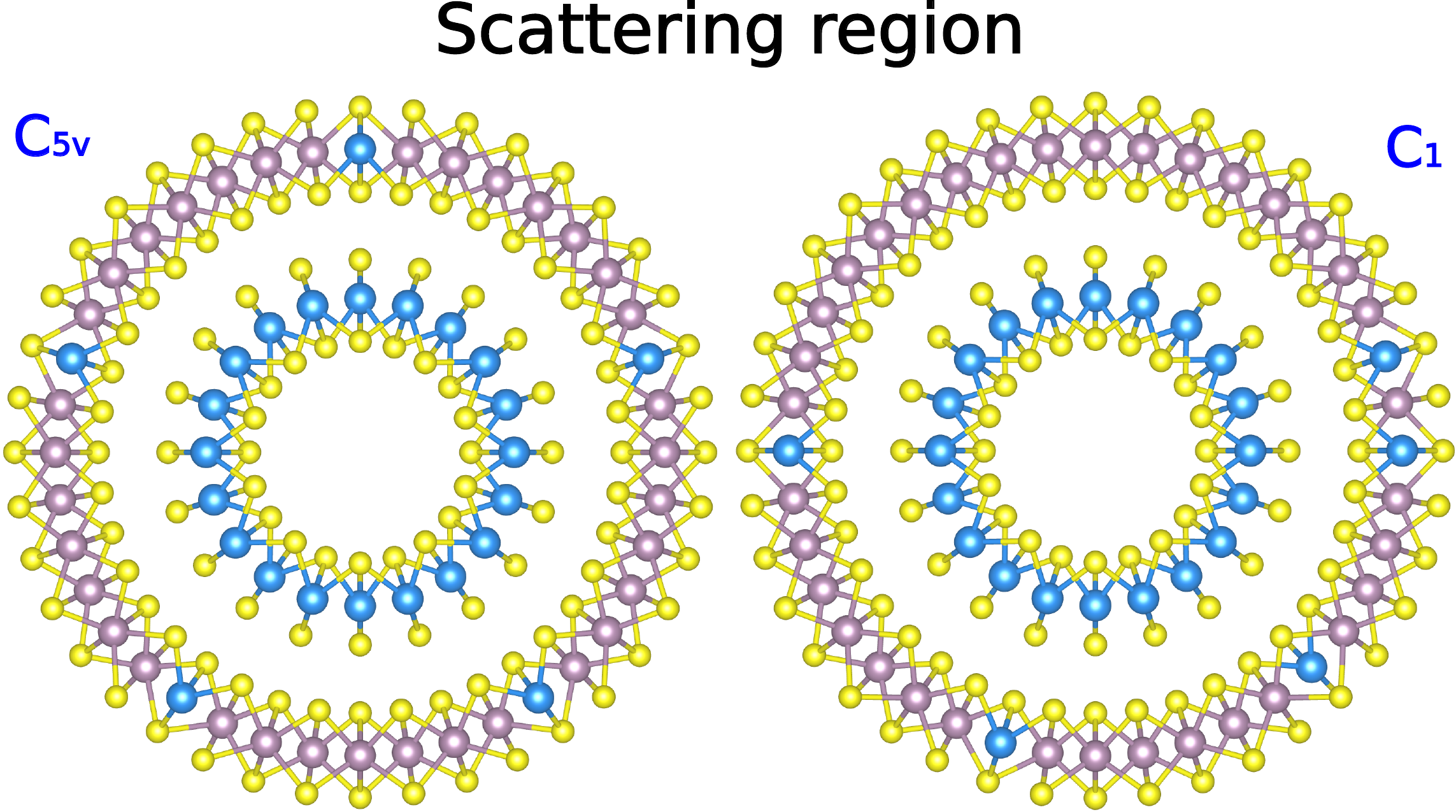}
        \caption{Cross-sectional views of the two defect-laden configurations with different symmetries.}
    \end{subfigure}
    \hfill
    \begin{subfigure}[b]{0.48\linewidth}
        \centering
        \includegraphics[width=\linewidth]{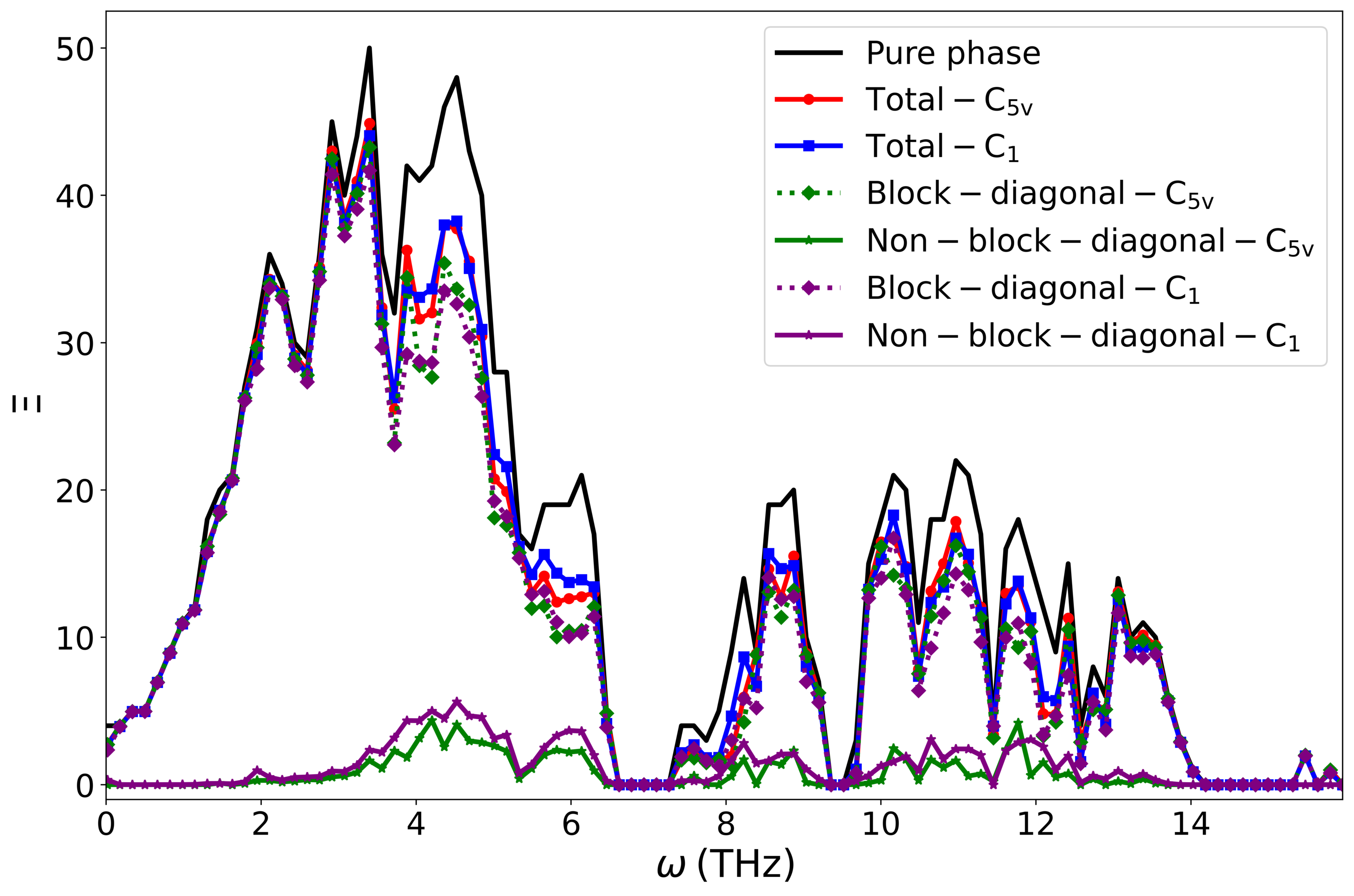}
        \caption{Contributions to the transmission from diagonal and off-diagonal blocks in the transmission matrix for different pristine or defect-laden configurations.}
    \end{subfigure}
    \vskip\baselineskip
        \begin{subfigure}{0.6\textwidth}
    \centering
    \includegraphics[width=1\linewidth]{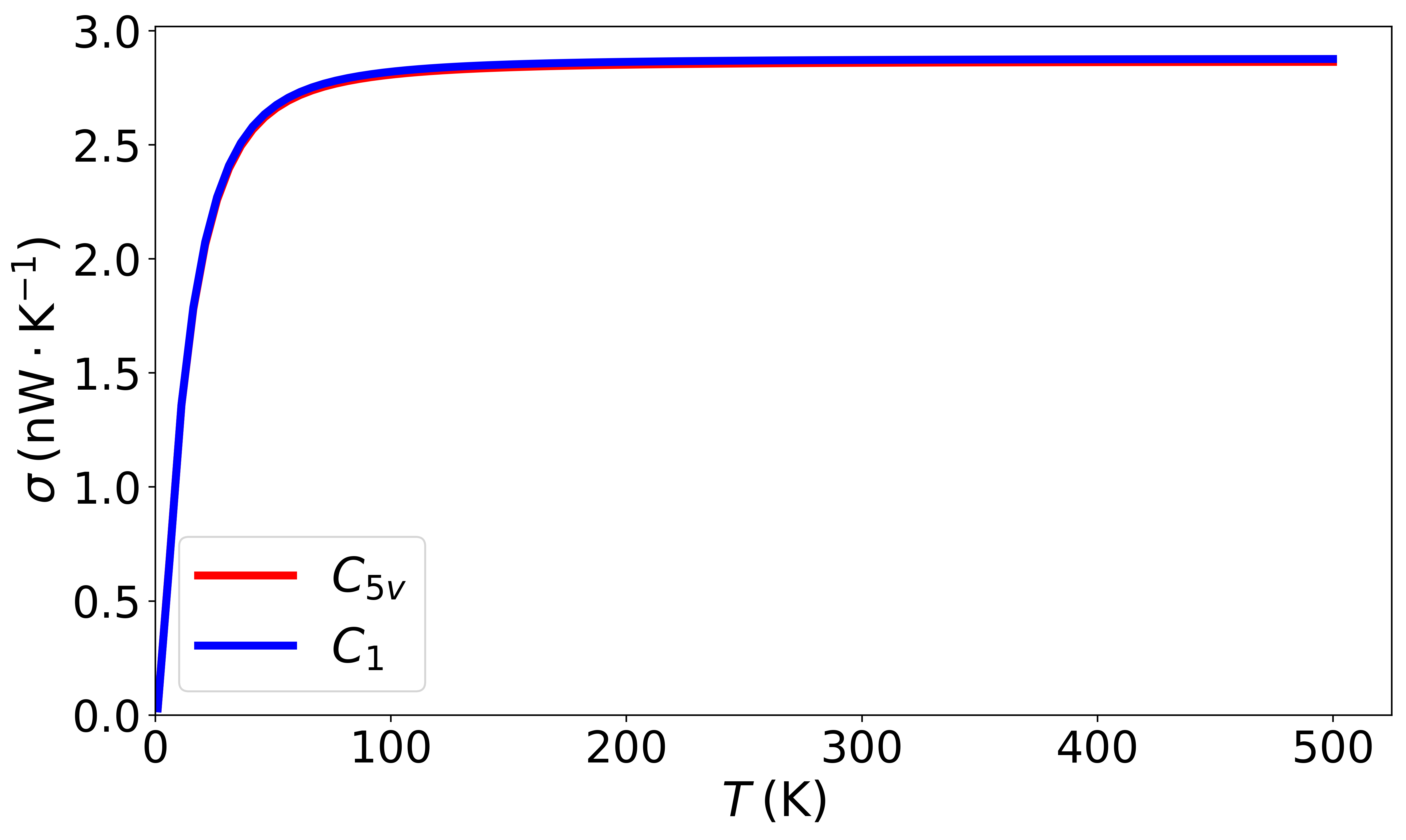}
    \caption{Temperature dependence of the thermal conductances for the two defect-laden structures.}
    \end{subfigure}
    \caption{Additional case 5: 5 Mo atoms are substituted by W atoms in the outer-layer nanotube to form the $C_{5v}$ and $C_{1}$ configurations.}
    \label{fig:case5}
\end{figure}

Here we show results for eight additional pairs of defect-laden configurations, in order to demonstrate that the results presented in the main text are not a coincidence. We consider two kinds of defects: (1) Mo-W substitutional defects and (2) S vacancies. In the case shown in Fig.~\ref{fig:case1}, 10 W atoms are substituted for Mo atoms in the outer nanotube to build $C_{10v}$ and $C_{1}$ structures. Since the number of defects is relatively small, the resulting decrease in transmission compared with the pristine nanotube is also minor, and the influence of symmetry on thermal conductance is not as apparent. In the case shown in Fig.~\ref{fig:case2}, 10 Mo atoms are used to replace W atoms in the inner layer nanotube. It can be observed that the effect of the inner nanotube is relatively insignificant compared with that of the outer nanotube. In the cases shown in Fig.~\ref{fig:case3} and Fig.~\ref{fig:case4}, there are 20 defect atoms and the transmission decreases more substantially. These two cases exhibit similar behavior to that discussed in the main text. In Fig.~\ref{fig:case5}, the symmetry of the structure is partially broken, leading to slightly lower off-diagonal transmission blocks in the $C_{5v}$ structure than in the $C_1$ structure. This indicates that the selection rules are partially relaxed in the $C_{5v}$ configuration.

\begin{figure}[htbp]
    \centering
    \begin{subfigure}[b]{0.48\linewidth}
        \centering
        \includegraphics[width=\linewidth]{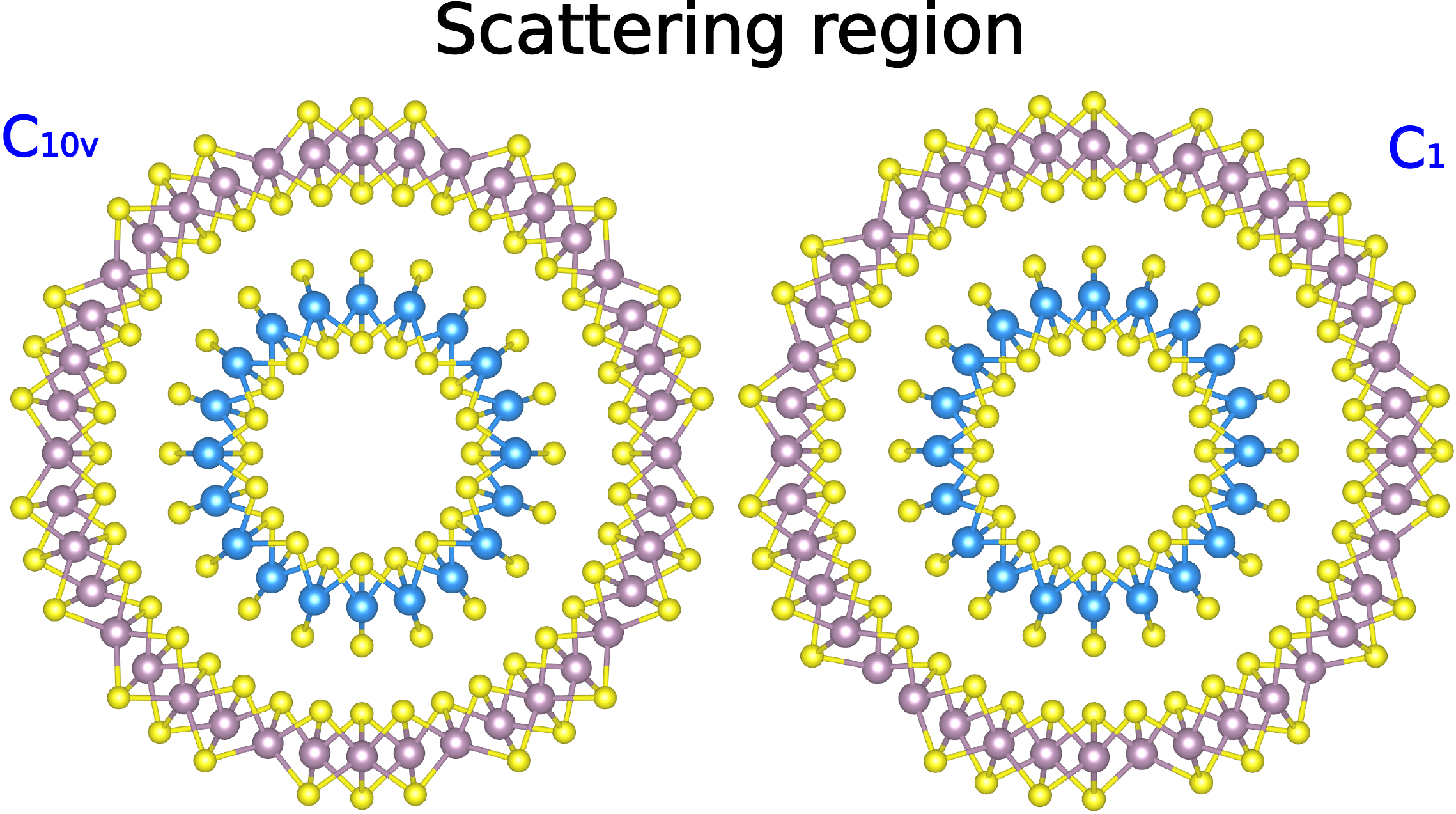}
        \caption{Cross-sectional views of the two defect-laden configurations with different symmetries.}
    \end{subfigure}
    \hfill
    \begin{subfigure}[b]{0.48\linewidth}
        \centering
        \includegraphics[width=\linewidth]{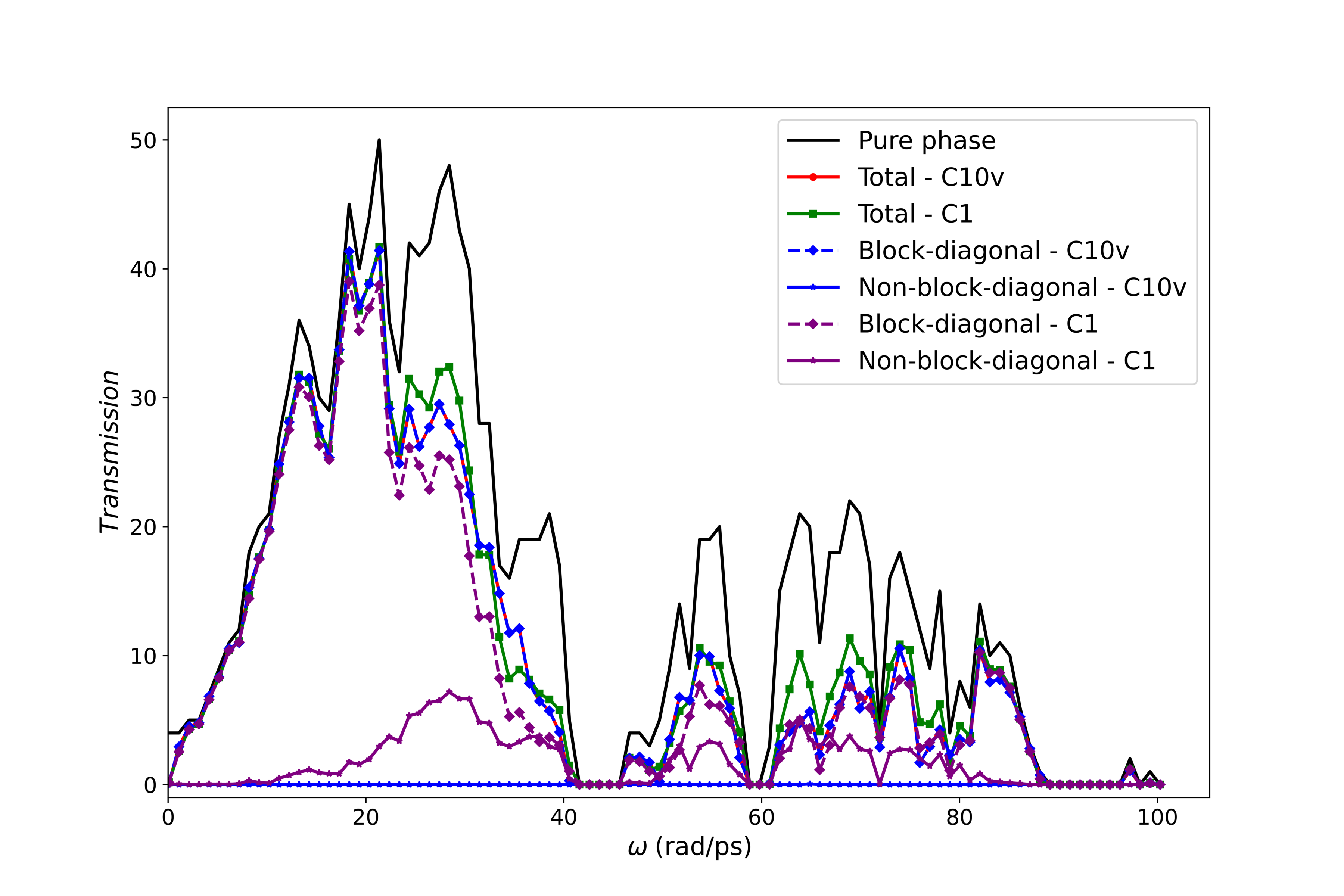}
        \caption{Contributions to the transmission from diagonal and off-diagonal blocks in the transmission matrix for different pristine or defect-laden configurations.}
    \end{subfigure}
    \vskip\baselineskip
        \begin{subfigure}{0.6\textwidth}
    \centering
    \includegraphics[width=1\linewidth]{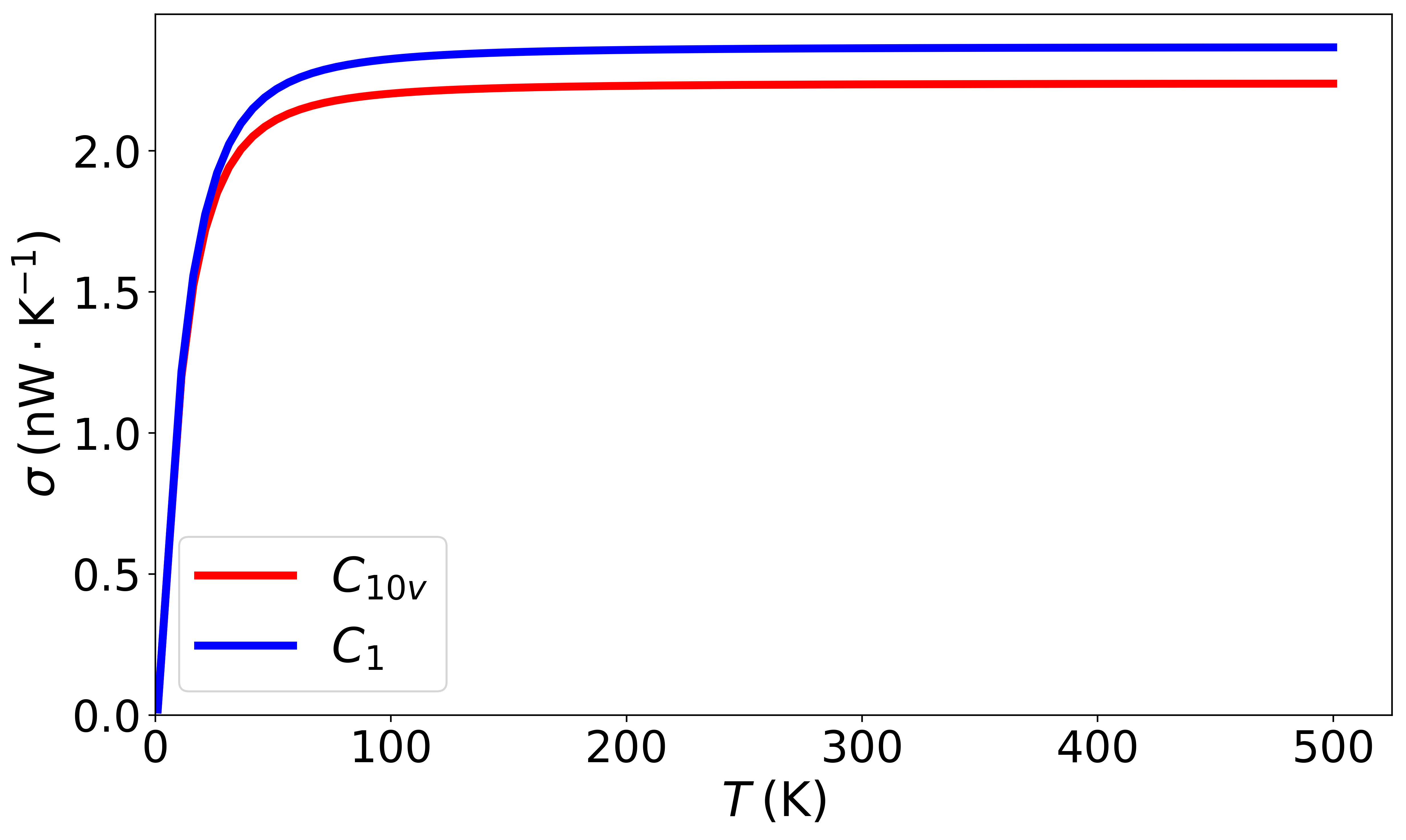}
    \caption{Temperature dependence of the thermal conductances for the two defect-laden structures.}
    \end{subfigure}
    \caption{Additional case 6: 10 S atoms are removed from the outermost layer connected to Mo atoms to form the $C_{10v}$ and $C_{1}$ configurations.}
    \label{fig:case6}
\end{figure}

\begin{figure}[htbp]
    \centering
    \begin{subfigure}[b]{0.48\linewidth}
        \centering
        \includegraphics[width=\linewidth]{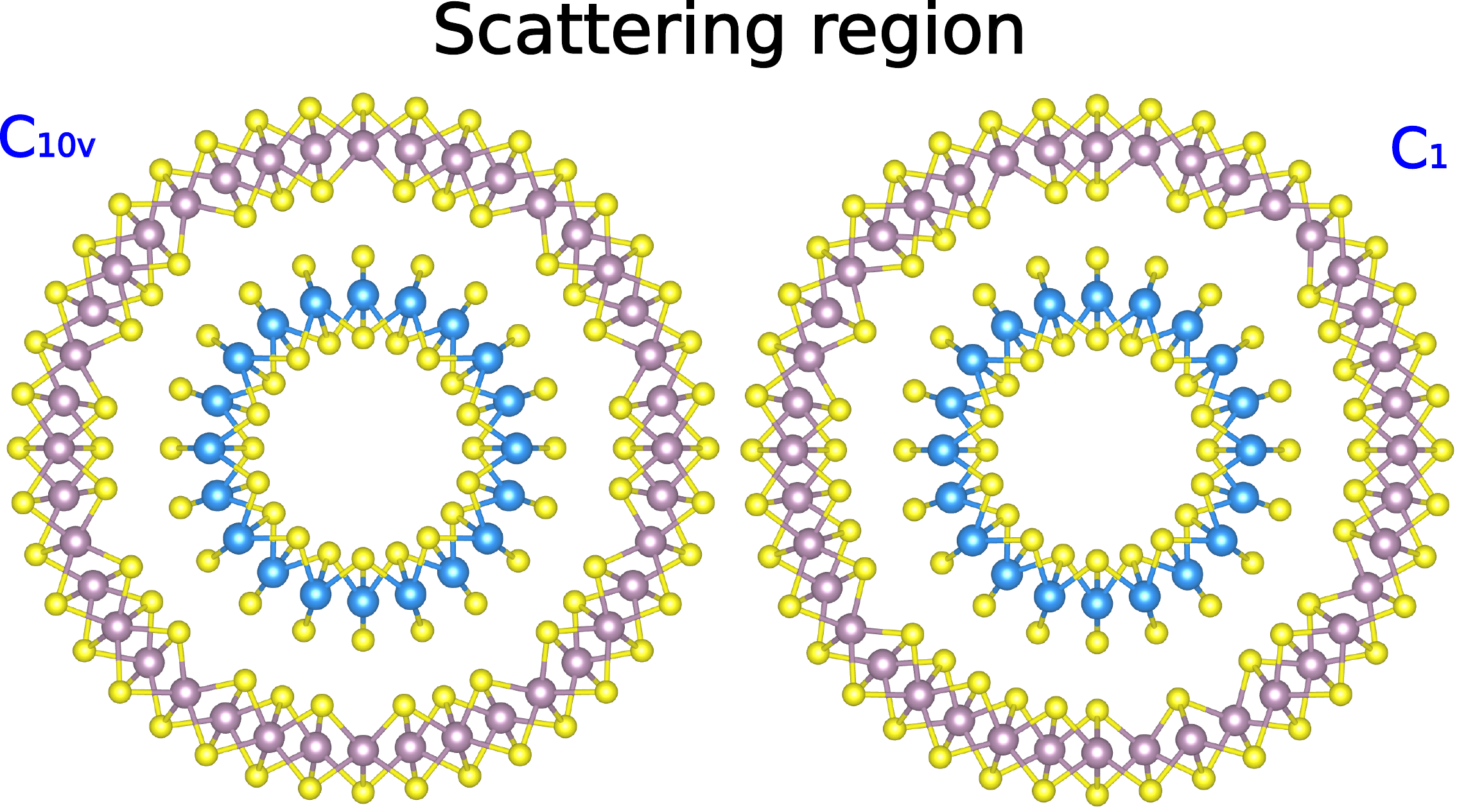}
        \caption{Cross-sectional views of the two defect-laden configurations with different symmetries.}
    \end{subfigure}
    \hfill
    \begin{subfigure}[b]{0.48\linewidth}
        \centering
        \includegraphics[width=\linewidth]{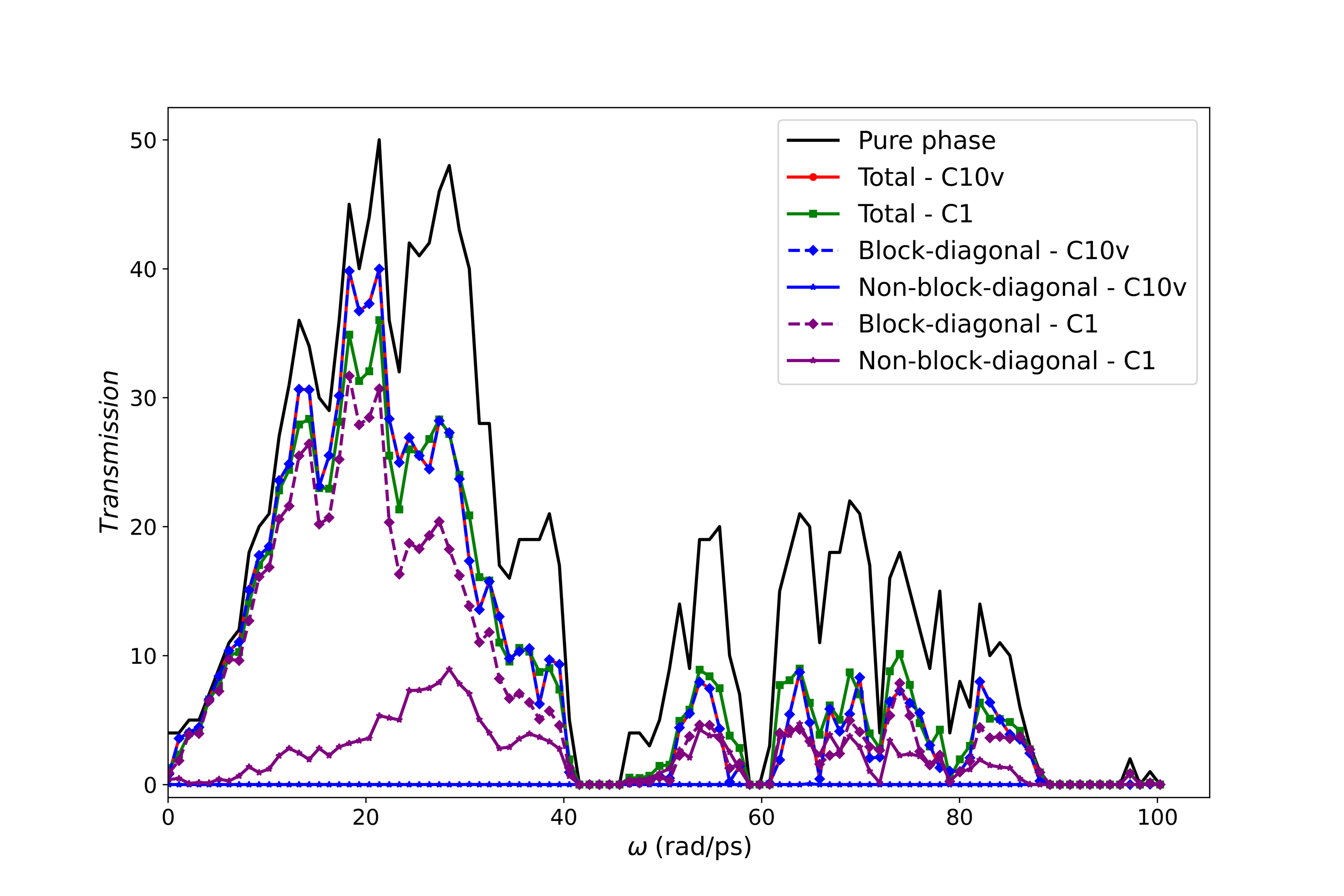}
        \caption{Contributions to the transmission from diagonal and off-diagonal blocks in the transmission matrix for different pristine or defect-laden configurations.}
    \end{subfigure}
    \vskip\baselineskip
        \begin{subfigure}{0.6\textwidth}
    \centering
    \includegraphics[width=1\linewidth]{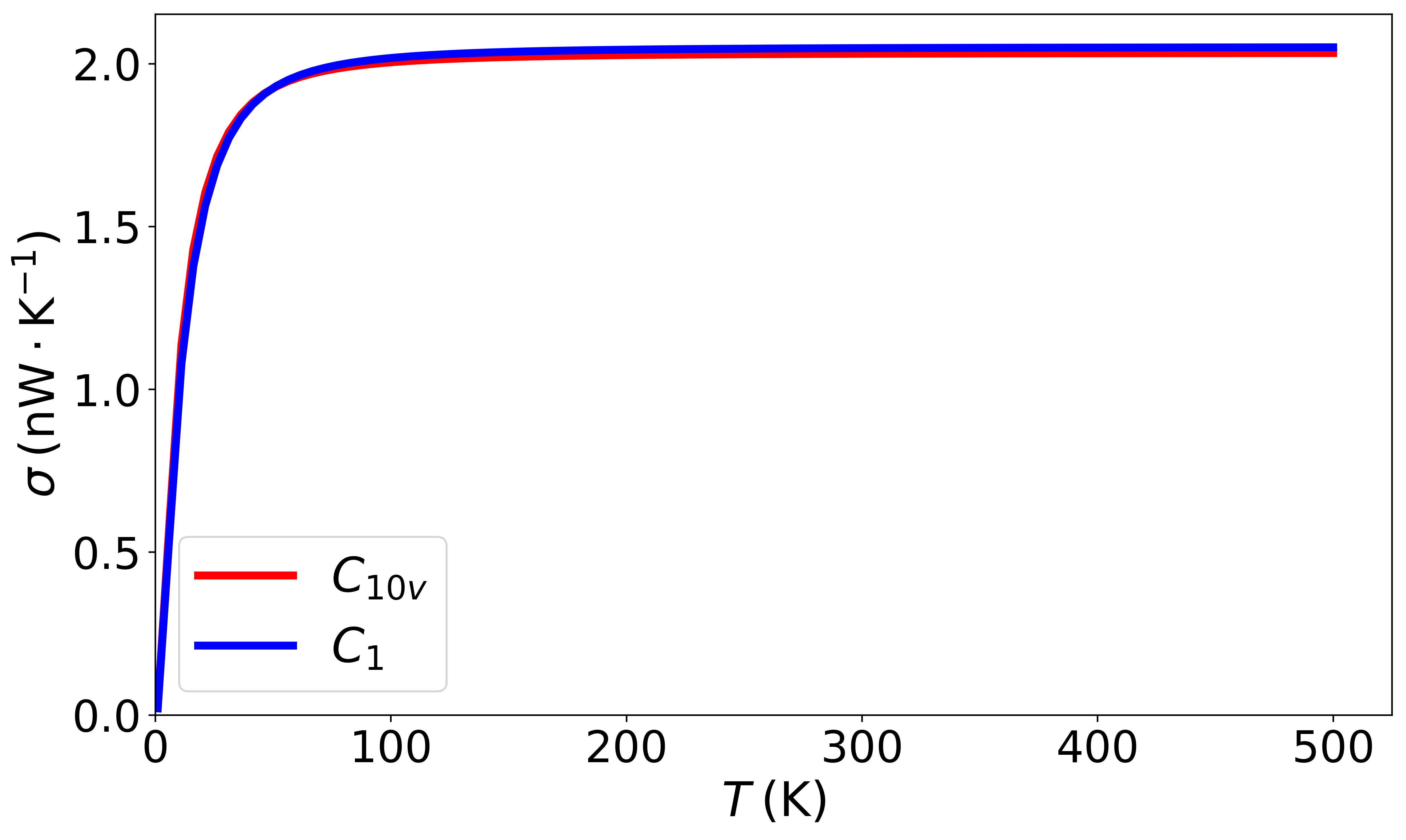}
    \caption{Temperature dependence of the thermal conductances for the two defect-laden structures.}
    \end{subfigure}
    \caption{Assiditional case 7: 10 S atoms are removed from the inner layer connected to Mo atoms to form the $C_{10v}$ and $C_{1}$ configurations.}
    \label{fig:case7}
\end{figure}

\begin{figure}[htbp]
    \centering
    \begin{subfigure}[b]{0.48\linewidth}
        \centering
        \includegraphics[width=\linewidth]{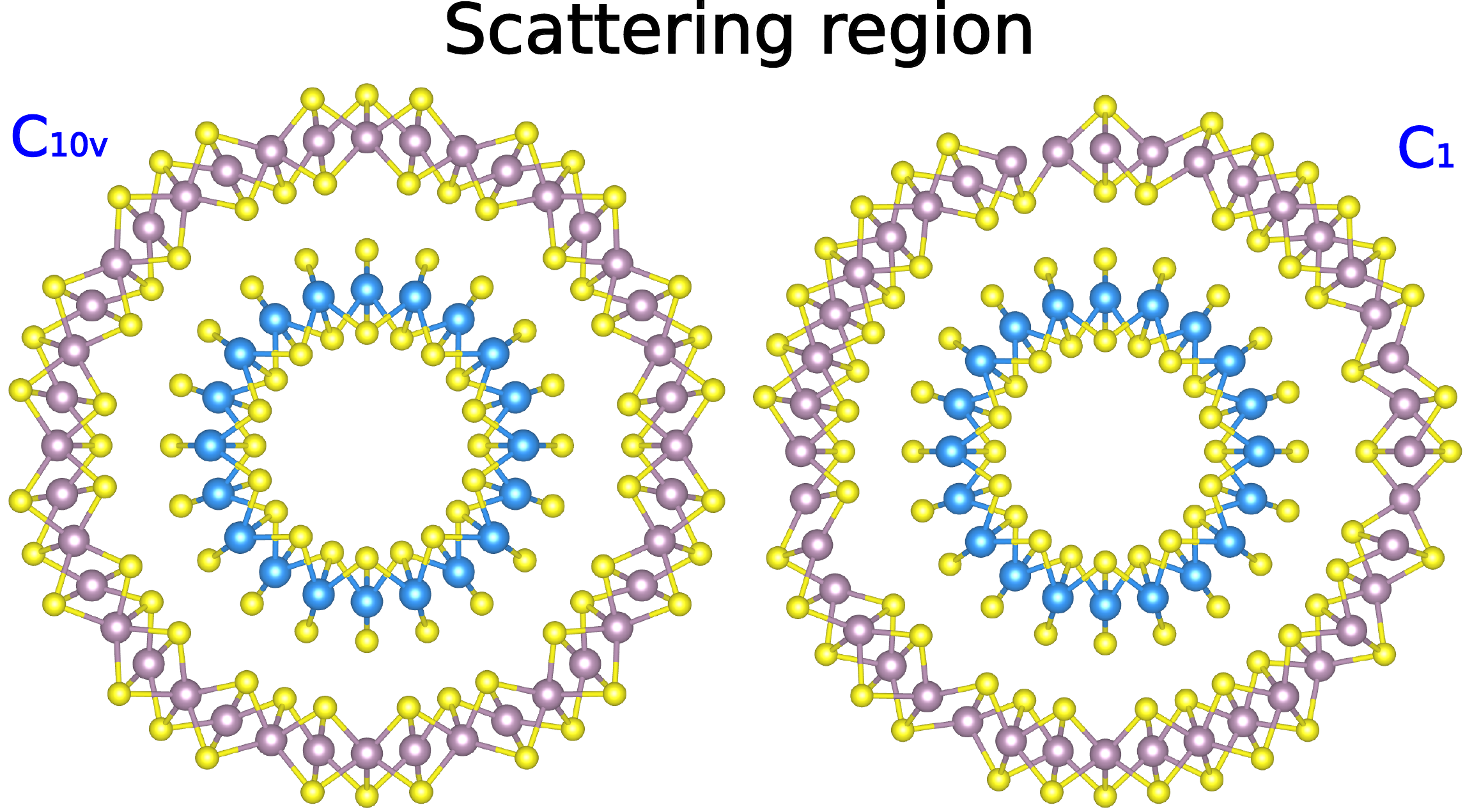 }
        \caption{Cross-sectional views of the two defect-laden configurations with different symmetries.}
    \end{subfigure}
    \hfill
    \begin{subfigure}[b]{0.48\linewidth}
        \centering
        \includegraphics[width=\linewidth]{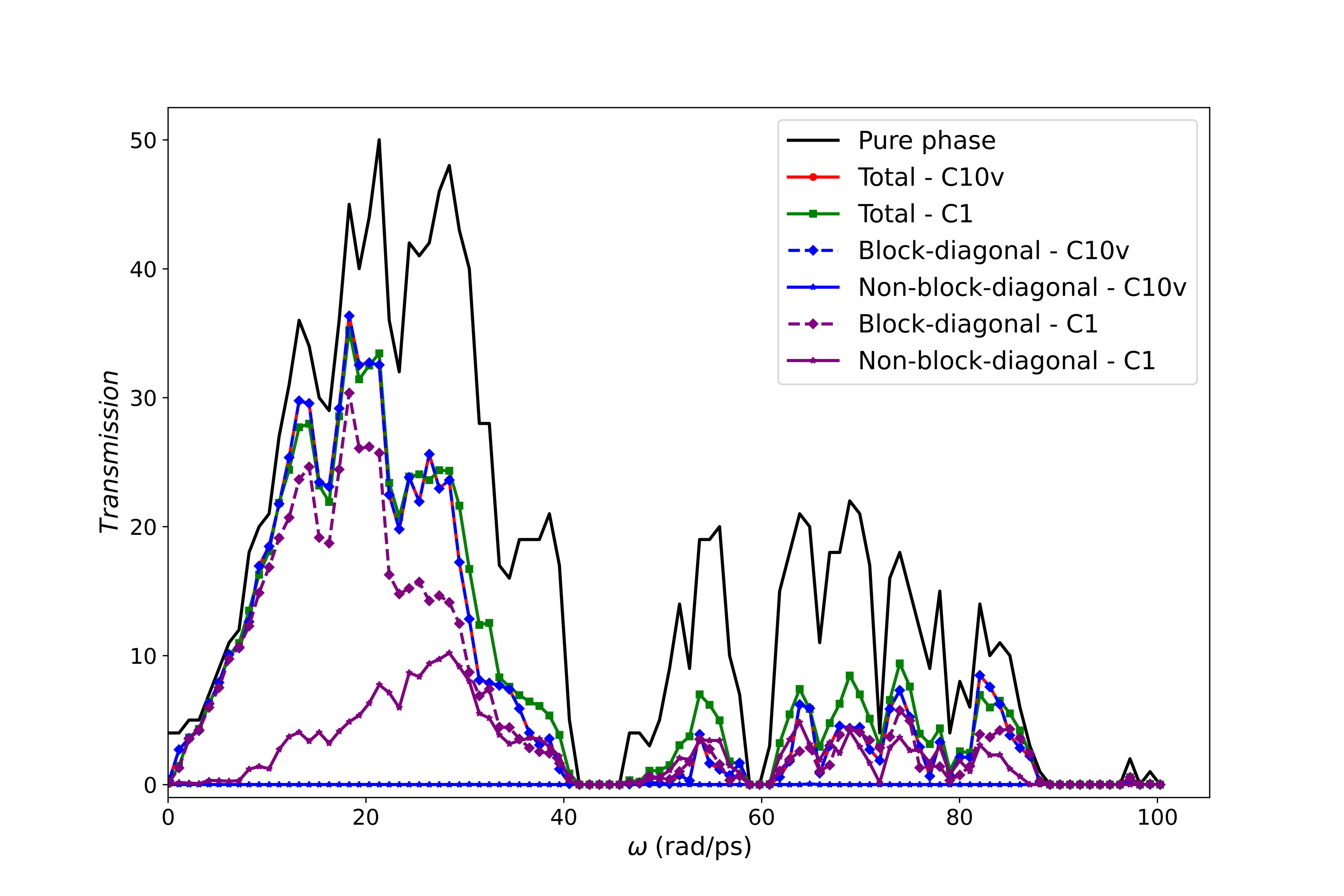}
        \caption{Contributions to the transmission from diagonal and off-diagonal blocks in the transmission matrix for different pristine or defect-laden configurations.}
    \end{subfigure}
    \vskip\baselineskip
        \begin{subfigure}{0.6\textwidth}
    \centering
    \includegraphics[width=1\linewidth]{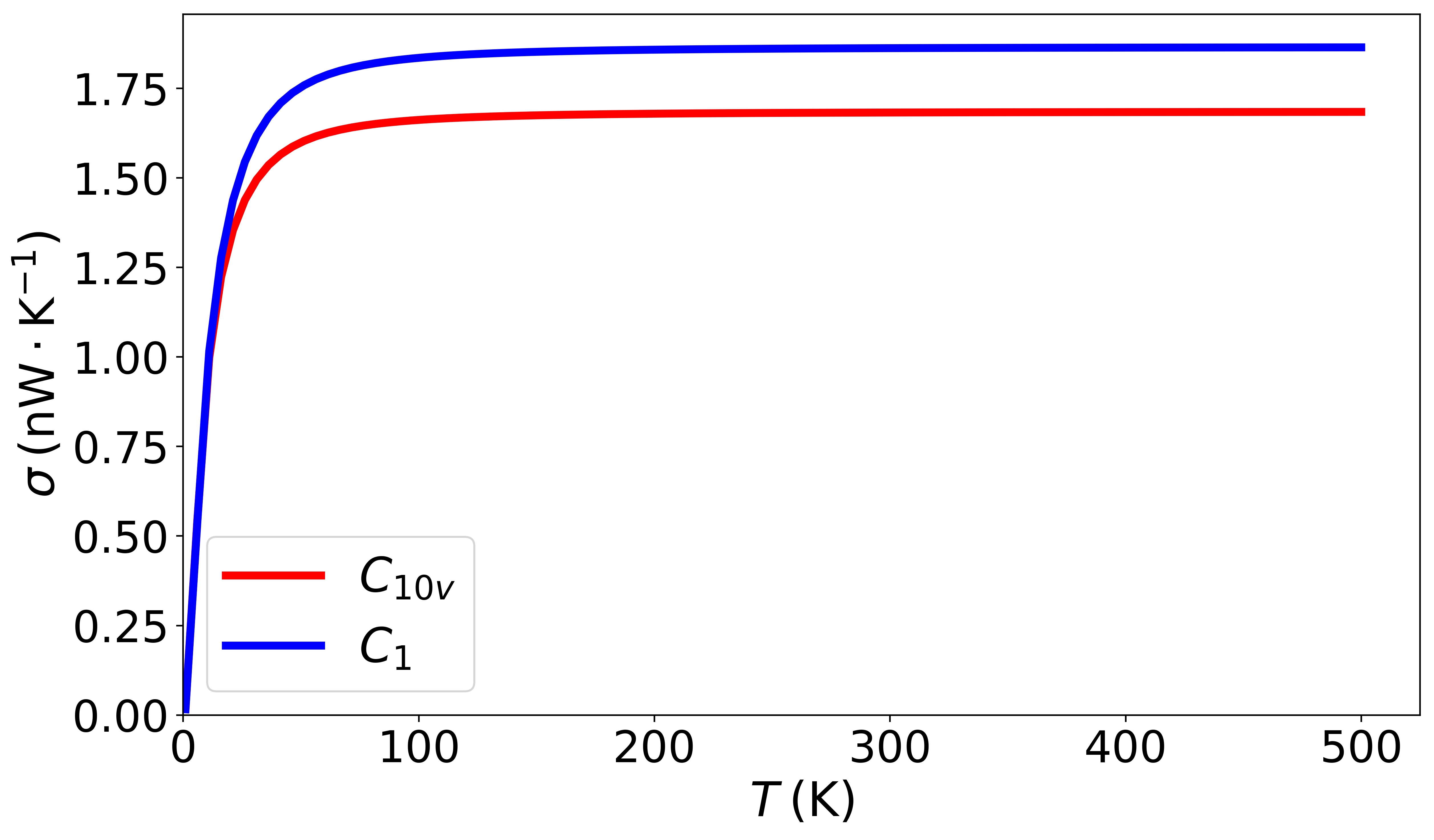}
    \caption{Temperature dependence of the thermal conductances for the two defect-laden structures.}
    \end{subfigure}
    \caption{Additional case 8: 10 S atoms are removed from the outermost and inner layer individually connected to Mo atoms to form the $C_{10v}$ and $C_{1}$ configurations.}
    \label{fig:case8}
\end{figure}

As for the cases involving S vacancies, some configurations lead to severe distortion or even structural collapse. We only show the examples that can be stably maintained under the MLIP optimization. In the case shown in Fig.~\ref{fig:case6}, 10 S atoms connected to Mo atoms are removed from the outermost layer; the difference in thermal conductance between A and B is not significant.
In the case shown in Fig.~\ref{fig:case7}, when 10 S atoms connected to Mo atoms are removed from the inner layer, the thermal conductance does not change significantly either.
When 10 S atoms are removed individually from the outermost and inner layers, as shown in Fig.~\ref{fig:case8}, the gap in thermal conductance between the $C_{10v}$ and $C_{1}$ configurations increases further.

Overall, in the (10,0)-(20,0) \ce{WS2}-\ce{MoS2} double-layer nanotube, heat transfer is mainly influenced by the outer \ce{MoS2} layer. Moreover, the effects of symmetry become significant only when defect density is high and transmission is reduced to a noticeable degree. However, in every case the higher-symmetry configuration has a lower conductance (even if only slightly so) and that discrepancy can be traced to the contribution of the off-diagonal blocks to the transmission, supporting the conclusions presented in the main text.

\section{Relaxation of the double-layer structure}
\begin{figure}[htbp]
    \centering
    \begin{subfigure}[b]{0.8\linewidth}
        \centering
        \includegraphics[width=\linewidth]{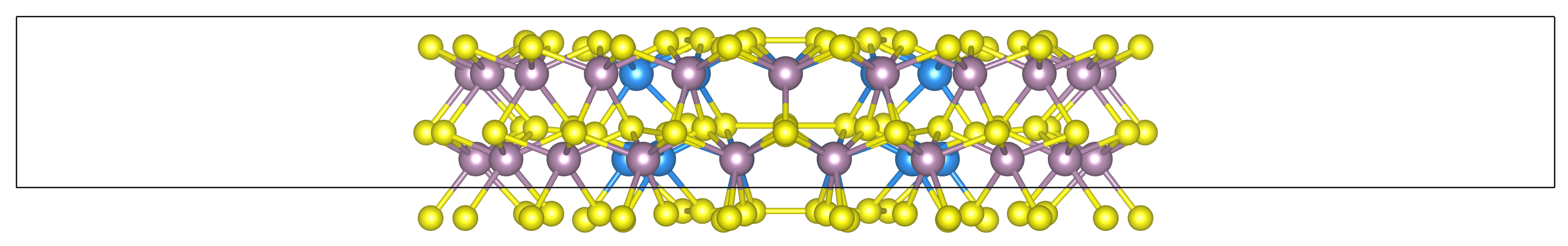}
        \caption{Before DFT relaxation}
        \label{fig:poscar}
    \end{subfigure}
    \hfill
    \begin{subfigure}[b]{0.8\linewidth}
        \centering
        \includegraphics[width=\linewidth]{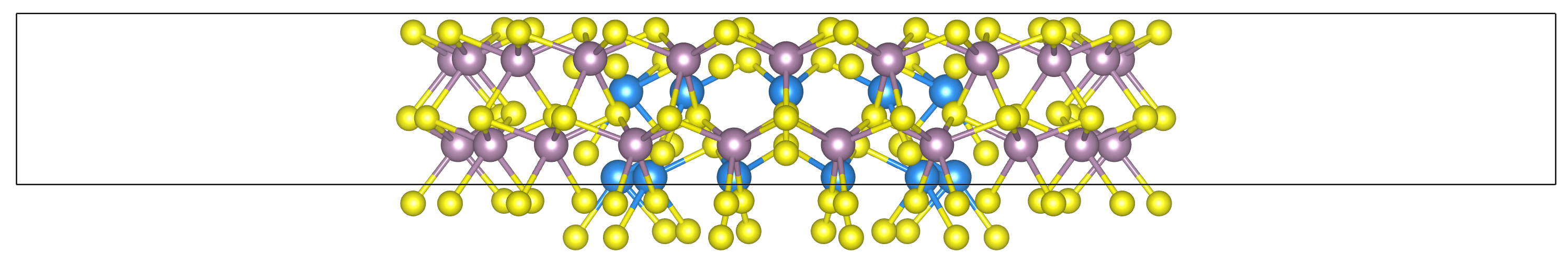}
        \caption{DFT-optimized structure}
        \label{fig:contcar}
    \end{subfigure}
    \caption{View of the double-walled nanotube before and after relaxing the atomic positions.}
    \label{fig:relaxation}
\end{figure}

All calculations for the double-walled nanotube were performed on a relaxed structure, obtained using the DFT code and parameters discussed in the main text. As shown in Fig.~\ref{fig:relaxation}, a relative displacement between the inner and outer nanotube along the long axis of the structure as a result of energy minimization was detected.

\section{Performance of the Allegro MLIP on structures from finite-T trajectories}
\FloatBarrier

\begin{figure}[h]
    \centering
    \includegraphics[width=0.75\linewidth]{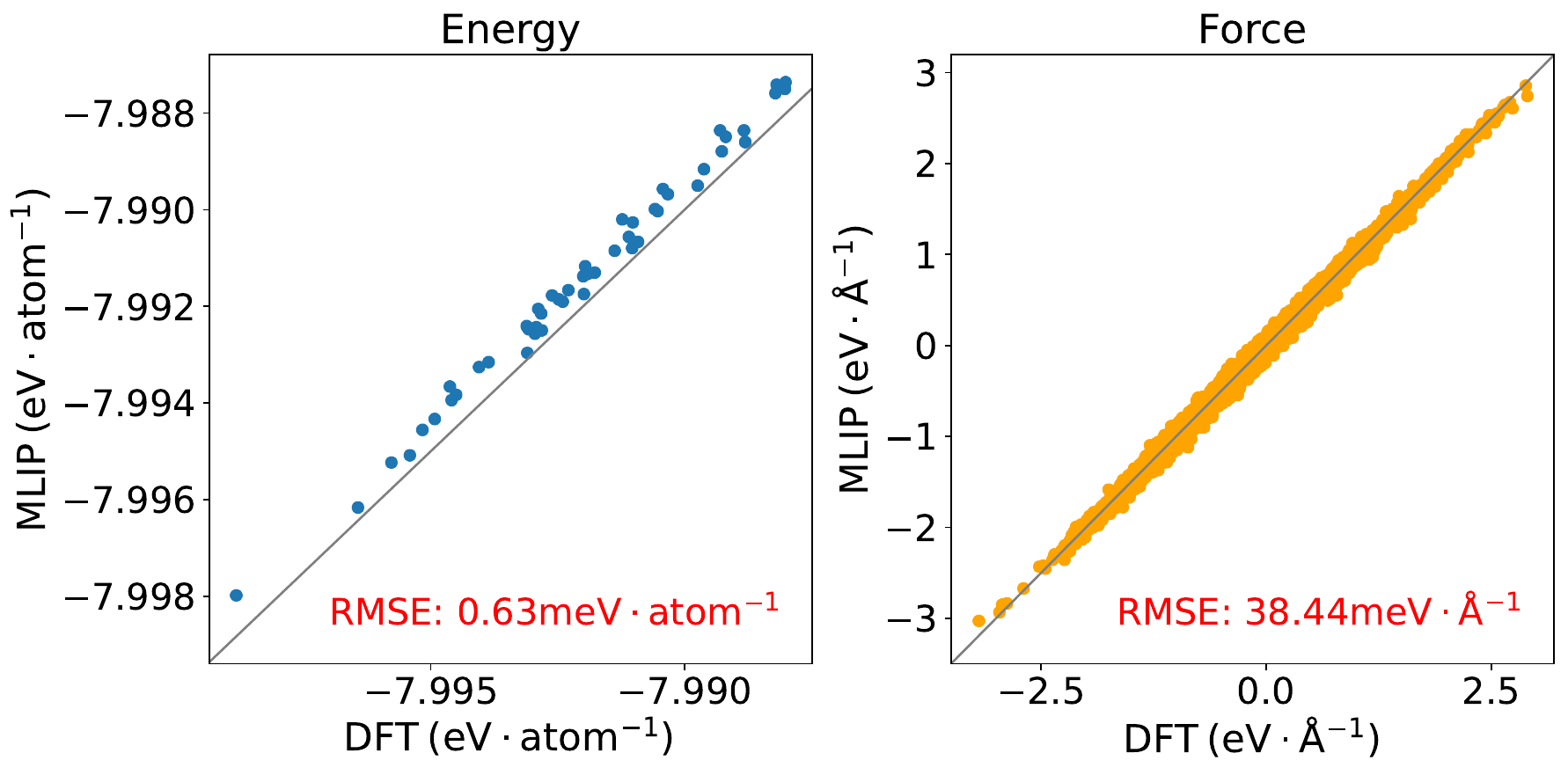}
    \caption{Potential energy and forces for the structures sampled from the room-temperature (\SI{300}{\kelvin}) MD trajectory structures of the pristine double-walled \ce{WS2}-\ce{MoS2} nanotube: MLIP vs DFT.}
    \label{fig:parity_pristine}
\end{figure}

\begin{figure}[h]
    \centering
    \includegraphics[width=0.75\linewidth]{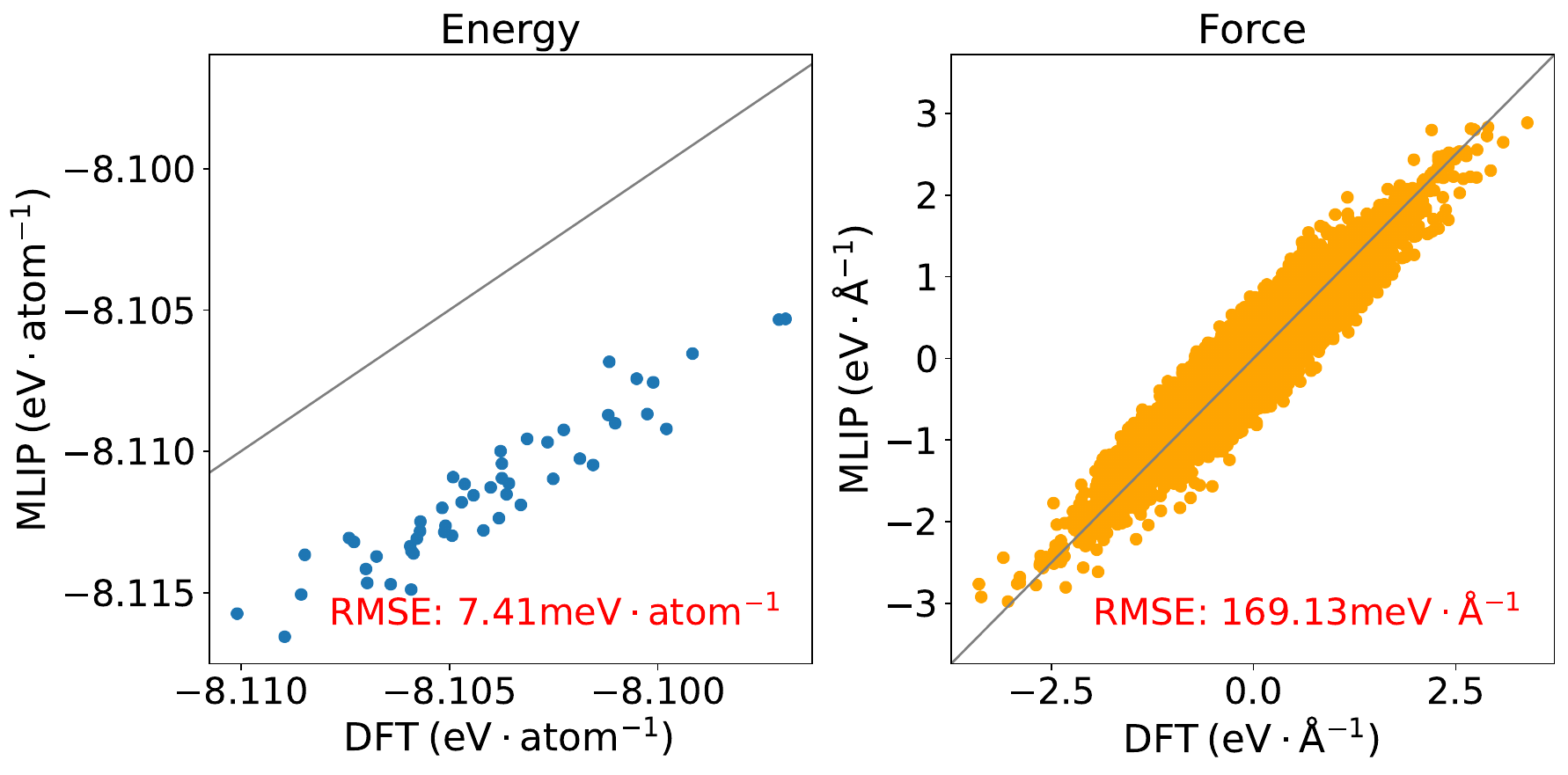}
    \caption{Potential energy and forces for the structures sampled from the room-temperature (\SI{300}{\kelvin}) MD trajectory structures of the defect-laden double-walled \ce{WS2}-\ce{MoS2} nanotube with C10v symmetry: MLIP vs DFT.}
    \label{fig:parity_defect_laden}
\end{figure}

Using the pristine $C_{10v}$ defect-laden structures of the double-walled \ce{WS2}-\ce{MoS2} nanotube presented in the main text as initial conditions, we generated MD trajectories at \SI{300}{\kelvin} with an integration time step of \SI{1}{\femto\second}. Starting with random velocities from a Maxwell-Boltzmann distribution, we generated trajectories in the NVT ensemble using the Langevin thermostat for 10000 steps. 50 structures were sampled evenly between 5000 and 10000 steps for each of the two structures. We then calculated the potential energies and forces for each of those configurations using the MLIP and a direct DFT run. The results for the pristine structure are shown in Fig.~\ref{fig:parity_pristine}, with those for the defect-laden case displayed in Fig.~\ref{fig:parity_defect_laden}.

\section{Green-Kubo simulations to obtain the thermal conductivity}

\begin{figure}[htbp]
    \centering
    \begin{subfigure}[b]{0.48\linewidth}
        \centering
        \includegraphics[width=\linewidth]{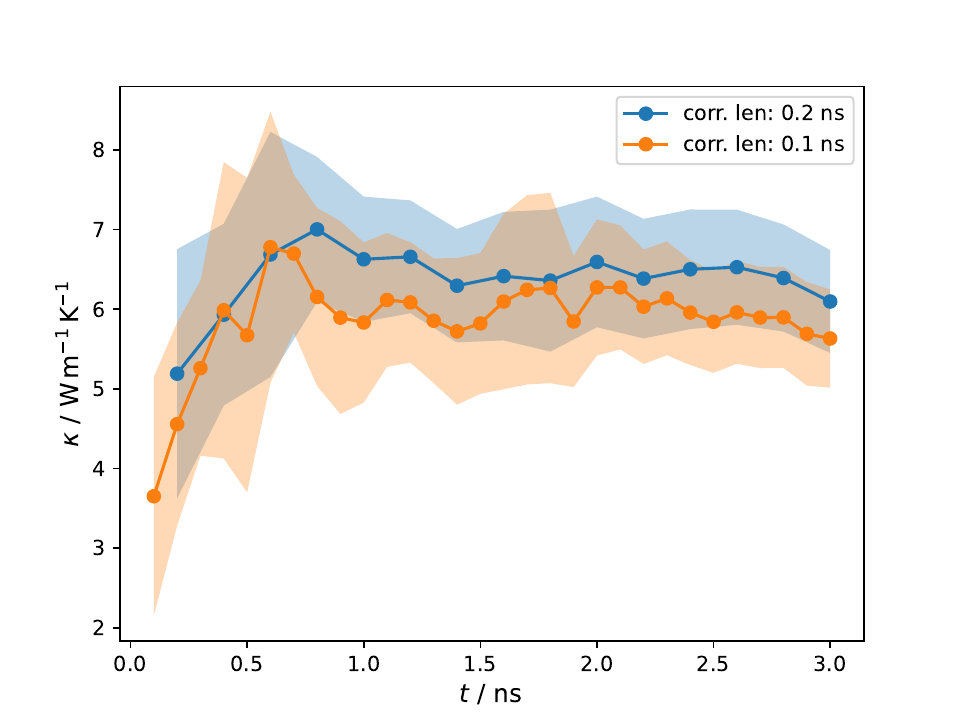}
        \caption{Time convergence for the double-walled nanotube system containing the $C_{10v}$-symmetric defect.}
    \end{subfigure}
    \hfill
    \begin{subfigure}[b]{0.48\linewidth}
        \centering
        \includegraphics[width=\linewidth]{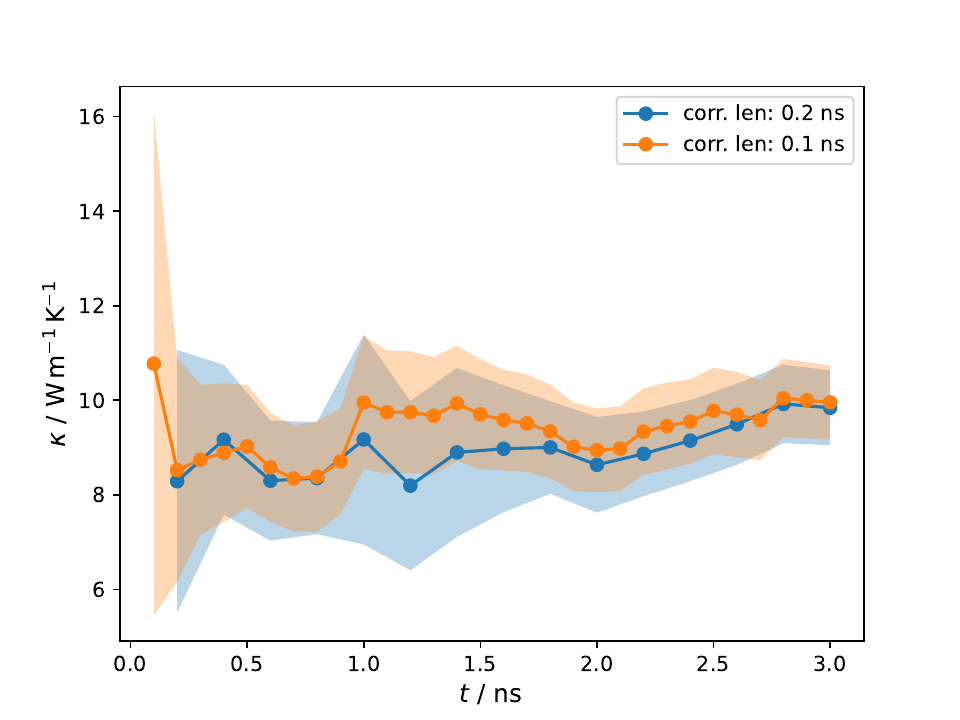}
        \caption{Time convergence for the double-walled nanotube system containing the asymmetric defect.}
    \end{subfigure}
    \vskip\baselineskip
    \begin{subfigure}[b]{0.48\linewidth}
        \centering
        \includegraphics[width=\linewidth]{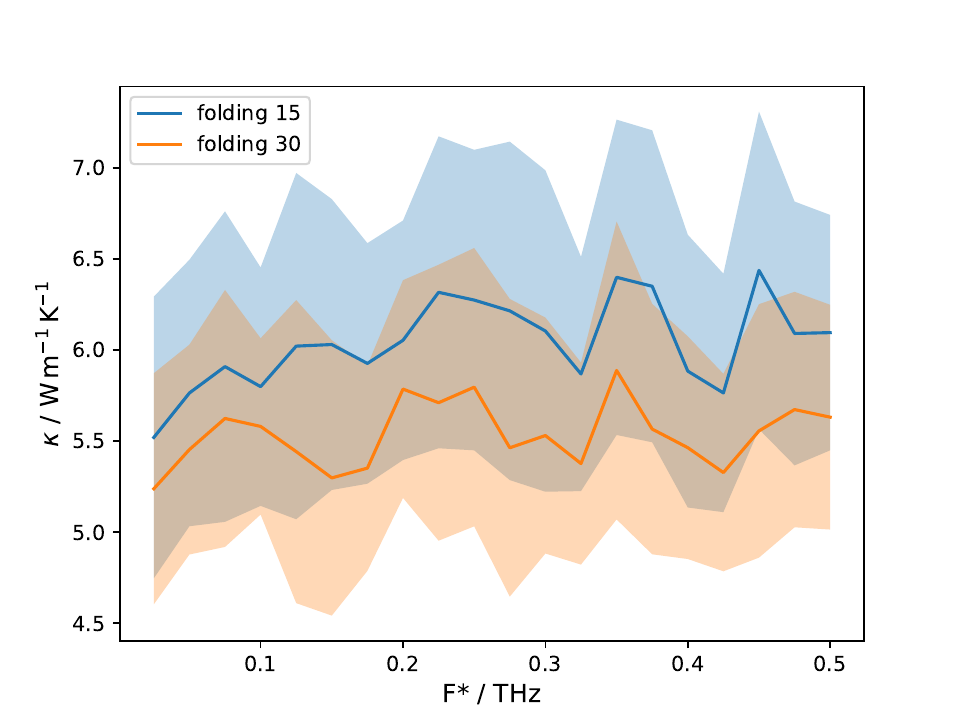}
        \caption{Differences based on the choice of cutoff for the double-walled nanotube system containing the $C_{10v}$-symmetric defect.}
    \end{subfigure}
    \hfill
    \begin{subfigure}[b]{0.48\linewidth}
        \centering
        \includegraphics[width=\linewidth]{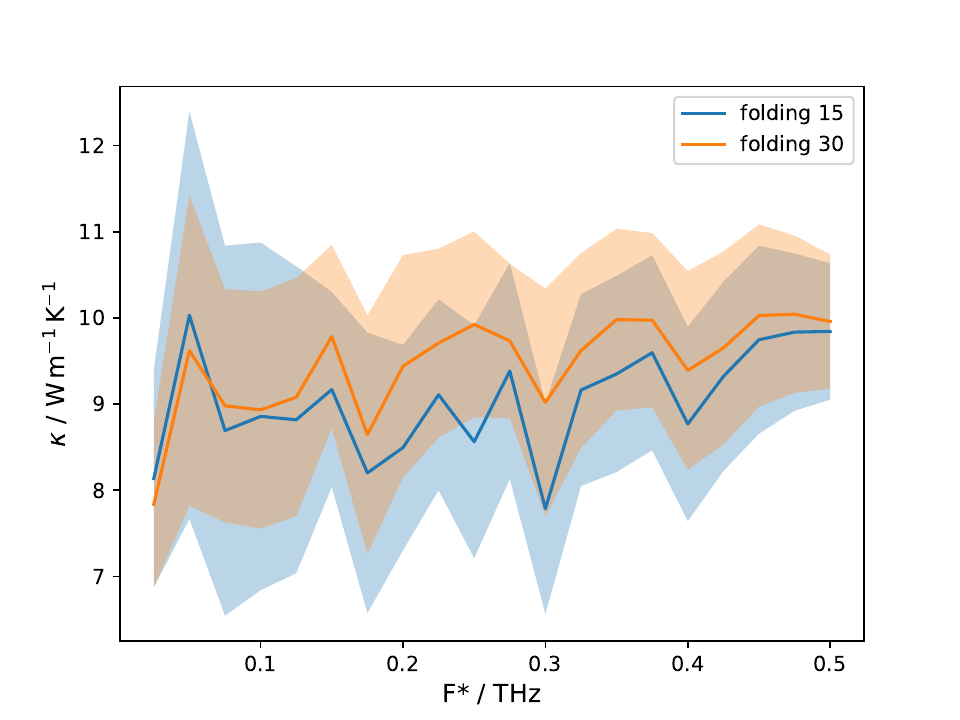}
        \caption{Differences based on the choice of cutoff for the double-walled nanotube system containing the asymmetric defect.}
    \end{subfigure}
    \caption{Convergence figures for the thermal conductivity from Green-Kubo simulations.}
    \label{fig:gk_convergence}
\end{figure}

Figure \ref{fig:gk_convergence} shows the convergence behavior of the Green-Kubo simulations. The simulations were performed on the double-walled nanotubes, employing supercells containing $5$ layers of their unit cell, where one of the layers contained the symmetric or asymmetric defect configurations equivalent to those used for the transmission calculations. This led to a lattice parameter in the periodic direction of \SI{27.73}{\angstrom} and to a simulation cell with \SI{900}{atoms}. The autocorrelation functions were averaged over the 15 independent simulations (correlation time of \SI{200}{\pico\second}) or for twice that number after splitting each trajectory into two pieces (correlation time of \SI{100}{\pico\second}). For the final values in the main manuscript, a correlation time of \SI{200}{\pico\second} was chosen. The same correlation-time dependence is reflected in the cutoff analysis in Fig. \ref{fig:gk_convergence}c and d, where the number of folds indicates over how many pieces the averaging was performed over. The results are given for the entire simulation time and it is clearly observed that the choice of cutoff is of limited significance within the predicted error. The cutoff was set to low values, as the objective is to obtain the denoised value at a frequency of \SI{0}{\tera\hertz}.

\FloatBarrier